\documentclass[showpacs,preprintnumbers,aps,amsmath,amssymb]{revtex4}
\usepackage{doublespace}
\begin{document}
\bibliographystyle{try}

\newcounter{univ_counter}
\setcounter{univ_counter} {0}

\addtocounter{univ_counter} {1} 
\edef\JLAB{$^{\arabic{univ_counter}}$ } 

\addtocounter{univ_counter} {1} 
\edef\FIU{$^{\arabic{univ_counter}}$ } 

\addtocounter{univ_counter} {1} 
\edef\ANL{$^{\arabic{univ_counter}}$ } 

\addtocounter{univ_counter} {1} 
\edef\ASU{$^{\arabic{univ_counter}}$ } 

\addtocounter{univ_counter} {1} 
\edef\SACLAY{$^{\arabic{univ_counter}}$ } 

\addtocounter{univ_counter} {1} 
\edef\CSU{$^{\arabic{univ_counter}}$ } 

\addtocounter{univ_counter} {1} 
\edef\CMU{$^{\arabic{univ_counter}}$ } 

\addtocounter{univ_counter} {1} 
\edef\CUA{$^{\arabic{univ_counter}}$ } 

\addtocounter{univ_counter} {1} 
\edef\CNU{$^{\arabic{univ_counter}}$ } 

\addtocounter{univ_counter} {1} 
\edef\UCONN{$^{\arabic{univ_counter}}$ } 

\addtocounter{univ_counter} {1} 
\edef\EDINBURGH{$^{\arabic{univ_counter}}$ } 

\addtocounter{univ_counter} {1} 
\edef\FU{$^{\arabic{univ_counter}}$ } 

\addtocounter{univ_counter} {1} 
\edef\FSU{$^{\arabic{univ_counter}}$ } 

\addtocounter{univ_counter} {1} 
\edef\INFNFR{$^{\arabic{univ_counter}}$ } 

\addtocounter{univ_counter} {1} 
\edef\INFNGE{$^{\arabic{univ_counter}}$ } 

\addtocounter{univ_counter} {1} 
\edef\GWU{$^{\arabic{univ_counter}}$ } 

\addtocounter{univ_counter} {1} 
\edef\GLASGOW{$^{\arabic{univ_counter}}$ } 

\addtocounter{univ_counter} {1} 
\edef\ISU{$^{\arabic{univ_counter}}$ } 

\addtocounter{univ_counter} {1} 
\edef\ITEP{$^{\arabic{univ_counter}}$ } 

\addtocounter{univ_counter} {1} 
\edef\JMU{$^{\arabic{univ_counter}}$ } 

\addtocounter{univ_counter} {1} 
\edef\KYUNGPOOK{$^{\arabic{univ_counter}}$ } 

\addtocounter{univ_counter} {1} 
\edef\MOSCOW{$^{\arabic{univ_counter}}$ } 

\addtocounter{univ_counter} {1} 
\edef\MSU{$^{\arabic{univ_counter}}$ } 

\addtocounter{univ_counter} {1} 
\edef\UNH{$^{\arabic{univ_counter}}$ } 

\addtocounter{univ_counter} {1} 
\edef\NSU{$^{\arabic{univ_counter}}$ } 

\addtocounter{univ_counter} {1} 
\edef\ODU{$^{\arabic{univ_counter}}$ } 

\addtocounter{univ_counter} {1} 
\edef\OHIOU{$^{\arabic{univ_counter}}$ } 

\addtocounter{univ_counter} {1} 
\edef\ORSAY{$^{\arabic{univ_counter}}$ } 

\addtocounter{univ_counter} {1} 
\edef\RPI{$^{\arabic{univ_counter}}$ } 

\addtocounter{univ_counter} {1} 
\edef\SCAROLINA{$^{\arabic{univ_counter}}$ } 

\addtocounter{univ_counter} {1}
\edef\CHILE{$^{\arabic{univ_counter}}$ }

\addtocounter{univ_counter} {1} 
\edef\UNION{$^{\arabic{univ_counter}}$ } 

\addtocounter{univ_counter} {1} 
\edef\VIRGINIA{$^{\arabic{univ_counter}}$ } 

\addtocounter{univ_counter} {1} 
\edef\YEREVAN{$^{\arabic{univ_counter}}$ }

\boldmath
\title{\large Beam-Recoil Polarization Transfer in the Nucleon
Resonance Region in the Exclusive $\vec{e}p \to e'K^+\vec{\Lambda}$ and
$\vec{e}p \to e'K^+\vec{\Sigma}^0$ Reactions at CLAS}
\unboldmath

%%%%%%%%%%%%%%% authors %%%%%%%%% 

\author{
D.S.~Carman,\JLAB\
B.A.~Raue,\FIU\
K.P.~Adhikari,\ODU\
M.J.~Amaryan,\ODU\
M.~Anghinolfi,\INFNGE\
H.~Baghdasaryan,\VIRGINIA\ 
M.~Battaglieri,\INFNGE\
M.~Bellis,\CMU\
A.S.~Biselli,\FU\
C.~Bookwalter,\FSU\
D.~Branford,\EDINBURGH\
W.J.~Briscoe,\GWU\
W.K.~Brooks,\CHILE$\!\!^,$\JLAB\
V.D.~Burkert,\JLAB\
P.L.~Cole,\ISU$\!\!^,$\JLAB\
P.~Collins,\ASU\
V.~Crede,\FSU\
A.~Daniel,\OHIOU\
N.~Dashyan,\YEREVAN\
R.~De~Vita,\INFNGE\
E.~De~Sanctis,\INFNFR\
A.~Deur,\JLAB\
S.~Dhamija,\FIU\
C.~Djalali,\SCAROLINA\
G.E.~Dodge,\ODU\
P.~Eugenio,\FSU\
G.~Fedotov,\MOSCOW\
S.~Fegan,\GLASGOW\
A.~Fradi,\ORSAY\
M.Y.~Gabrielyan,\FIU\
K.L.~Giovanetti,\JMU\
F.X.~Girod,\JLAB\ 
W.~Gohn,\UCONN\
A.~Gonenc,\FIU\
R.W.~Gothe,\SCAROLINA\
H.~Hakobyan,\CHILE$\!\!^,$\YEREVAN\
C.~Hanretty,\FSU\
N.~Hassall,\GLASGOW\
K.~Hicks,\OHIOU\
M.~Holtrop,\UNH\
Y. Ilieva,\SCAROLINA\
D.G.~Ireland,\GLASGOW\
H.S.~Jo,\ORSAY\
J.R.~Johnstone,\GLASGOW\
P.~Khetarpal,\RPI\
W.~Kim,\KYUNGPOOK\
V.~Kubarovsky,\JLAB\
V.~Kuznetsov,\KYUNGPOOK\
K.~Livingston,\GLASGOW\
M.~Mayer,\ODU\
M.E.~McCracken,\CMU\
C.A.~Meyer,\CMU\
K.~Mikhailov,\ITEP\
T.~Mineeva,\UCONN\
M.~Mirazita,\INFNFR\
V.~Mokeev,\MOSCOW$\!\!^,$\JLAB\
B.~Moreno,\ORSAY\
K.~Moriya,\CMU\
M.~Moteabbed, \FIU\
P.~Nadel-Turonski,\CUA\
S.~Niccolai,\ORSAY\
M.R.~Niroula,\ODU\
M.~Osipenko,\INFNGE$\!\!^,$\MOSCOW\
A.I.~Ostrovidov,\FSU\
K.~Park,\JLAB$\!\!^,$\SCAROLINA\
S.~Park,\FSU\
E.~Pasyuk,\ASU\
O.~Pogorelko,\ITEP\
J.W.~Price,\CSU\
D.~Protopopescu,\GLASGOW\
G.~Ricco,\INFNGE\
M.~Ripani,\INFNGE\
B.G.~Ritchie,\ASU\
G.~Rosner,\GLASGOW\
P.~Rossi,\INFNFR\
F.~Sabati\'e,\SACLAY\
M.S.~Saini,\FSU\
C.~Salgado,\NSU\
D.~Sayre,\OHIOU\
D.~Schott,\FIU\
R.A.~Schumacher,\CMU\
H.~Seraydaryan,\ODU\
Y.G.~Sharabian,\JLAB\
D.I.~Sober,\CUA\
D.~Sokhan,\EDINBURGH\
S.~Stepanyan,\JLAB\
S.S.~Stepanyan,\KYUNGPOOK\
S.~Strauch,\SCAROLINA\
M.~Taiuti,\INFNGE\
D.J.~Tedeschi,\SCAROLINA\
S.~Tkachenko,\ODU\
M.~Ungaro,\UCONN\
M.F.~Vineyard,\UNION\
E.~Wolin,\JLAB\
M.H.~Wood,\SCAROLINA\
J.~Zhang,\ODU\
B.~Zhao,\UCONN\
\\
(CLAS Collaboration)
}

\affiliation{\JLAB Thomas Jefferson National Accelerator Laboratory, 
Newport News, Virginia 23606}
\affiliation{\FIU Florida International University, Miami, Florida 33199}
\affiliation{\ANL Argonne National Laboratory, Argonne, Illinois 60439}
\affiliation{\ASU Arizona State University, Tempe, Arizona 85287}
\affiliation{\SACLAY CEA-Saclay, DAPNIA-SPhN, F91191 Gif-sur-Yvette Cedex, 
France}
\affiliation{\CSU California State University, Dominguez Hills, Carson, CA 
90747}
\affiliation{\CMU Carnegie Mellon University, Pittsburgh, Pennsylvania 15213}
\affiliation{\CUA Catholic University of America, Washington, D.C. 20064}
\affiliation{\CNU Christopher Newport University, Newport News, Virginia 23606}
\affiliation{\UCONN University of Connecticut, Storrs, Connecticut 06269}
\affiliation{\EDINBURGH Edinburgh University, Edinburgh EH9 3JZ, United 
Kingdom}
\affiliation{\FU Fairfield University, Fairfield CT 06824}
\affiliation{\FSU Florida State University, Tallahasee, Florida 32306}
\affiliation{\INFNFR INFN, Laboratori Nazionali di Frascati, 00044 Frascati, Italy}
\affiliation{\INFNGE INFN, Sezione di Genova and Dipartimento di Fisica, 
Universit\`a di Genova, 16146 Genova, Italy}
\affiliation{\GWU The George Washington University, Washington, DC 20052}
\affiliation{\GLASGOW University of Glasgow, Glasgow G12 8QQ, United Kingdom}
\affiliation{\ISU Idaho State University, Pocatello, Idaho 83209}
\affiliation{\ITEP Institute of Theoretical and Experimental Physics, Moscow, 
117259, Russia}
\affiliation{\JMU James Madison University, Harrisonburg, Virginia 22807}
\affiliation{\KYUNGPOOK Kyungpook National University, Daegu 702-701, South 
Korea}
\affiliation{\MOSCOW Skobeltsyn Nuclear Physics Institute, 119899 Moscow, Russia}
\affiliation{\MSU Moscow State University, 119899 Moscow, Russia}
\affiliation{\UNH University of New Hampshire, Durham, New Hampshire 03824}
\affiliation{\NSU Norfolk State University, Norfolk, Virginia 23504}
\affiliation{\ODU Old Dominion University, Norfolk, Virginia 23529}
\affiliation{\OHIOU Ohio University, Athens, Ohio  45701}
\affiliation{\ORSAY Institut de Physique Nucleaire d'ORSAY, IN2P3, BP1, 
91406 Orsay, France}
\affiliation{\RPI Rensselaer Polytechnic Institute, Troy, New York 12180}
\affiliation{\SCAROLINA University of South Carolina, Columbia, South 
Carolina 29208}
\affiliation{\CHILE Universidad T\'ecnica Federico Santa Mar\'ia, Valparaiso, Chile}
\affiliation{\UNION Union College, Schenectady, NY 12308}
\affiliation{\VIRGINIA University of Virginia, Charlottesville, Virginia 22901}
\affiliation{\YEREVAN Yerevan Physics Institute, 375036 Yerevan, Armenia}

\date{\today}

\begin{abstract}

Beam-recoil transferred polarizations for the exclusive $\vec{e}p \to e'K^+ 
\vec{\Lambda},\vec{\Sigma}^0$ reactions have been measured using the CLAS 
spectrometer at Jefferson Laboratory. New measurements have been completed 
at beam energies of 4.261 and 5.754~GeV that span a range of momentum transfer 
$Q^2$ from 0.7 to 5.4~GeV$^2$, invariant energy $W$ from 1.6 to 2.6~GeV, and 
the full center-of-mass angular range of the $K^+$ meson.  These new data add 
to the existing CLAS $K^+\Lambda$ measurements at 2.567~GeV, and provide the 
first-ever data for the $K^+\Sigma^0$ channel in electroproduction. Comparisons 
of the data with several theoretical models are used to study the sensitivity 
to $s$-channel resonance contributions and the underlying reaction mechanism. 
Interpretations within two semi-classical partonic models are made to probe the 
underlying reaction mechanism and the $s\bar{s}$ quark-pair creation dynamics.
\end{abstract}

\pacs{13.88.+e, 14.40.aq, 14.20.Gk, 14.20.Jn}

\maketitle

%
%%%%%%%%%%%%%%%%%%%%%%%%%%%%%%%%%%%%%%%%%%%%%%%%%%%%%%%%%%%%%%%%%%%%%%%%
%
\section{INTRODUCTION}
\label{sec:intro}

An important requirement to better understand the structure of the nucleon 
is to map out its spectrum of excited states.  However, deciphering the 
data to understand the resonance excitations has been limited both by the 
data itself and the current state of existing theories.  Ideally we should 
expect the fundamental theory of the strong interaction, quantum 
chromodynamics (QCD), to provide a prediction for the nucleon excitation 
spectrum.  However, due to the non-perturbative nature of QCD at the relevant 
energies, this idea has not yet been fully realized.  Thus we have looked 
instead to effective models of QCD, such as constituent quark models, to 
gain some insight.  Present quark model calculations of the nucleon spectrum 
have predicted more states than have been seen experimentally~\cite{capstick}.  
This has been termed the ``missing'' resonance problem, and the existence of 
these states is tied in directly with the underlying degrees of freedom of the 
nucleon that govern hadronic production at moderate energies~\cite{isgur}.

Most of our current understanding of nucleon resonances comes from reactions
involving pions in the initial and/or final states.  Koniuk and Isgur
suggested that the missing states might be revealed in decays to channels 
where mesons other than pions or multiple pions are in the final state
\cite{kon_isg}.  Indeed, there are indications from theory that some 
missing states have a similar probability of decaying into channels such as 
$\omega N$, $\eta N$, $\pi \pi N$, and $KY$ ($Y = \Lambda, \Sigma$) compared 
to the $\pi N$ channel~\cite{capstick2,capstick}.  As baryon resonances have 
large widths and are often overlapping, studies of different final states 
provide important complementary cross checks in quantitatively understanding 
the contributing amplitudes. 

In this work we study the electroproduction of strange final states. While 
electromagnetic production of $KY$ final states has a much lower cross 
section than hadronic production reactions, the use of an electromagnetic 
probe has a distinct advantage, namely that all electromagnetic quantities in 
the reaction amplitude can be straightforwardly expressed in the context of 
quantum electrodynamics.  Furthermore, in addition to the different coupling 
constants compared to the $\pi N$ channel (e.g. $g_{KNY}$ vs. $g_{\pi NN}$), 
the study of the exclusive production of $KY$ final states has another 
advantage in the search for missing resonances.  The higher masses of the 
kaon and hyperons, compared to their non-strange counterparts, kinematically 
favor a two-body decay mode for states with masses near 2~GeV.  Not only is
this situation advantageous from an experimental viewpoint, but this also 
happens to be the mass region where the majority of the missing resonance 
states are expected to exist~\cite{capstick}.  

Although the two ground-state hyperons have the same valence quark structure 
($uds$), they differ in isospin, such that intermediate $N^*$ resonances can 
decay strongly to $K\Lambda$ final states, while both $N^*$ and $\Delta^*$ 
decays can couple to $K\Sigma$ final states.  Existing studies of 
$N^* \to K\Lambda, K\Sigma$ and $\Delta^* \to K\Sigma$ decays have not yet 
provided extensive or precise information on the $N^*,\Delta^* \to KY$ 
couplings.  To date, the Particle Data Group (PDG) only lists four $N^*$ 
states with known couplings to $K\Lambda$ and no $N^*$ states are listed that 
couple to $K\Sigma$~\cite{pdg}; only a single $\Delta^*$ state is listed with 
coupling strength to $K\Sigma$.  The current landscape as given by the PDG for 
$N^*,\Delta^* \to KY$ is given in Table~\ref{landscape}.

%%%%%%%%%%%%%%%%%%%%%%%%%%%%%%%%%%%%%%%%%%%%%%%%%%%%%%%%%%%%%%%%%%%%%%%%%%%%%
\begin{table}[htbp]
\begin{center}
\begin{tabular} {c|c|c|c|c|c|c} \hline
\multicolumn{4} {c} {\boldmath $N^* \to KY$ \unboldmath} & 
\multicolumn{3} {|c} {\boldmath $\Delta^* \to K \Sigma$ \unboldmath} \\ \hline
State & Rating & B.R. ($K \Lambda$) & B.R. ($K \Sigma$) & State & Rating & 
B.R. ($K \Sigma$) \\ \hline
$N^*(1650)$ $S_{11}$ & **** & 3 -- 11\% & -- & $\Delta^*(1700)$ $D_{33}$ & **** & --    \\ 
$N^*(1675)$ $D_{15}$ & **** & $<$ 1\%   & -- & $\Delta^*(1750)$ $P_{31}$ & *    & --    \\    
$N^*(1680)$ $F_{15}$ & **** & --        & -- & $\Delta^*(1900)$ $S_{31}$ & **   & --    \\   
$N^*(1700)$ $D_{13}$ & ***  & $<$ 3\%   & -- & $\Delta^*(1905)$ $F_{35}$ & **** & --    \\  
$N^*(1710)$ $P_{11}$ & ***  & 5 -- 25\% & -- & $\Delta^*(1910)$ $P_{31}$ & **** & --    \\  
$N^*(1720)$ $P_{13}$ & ***  & 1 -- 15\% & -- & $\Delta^*(1920)$ $P_{33}$ & ***  & 2.1\% \\ 
$N^*(1900)$ $P_{13}$ & **   & 2.4\%     & -- & $\Delta^*(1930)$ $D_{35}$ & ***  & -- 	\\
$N^*(1990)$ $F_{17}$ & **   & --        & -- & $\Delta^*(1940)$ $D_{33}$ & *    & -- 	\\
$N^*(2000)$ $F_{15}$ & **   & --        & -- & $\Delta^*(1950)$ $F_{37}$ & **** & -- 	\\
                     &      &           &    & $\Delta^*(2000)$ $F_{35}$ & **   & --    \\ \hline
\end{tabular}
\end{center}
\caption{PDG listings for the coupling of $N^*$ ($\Delta^*$) states below 
2~GeV to $K\Lambda$ and $K\Sigma$ ($K\Sigma$)~\cite{pdg}.  The {\it Rating} 
column gives the PDG star rating for the $N^*$ states and {\it B.R.} 
indicates the branching ratio.}
\label{landscape}
\end{table}
%%%%%%%%%%%%%%%%%%%%%%%%%%%%%%%%%%%%%%%%%%%%%%%%%%%%%%%%%%%%%%%%%%%%%%%%%%%%%

Theoretically, there has been considerable effort during the past two 
decades to develop models for $KY$ photo- and electroproduction. However, 
the present state of understanding is limited by a lack of precision data 
(Ref.~\cite{5st} contains a brief review).  Model fits to the cross section 
data are generally obtained at the expense of many free parameters, which 
makes it difficult to provide precise constraints. Moreover, cross section 
data alone are not sufficient to fully understand the reaction mechanism, 
as they represent only a portion of the full amplitude response.  In this 
regard, measurements of spin observables are essential for continued 
theoretical development in this field.  Fits to the limited available data 
lead to ambiguities and model dependence in interpreting the results.  
Polarization data can provide for improved constraints on the model parameters,
increasing their discriminatory power and allow for a quantitative measure 
of whether or not new resonance states are required to explain these and other 
hyperon production data.  One main issue involves discriminating resonant 
states from the non-resonant background and from effects caused by final-state 
interactions or channel-couplings instead of $N^*$ and $\Delta^*$ contributions
\cite{corthals}.

CLAS at Jefferson Laboratory (JLab) has provided photoproduction $K^+\Lambda$ 
and $K^+\Sigma^0$ recoil polarization data from the proton~\cite{mcnabb}. In 
addition, beam-recoil polarization transfer data from CLAS have been published 
for both $K^+\Lambda$ and $K^+\Sigma^0$ photoproduction~\cite{bradford1} and 
$K^+\Lambda$ electroproduction \cite{carman03} reactions on the proton. Data 
such as these that span both a wide energy and angular range and are essential to 
disentangle the resonant and non-resonant contributions to the $KY$ spectrum
\cite{corthals,nikonov}. This has been demonstrated in several recent 
amplitude-level analyses with channel couplings based on photoproduction data
\cite{ani05a,ani05b,penner,chiang,shklyar,diaz,sarantsev,anisovich}. Further 
progress is expected as data with broad coverage and smaller experimental 
uncertainties are made available (which includes new CLAS data with linearly 
polarized photon beams and polarized targets~\cite{g13,frost,hdice}). 

In this work, we focus on measurements of spin transfer from a 
longitudinally polarized electron beam to the ground-state hyperons produced 
in the reactions $p(\vec{e},e'K^+)\vec{\Lambda}$ and 
$p(\vec{e},e'K^+)\vec{\Sigma}^0$ at beam energies $E_b$ of 4.261 and 
5.754~GeV.  This work represents a higher-statistics follow-up to the first 
data presented by CLAS in the $K^+\Lambda$ channel for an electron beam 
energy of 2.567~GeV~\cite{carman03}, where the transferred $\Lambda$ 
polarization was studied as a function of the invariant energy $W$ and 
$\cos \theta_K^{c.m.}$ (the $K^+$ center-of-mass angle).  The transferred 
polarization data for the $K^+\Sigma^0$ final state included here represent 
the first-ever published data for this observable in electroproduction.  

From the polarization data in Ref.~\cite{carman03}, the ratio of the 
longitudinal to transverse structure functions $\sigma_L/\sigma_T$ for the 
$K^+\Lambda$ final state at $\theta_K^{c.m.}=0^{\circ}$ was extracted
for several $W$ points near 1.8~GeV and $Q^2 \sim$0.7~GeV$^2$~\cite{raue05}.
These results indicated a ratio that was systematically smaller than 
previously published results using a Rosenbluth separation performed in 
Hall~C at Jefferson Laboratory~\cite{mohring}, albeit with large statistical 
uncertainties.  In fact, the data were consistent with zero within the 
experimental uncertainties, which would imply a small longitudinal structure 
function, and hence, a small longitudinal coupling of the virtual photon.  The 
results of Ref.~\cite{raue05} are expanded upon in this work with larger data 
sets that reduce uncertainties in the extrapolation to 
$\theta_K^{c.m.}=0^\circ$.  The new data presented include three data points 
near $W$=1.9~GeV with an average $Q^2$ of $\sim$1.6~GeV$^2$ and three data 
points near $W$=2.0~GeV with an average $Q^2$ of $\sim$2.5~GeV$^2$.

Using a semi-classical partonic framework, the CLAS polarization data in 
Ref.~\cite{carman03} were shown to support a description where the spin 
properties of the quark-pair creation operator might be responsible for the 
observed trends in the $\Lambda$ polarization. This framework indicated that
the quark-pair creation operator dominating the reaction produces the 
$s\bar{s}$ pair with spins anti-aligned.  This finding, if confirmed, has 
important implications since many, if not most, calculations of hadron 
spectroscopy use a $^3\!P_0$ operator to calculate the transition to the 
final-state particles~\cite{barnes}.  In this work, the angular distribution 
of the transferred polarization is studied with greater precision than in 
Ref.~\cite{carman03} and compared against two semi-classical partonic models 
that lead to quite different predictions regarding the $K^+\Lambda$ reaction
mechanism and the quark-pair creation dynamics. The first is the model from 
Ref.~\cite{carman03}. The second assumes the reaction proceeds from an 
$s\bar{s}$ quark-pair with the quark spins aligned.  The main differences 
between the models are discussed and a possible experiment to discriminate 
between them is proposed.

The organization for the remainder of this paper is as follows.  In
Sections~\ref{sec:theory} and \ref{sec:formalism}, the theoretical models to 
be compared with the measurements are briefly introduced and the relevant 
formalism for the polarization measurements is provided.  
Section~\ref{sec:polext} gives a detailed description of how the polarization 
is extracted and Section~\ref{sec:analysis} gives details regarding the 
analysis cuts and corrections to the data. Section~\ref{systematics} details 
the sources of systematic uncertainty. Section~\ref{sec:results} contains 
the physics results, with the presentation of the $K^+\Lambda$ and
$K^+\Sigma^0$ polarization transfer data in Sections~\ref{lambda} and
\ref{sigma}, respectively, the new $\sigma_L/\sigma_T$ extraction in
Section~\ref{rat_ext}, and comparisons of the data to the newly developed
partonic models in Section~\ref{quark_models}.  Finally, we present a summary 
of this work and our conclusions in Section~\ref{sec:conclusions}.

%
%%%%%%%%%%%%%%%%%%%%%%%%%%%%%%%%%%%%%%%%%%%%%%%%%%%%%%%%%%%%%%%%%%%%%%%%%%%
%

\section{THEORETICAL MODELS}
\label{sec:theory}

While the QCD description of quark interactions and pair creation is well 
accepted at high energies, the situation is considerably more complex in
the low-energy nucleon resonance region due to the non-perturbative
nature of the theory.  In order to arrive at any theoretical expectations 
for the transferred polarization, effective models must be employed that 
ultimately represent approximations to QCD.  This analysis highlights three 
different theoretical model approaches.  The first is a traditional 
hadrodynamic model, the second is based on kaon Regge trajectory exchange, 
and the third is a hybrid Regge plus resonance approach.

\subsection{Hadrodynamic models}

\begin{figure}[htbp]
\vspace{4.0cm}
\includegraphics{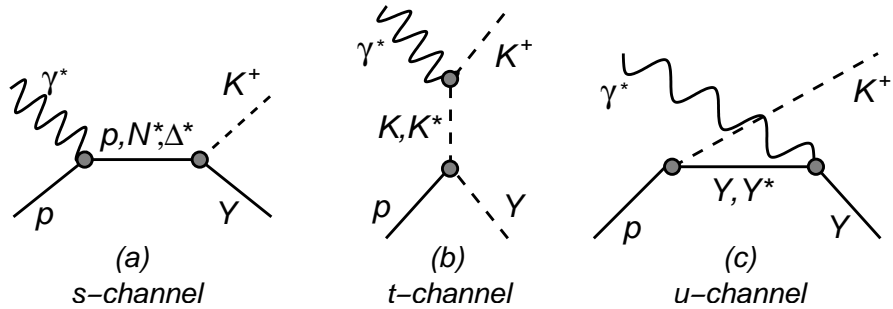}
\caption{Tree-level diagrams contributing to the $KY$ reactions: (a) 
$s$-channel exchanges, (b) $t$-channel exchanges, and (c) $u$-channel 
exchanges.}
\label{born}
\end{figure}

Hadrodynamic models provide a description of the reaction based on an
effective Lagrangian constructed from tree-level Born and extended Born
terms in the $s$, $t$, and $u$ reaction channels (see Fig.~\ref{born}).
The Born diagrams include the exchange of the proton, kaon, and
ground-state hyperons, while the extended Born diagrams include the
exchange of the associated excited states.  This description of the
interaction, which involves only first-order terms, is sensible as the
incident and outgoing electrons interact rather weakly with the hadrons.  
A complete description of the physics processes requires taking into
account all possible channels that could couple to the initial and final
states, but the advantages of the tree-level approach are to limit
complexity and to identify the dominant trends.  The drawback in this class
of models is the large number of hadrons that can contribute in the
intermediate state of the reaction.  Depending on which set of resonances
a given model includes, very different conclusions about the strengths of 
the contributing diagrams may be reached.

%%%%%%%%%%%%%%%%%%%%%%%%%%%%%%%%%%%%%%%%%%%%%%%%%%%%%%%%%%%%%%%%%%%%%%%%%
%%%%%%%%%%%%%%%%%%%%%%%%%%%%%%%%%%%%%%%%%%%%%%%%%%%%%%%%%%%%%%%%%%%%%%%%%

The hadrodynamic model employed in this work is from Mart and Bennhold
\cite{mart} (referred to here as MB).  The $s$-channel terms included in 
this model are listed in Table~\ref{ext_states}.  The coupling 
strengths have been determined by a simultaneous fit to low-energy 
$K^- p \to \gamma Y$ and $\gamma^{(*)} p \to K^+ Y$ data, by adding the 
non-resonant Born terms with a number of resonances, leaving the coupling 
constants as free parameters.  The coupling constants are required to respect 
the limits imposed by SU(3), allowing for a symmetry breaking at the level of 
about 20\%.  In this model, the inclusion of hadronic form factors leads to a 
breaking of gauge invariance that is restored by the inclusion of counter 
terms~\cite{mart}.  The model has been compared to the CLAS photoproduction
\cite{mcnabb,bradford} and electroproduction data~\cite{5st} and provides for 
a fair description of those results, although no CLAS data were employed in 
the model fits.

%%%%%%%%%%%%%%%%%%%%%%%%%%%%%%%%%%%%%%%%%%%%%%%%%%%%%%%%%%%%%%%%%%%%%%%%%%%%%%
\begin{table}[htbp]
\begin{center}
\begin{tabular} {c|c|c|c|c} \hline
      & \multicolumn{2} {c|} {MB} & \multicolumn{2} {|c} {RPR} \\ \cline{2-5} 
State & $K^+ \Lambda$ & $K^+ \Sigma^0$  & $K^+ \Lambda$ & $K^+ \Sigma^0$ \\ 
\hline
$N^*(1650)$ ($S_{11}$)      & * & * & *  & * \\
$N^*(1710)$ ($P_{11}$)      & * & * & *  & * \\
$N^*(1720)$ ($P_{13}$)      & * & * & *  & * \\
$N^*(1900)$ ($P_{13}$)      &   &   & *  & * \\
$N^*(1900)$ ($D_{13}$)      & * & * & *  &   \\
$N^*(1900)$ ($P_{11}$)      &   &   & *  &   \\ \hline
$\Delta^*(1700)$ ($D_{33}$) &   &   &    & *  \\ 
$\Delta^*(1900)$ ($S_{31}$) &   & * &    & *  \\
$\Delta^*(1910)$ ($P_{31}$) &   & * &    & *  \\ 
$\Delta^*(1920)$ ($P_{31}$) &   &   &    & *  \\ \hline
\end{tabular}
\end{center}
\caption{List of $s$-channel resonant terms included in the MB model
\cite{mart} and the Regge plus resonance (RPR) model~\cite{corthals} 
included in this work. Note that the RPR model has two variants that 
include either a $D_{13}(1900)$ or $P_{11}(1900)$ state.}
\label{ext_states}
\end{table}
%%%%%%%%%%%%%%%%%%%%%%%%%%%%%%%%%%%%%%%%%%%%%%%%%%%%%%%%%%%%%%%%%%%%%%%%%%%%%%

For $K^+\Lambda$ production, the MB model includes four baryon resonance 
terms.  Near threshold, the steep rise of the cross section is accounted for 
with a core set of $N^*$ states: $S_{11}(1650)$, $P_{11}(1710)$, 
$P_{13}(1720)$.  To explain the broad bump in the energy dependence of the 
cross section seen by SAPHIR~\cite{saphir1} and CLAS~\cite{mcnabb,bradford,5st},
the MB model includes a spin-3/2 $D_{13}(1900)$ resonance that was predicted 
in the quark model of Capstick and Roberts~\cite{capstick} to have a strong 
coupling to the $K^+\Lambda$ channel, but which was not well established 
from existing pion-production data. For $K^+\Sigma^0$ production, the MB model 
includes the core $N^*$ states and the $\Delta^*$ resonances $S_{31}(1900)$ 
and $P_{31}(1910)$.  The model also includes $K^*(892)$ and $K_1(1270)$ 
exchanges for both $KY$ final states, but does not include any $u$-channel 
diagrams.  

The $N^*$ states $S_{11}(1650)$, $P_{11}(1710)$, and $P_{13}(1720)$ are the 
only states listed by the Particle Data Group~\cite{pdg} with coupling 
strengths to $K\Lambda$ (see Table~\ref{landscape}).  While the relevance 
of these core states in the $\gamma^{(*)} p \to K^+ \Lambda$ reaction has 
long been considered a well-established fact, this set of states falls short 
of reproducing the experimental results in the region below $W$=2.0~GeV.  
Furthermore, two recent analyses have called the importance of the 
$P_{11}(1710)$ state into question~\cite{shklyar,sarantsev}.  Beyond the core 
states, the PDG lists a two-star $P_{13}(1900)$ state as the sole established 
$N^*$ near 1900~MeV.  However, with a 500-MeV width, it appears unlikely that 
this state by itself can explain the structure(s) visible in the CLAS and 
SAPHIR cross sections, unless its parameters are significantly different than 
those given by the PDG. This has led to suggestions of a new (unconfirmed) 
$N^*$ state in this mass region (e.g. the $D_{13}(1900)$ state in the MB model).
However, the analysis of Saghai~\cite{saghai}, using the same data sets 
employed for the MB model fits, concluded that by tuning the $u$-channel 
background processes involved in the $K^+\Lambda$ reaction, the need to 
include any states beyond the core set was removed.  Note that the 
investigation of contributing $N^*$ states to the $KY$ reactions has typically 
been limited to spin $j < 5/2$ due to the expectations that higher-spin 
resonances do not significantly contribute to the reaction dynamics.
\cite{shklyar,shklyar2}.

Moving beyond tree-level approaches to consider recent multipole and
coupled-channels models has not led to dramatic new insights to the $N^*$
spectrum.  The multipole analysis by Mart and Sulaksono~\cite{mart_sul}, as 
well as the coupled-channels models of Julia-Diaz {\it et al.}~\cite{diaz} 
and Sarantsev {\it et al.}~\cite{sarantsev} (which all employ CLAS 
photoproduction data in their fits), claim that a $D_{13}(1900)$ state is 
required by both the CLAS and SAPHIR $\gamma p \to K^+ \Lambda$ data.  
However, the coupled-channels model of Ireland {\it et al.}~\cite{ireland} 
points to a $P_{11}(1840)$ state as a more likely candidate (although one or 
more of $S_{11}$, $P_{11}$, $P_{13}$, $D_{13}$ are not ruled out).  The fits 
of Julia-Diaz {\it et al.}~\cite{diaz} suggests a third $S_{11}$ resonance 
might also be playing a role, while Sarantsev {\it et al.}~\cite{sarantsev} 
also require (in addition to a $D_{13}(1900)$) the presence of a 
$P_{11}(1840)$ and another $D_{13}$ state at 2170~MeV.  An extension of the 
coupled-channels model of Sarantsev {\it et al.}~\cite{sarantsev} by 
Avisovich {\it et al.}~\cite{anisovich}, which was the first model to include 
the CLAS photoproduction hyperon polarization transfer observables $C_x$ and 
$C_z$~\cite{bradford1}, concluded that a $P_{13}(1900)$ state was also 
required to satisfactorily fit the data.

In the recent fits of the $\gamma p \to K^+ \Sigma^0$ data, all $N^*$
resonances found to be necessary to fit the $K^+\Lambda$ data have been
included.  However, the existing $K^+\Sigma^0$ database is much smaller
than the $K^+\Lambda$ database, with significantly larger statistical 
uncertainties.  Even with this situation, the recent coupled-channels
models~\cite{diaz,sarantsev,anisovich} indicate important resonant
contributions to the $K^+\Sigma^0$ final state from the $N^*$ states
$P_{11}(1840)$, $D_{13}(1870)$, $P_{13}(1885)$, and $D_{13}(2170)$, and
from the $\Delta^*$ states $F_{35}(1905)$, $P_{33}(1940)$, and
$F_{37}(1950)$.

Each different model has ambiguities that can be better constrained only by 
incorporating better quality data or including new experimental observables.  
Comparison of the models to the data can be used to provide indirect support 
for the existence of the different baryonic resonances and their branching 
ratios into the strange channels, as well as improved constraints on the 
phenomenology of the different strangeness production reactions.

\subsection{Regge and regge plus resonance models}

Our $KY$ electroproduction data are also compared to the Reggeon-exchange 
model from Guidal, Laget, and Vanderhaeghen~\cite{glv} (referred to here as 
GLV).  This calculation includes no baryon resonance terms at all.  Instead, 
it is based only on gauge-invariant $t$-channel $K$ and $K^*$ Regge-trajectory 
exchange.  It therefore provides a complementary basis for studying the 
underlying dynamics of strangeness production.  It is important to note that 
the Regge approach has far fewer parameters compared to the hadrodynamic 
models.  These include the $K$ and $K^*$ form factors (assumed to be of a 
monopole form) and the coupling constants $g_{KYN}$ and $g_{K^*YN}$ (taken 
from photoproduction studies).  The GLV model was fit to higher-energy 
photoproduction data where kaon exchanges dominate and extrapolated down to 
JLab energies.  Furthermore, the use of Regge propagators eliminates the need 
to introduce strong form factors in the background terms, thus avoiding the 
gauge-invariance issues associated with traditional effective Lagrangian 
models.  

The GLV Regge model reasonably accounts for the strength in the CLAS 
$K^+\Lambda$ differential cross sections and separated structure functions
\cite{5st,bradford}.  Although the reasonable performance of a pure Regge 
description in this channel suggests a $t$-channel dominated process, there 
are obvious discrepancies with the data, indicative of $s$-channel strength.  
In the $K^+\Sigma^0$ channel, the same Regge description significantly 
underpredicts the differential cross sections and separated structure 
functions~\cite{5st,bradford}.  The fact that the Regge model fares poorly 
when compared to the $K^+\Sigma^0$ data is indicative that this process has a 
much larger $s$-channel content compared to $K^+\Lambda$ production.

The final model included in this work is based on a tree-level effective 
field model for $\Lambda$ and $\Sigma^0$ photoproduction from the proton. It 
differs from traditional isobar approaches in its description of the 
non-resonant diagrams, which involve the exchange of $K$ and $K^*$ Regge 
trajectories.  A selection of $s$-channel resonances are then added to this 
background.  This ``Regge plus resonance'' model (referred to here as RPR) 
\cite{corthals} has the advantage that the background diagrams contain only a 
few parameters that are constrained by high-energy data where the $t$-channel 
processes dominate.  In addition to the kaonic trajectories, the RPR model 
includes the $s$-channel resonances $S_{11}(1650)$, $P_{11}(1710)$, 
$P_{13}(1720)$, and $P_{13}(1900)$ (see Table~\ref{ext_states}). The model 
also includes either a $D_{13}(1900)$ or $P_{11}(1900)$ state in the 
$K^+\Lambda$ channel.  In detailed comparisons with the separated structure 
functions~\cite{5st,sltp} and transferred polarization data from CLAS
\cite{carman03}, only the $D_{13}(1900)$ assumption could be reconciled with 
the data, whereas the $P_{11}(1900)$ option could clearly be rejected
\cite{corthals}.  In the $K^+\Sigma^0$ channel, four $\Delta^*$ states, 
$D_{33}(1700)$, $S_{31}(1900)$, $P_{31}(1910)$, and $P_{31}(1920)$, have been 
included (see Table~\ref{ext_states}).

%
%%%%%%%%%%%%%%%%%%%%%%%%%%%%%%%%%%%%%%%%%%%%%%%%%%%%%%%%%%%%%%%%%%%%%%%%%%%
%

\section{POLARIZATION FORMALISM}
\label{sec:formalism}

\subsection{Polarization component definitions}

The differential cross section for kaon electroproduction can be written 
as the product of a virtual photon flux factor $\Gamma_v$ and the kaon 
virtual differential cross section, expressed in the kaon center-of-mass 
(c.m.) frame as
\begin{equation}
\frac{d\sigma}{d\Omega_{e'} d\Omega_K^{c.m.} dE_{e'}} = \Gamma_v 
\frac{d\sigma_v}{d\Omega_K^{c.m.}}.
\end{equation}

\noindent
The most general form for the differential cross section of a kaon from a 
proton target, allowing for a polarized electron beam, target proton, and 
recoil hyperon, is given by~\cite{knochlein}
\begin{eqnarray}
\label{csec1}
\frac{d\sigma_v}{d\Omega_K^{c.m.}} &=& {\cal K} \sum\limits_{\alpha,\beta} 
S_{\alpha} S_{\beta} \Bigl[ R_T^{\beta\alpha} + \epsilon R_L^{\beta\alpha}
+ c_+( ^c\!R_{LT}^{\beta\alpha} \cos{\Phi}
+ \!^s\!R_{LT}^{\beta\alpha} \sin{\Phi}) \nonumber \\
&+& \epsilon(^c\!R_{TT}^{\beta\alpha} \cos{2\Phi} + \!^s\!R_{TT}^{\beta\alpha} 
\sin{2\Phi}) \nonumber \\
&+& h c_- (^c\!R_{LT'}^{\beta\alpha} \cos{\Phi}
+ \!^s\!R_{LT'}^{\beta\alpha} \sin{\Phi}) + h c_0R_{TT'}^{\beta\alpha} \Bigr].
\end{eqnarray}

\begin{figure}[tbp]
\vspace{5.0cm}
\includegraphics{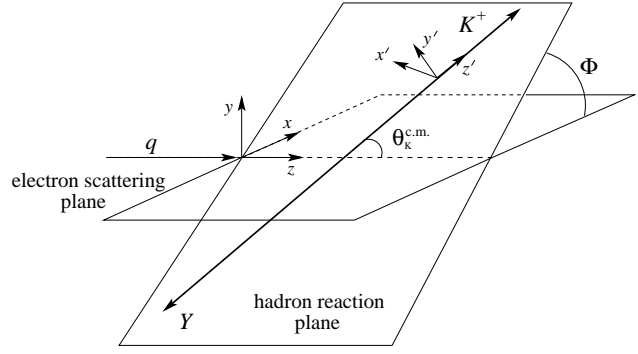}
\caption{Kinematics for $KY$ electroproduction defining the c.m. angles 
and coordinate systems used in the analysis.  This figure shows a 
positive $\Phi$ angle.}
\label{coor4}
\end{figure}

\noindent
The $R^{\beta \alpha}$ terms represent the response functions that account 
for the structure of the hadronic system and, in general, are functions of 
$Q^2$, $W$, and $\cos \theta_K^{c.m.}$ only.  The superscripts $\alpha$ and 
$\beta$ refer to the target and hyperon polarization axes, respectively, and 
the $c$ and $s$ superscripts indicate a cosine or sine dependence on the angle 
$\Phi$, where $\Phi$ is the angle between the electron and hadron planes. Here 
$\epsilon$ is the transverse polarization of the virtual photon, $h$ is the 
electron-beam helicity, and ${\cal K}$ is the ratio of the momentum of the kaon 
to the virtual photon in the c.m. frame.  The factors $c_{\pm}$ are given by 
$\sqrt{\epsilon(1\pm\epsilon)}$ and $c_0=\sqrt{1-\epsilon^2}$. Fig.~\ref{coor4} 
defines the angles of the scattering process in the c.m. system.

%%%%%%%%%%%%%%%%%%%%%%%%%%%%%%%%%%%%%%%%%%%%%%%%%%%%%%%%%%%%%%%%%%%%%%%%%
%%%%%%%%%%%%%%%%%%%%%%%%%%%%%%%%%%%%%%%%%%%%%%%%%%%%%%%%%%%%%%%%%%%%%%%%%

The operators $S_\alpha$ and $S_\beta$ project out the target polarization
vector in the $(x,y,z)$ system and the hyperon polarization in the
$(x',y',z')$ system, respectively (see Fig.~\ref{coor4}). The $(x,y,z)$ 
system is defined such that $\hat{z}$ is along the three momentum transfer 
$\vec{q}$ direction and $\hat{y}$ is normal to the electron-scattering plane.  
The $(x',y',z')$ system is defined such that $\hat{z}'$ is along the kaon 
momentum vector and $\hat{y}'$ is normal to the hadronic plane.  

In the case where there is no beam, target, or recoil polarization
($h$, $\alpha$, $\beta$ = 0), Eq.(\ref{csec1}) reduces to 
\begin{equation}
\label{csec2}
\frac{d\sigma_v}{d\Omega_K^{c.m.}} \equiv \sigma_0 = 
{\cal K} \left[ R_T^{00} + \epsilon R_L^{00} + c_+R_{LT}^{00} 
\cos{\Phi} + \epsilon R_{TT}^{00} \cos{2\Phi} \right].
\end{equation}

\noindent
For the case of a polarized-electron beam incident on an unpolarized target 
producing a polarized recoiling hyperon, Eq.(\ref{csec1}) becomes~\cite{laget}
\begin{equation}
\label{csec3}
\frac{d\sigma_v}{d\Omega_K^{c.m.}} = \sigma_0 (1 + h A_{LT'} +  P_{x'} 
\hat{x}' \cdot \hat{S}_{x'} + P_{y'} \hat{y}' \cdot \hat{S}_{y'} + 
P_{z'} \hat{z}' \cdot \hat{S}_{z'}),
\end{equation}

\noindent
where $A_{LT'} = \frac{\cal K}{\sigma_0} c_- R_{LT'}^{00} \sin{\Phi}$ is the 
polarized beam asymmetry defined in terms of the fifth response function 
$R_{LT'}^{00}$.  

Each of the recoil-hyperon polarization components can be split into a
beam-helicity-independent part $P_i^0$, called the {\em recoil}
polarization, and a beam-helicity-dependent part $P_i'$, called the {\em
transferred} polarization.  The components of the hyperon polarization
vector can be written as $P_i = P_i^0 + h P_i'$.  The three recoil
polarization components are given in terms of the response functions in the
$(x',y',z')$ system as
\begin{eqnarray}
\label{pind1}
P_{x'}^0 &=& \frac{\cal K}{\sigma_0} \left( c_+R_{LT}^{x'0}\sin{\Phi} + 
                      \epsilon\ R_{TT}^{x'0}\sin{2\Phi} \right) \nonumber\\
P_{y'}^0 &=& \frac{\cal K}{\sigma_0} \left( R_T^{y'0} + \epsilon R_L^{y'0}
                      + c_+R_{LT}^{y'0} \cos{\Phi} + \epsilon R_{TT}^{y'0} 
                      \cos{2\Phi} \right) \nonumber \\
P_{z'}^0 &=& \frac{\cal K}{\sigma_0} \left( c_+R_{LT}^{z'0} \sin{\Phi} + 
                      \epsilon R_{TT}^{z'0} \sin{2\Phi} \right),
\end{eqnarray}

\noindent
and the three transferred polarization components are written in the 
$(x',y',z')$ coordinate system as
\begin{eqnarray}
\label{ptran1}
P_{x'}' &=& \frac{\cal K}{\sigma_0} \left( c_- R_{LT'}^{x'0} \cos{\Phi} + 
                      c_0R_{TT'}^{x'0}\right)  \nonumber \\
P_{y'}' &=& \frac{\cal K}{\sigma_0} c_-R_{LT'}^{y'0} \sin{\Phi} \nonumber \\
P_{z'}' &=& \frac{\cal K}{\sigma_0} \left( c_- R_{LT'}^{z'0} \cos{\Phi} + 
                       R_{TT'}^{z'0}\right).
\label{eqn_Pzp}
\end{eqnarray}

To accommodate finite bin sizes and to improve statistics, our analysis sums
over all $\Phi$ angles.  The $\Phi$-integrated polarization components 
(represented by the ${\cal P}$ symbol) in the $(x',y',z')$ system are given 
in Table~\ref{big_def}.  In performing the $\Phi$ integration, the polarization 
components ${\cal P}_{x'}^0$, ${\cal P}_{z'}^0$, and ${\cal P}_{y'}'$ are equal 
to zero.  In Table~\ref{big_def}, the term $K_I = 1/(R_T^{00} + \epsilon
R_L^{00})$.

%%%%%%%%%%%%%%%%%%%%%%%%%%%%%%%%%%%%%%%%%%%%%%%%%%%%%%%%%%%%%%%%%%%%%%%%%%%
\begin{table}[htbp]
\begin{center}
\begin{tabular} {c|c|c|c} \hline
\multicolumn{4} {c} {$(x',y',z')$ Coordinate System} \\ \hline
%\rule{0mm}{6mm}
${\cal P}_{x'}^0$ & 0 & ${\cal P}_{x'}'$  & $K_I c_0 R_{TT'}^{x'0}$  \\ \hline
%\rule{0mm}{6mm}
${\cal P}_{y'}^0$ & $K_I (R_T^{y'0} + \epsilon R_L^{y'0})$ & ${\cal P}_{y'}'$  
& 0 \\  \hline
%\rule{0mm}{6mm}
${\cal P}_{z'}^0$ & 0 & ${\cal P}_{z'}'$  & $K_I c_0 R_{TT'}^{z'0}$  \\ \hline
\multicolumn{4} {c} {$(x,y,z)$ Coordinate System} \\ \hline
%\rule{0mm}{6mm}
${\cal P}_{x}^0$ & 0 & ${\cal P}_{x}'$  & $\frac{1}{2} K_I c_- (R_{LT'}^{x'0} 
\cos{\theta_K^{c.m.}} - R_{LT'}^{y'0} + R_{LT'}^{z'0} \sin{\theta_K^{c.m.}})$ 
\\  \hline
%\rule{0mm}{6mm}
${\cal P}_y^0$ & $\frac{1}{2} K_I c_+ (R_{LT}^{x'0} 
\cos{\theta_K^{c.m.}} + R_{LT}^{y'0} + R_{LT}^{z'0} \sin{\theta_K^{c.m.}})$  &
${\cal P}_y'$  & 0 \\ \hline
%\rule{0mm}{6mm}
${\cal P}_z^0$ & 0 & ${\cal P}_{z}'$  & $K_I c_0 (-R_{TT'}^{x'0} 
\sin{\theta_K^{c.m.}} + R_{TT'}^{z' 0} \cos{\theta_K^{c.m.}})$ \\ \hline
\end{tabular}
\end{center}
\caption{Polarization observables integrated over $\Phi$ in the two coordinate
systems used in this work.}
\label{big_def}
\end{table}
%%%%%%%%%%%%%%%%%%%%%%%%%%%%%%%%%%%%%%%%%%%%%%%%%%%%%%%%%%%%%%%%%%%%%%%%%%%

To define the polarization observables in the $(x,y,z)$ coordinate system, 
shown in Fig.~\ref{coor4}, the components defined for the $(x',y',z')$ system 
in Eqs.(\ref{pind1}) and (\ref{ptran1}) must undergo a simple transformation
that involves a rotation of $\theta_K^{c.m.}$ about $\hat{y}'$, followed by a 
rotation of $\Phi$ about $\hat{z}'$.  The $\Phi$-integrated recoil and 
transferred polarization components in the $(x,y,z)$ system are defined in 
Table~\ref{big_def}.

\boldmath
\subsection{$\sigma_L/\sigma_T$ ratio}
\label{ltrat}
\unboldmath

The $\Lambda$ polarization transfer data can be used to extract the ratio
of the longitudinal-to-transverse structure functions
$R_\sigma = \sigma_L/\sigma_T$ at $\cos\theta_K^{c.m.}=1$.  This ratio 
has been previously measured at $\cos\theta_K^{c.m.}=1$
\cite{bebek,mohring} and for $\cos\theta_K^{c.m.}\not=1$~\cite{5st} 
using the Rosenbluth separation technique.  Our previously published 
polarization transfer results~\cite{carman03} taken at a beam energy of 
2.567~GeV were used to extract $R_\sigma$ at $\cos\theta_K^{c.m.}=1$ for 
three points of $W$ and $Q^2$~\cite{raue05}.  In a similar way, $R_\sigma$
can be extracted from the 4.261 and 5.754~GeV CLAS data.

In parallel or anti-parallel kinematics ($\cos\theta_K^{c.m.}=\pm 1$),
the $z'$ and $z$ components of the transferred polarization integrated
over $\Phi$ (given in Table~\ref{big_def}) reduce to
\begin{equation}
\label{eqn_Pz1}
{\cal P}'_{z'} = \pm{\cal P}'_z = \pm {\frac{c_0 R_{TT'}^{z' 0}} {R_T^{00}
+ \epsilon R_L^{00}}} = \pm \frac{c_0 R_{TT'}^{z' 0}}{\sigma_U/{\cal K}},
\end{equation}

\noindent
where the plus (minus) sign is associated with the parallel (anti-parallel)
kinematics case and $\sigma_U = \sigma_T + \epsilon \sigma_L$.

The response functions used to express the $\Phi$-integrated components of 
${\cal P}'_{z'}$ and ${\cal P}'_z$ (see Table~\ref{big_def}) can be
written in terms of the Chew, Goldberger, Low, and Nambu (CGLN) amplitudes
\cite{chew} as shown in Ref.~\cite{knochlein}. For the case of 
$\theta_K^{c.m.}=0^\circ$, it can be shown that $R_{TT'}^{z' 0}=R_T^{00}$,
and Eq.(\ref{eqn_Pz1}) can be rewritten as
\begin{equation}
\label{eqn-Pz}
{\cal P}'_{z'}={\cal P}_z'= \frac{c_0 R_T^{00}} {R_T^{00} + \epsilon R_L^{00}}
= \frac{c_0 \sigma_T} {\sigma_T + \epsilon \sigma_L}.
\end{equation}

\noindent
Inverting this form and rearranging, the ratio $R_\sigma$ at 
$\cos \theta_K^{c.m.} = \pm 1$ can be written as
\begin{equation}
\label{eqn-ratio2}
R_\sigma = {\frac{\sigma_L}{\sigma_T}} =
{\frac{1}{\epsilon}}\left({\frac{c_0}{{\cal P}'_{z'}}}-1\right).
\end{equation}

\noindent
While the ${\cal P}'_{z'}$ and ${\cal P}'_z$ data presented here do not 
include data points at $\theta_K^{c.m.}=0^\circ$, an extrapolation to 
$\cos\theta_K^{c.m.}=\pm 1$ can be performed as shown in Section~\ref{rat_ext}.

%
%%%%%%%%%%%%%%%%%%%%%%%%%%%%%%%%%%%%%%%%%%%%%%%%%%%%%%%%%%%%%%%%%%%%%%%%%%%
%

\section{POLARIZATION EXTRACTION}
\label{sec:polext}

\subsection{Decay angular distributions}

The $\Lambda$ decays weakly into a pion and a nucleon with the decay nucleon 
constrained to move preferentially in the direction of the hyperon spin.  In 
the $\Lambda$ rest frame, the decay nucleon angular distribution is given 
by~\cite{bonner}
\begin{equation}
\label{ang_dist}
\frac{dN}{d \cos \theta_N^{RF}} = N(1 + \alpha_\Lambda P_{\Lambda} 
\cos \theta_N^{RF}),
\end{equation}

\noindent
where $P_{\Lambda}$ is the $\Lambda$ polarization and $\theta_N^{RF}$ is 
the angle between the polarization axis and the decay-nucleon momentum in 
the $\Lambda$ rest frame.  In this work we focus solely on the 
$\Lambda \to p \pi^-$ decay (B.R.=64\%) and explicitly replace $\theta_N^{RF}$ 
with $\theta_p^{RF}$.  The $\Lambda$ weak decay asymmetry parameter 
$\alpha_\Lambda$ has been measured to be 0.642$\pm$0.013~\cite{pdg}.  

The $\Sigma^0$ decays into a $\gamma$ and a $\Lambda$ (branching ratio 100\%).
A $\Sigma^0$ with polarization $P_\Sigma$ will yield a decay $\Lambda$ that 
retains some of the polarization of its parent.  As shown in Ref.~\cite{gatto}, 
$P_\Lambda = -\frac{1}{3} P_\Sigma$ on average for the decay $\Lambda$ in 
its rest frame.  For the case of a final-state $\Sigma^0$, the $\Lambda$ rest 
frame can be calculated only if four particles are detected in the final state.
In addition to the detection of the electron, kaon, and decay proton, either 
the decay pion of the $\Lambda$ or the decay $\gamma$ from the $\Sigma^0$ must 
be detected.  Due to the small CLAS acceptance for a four particle final state, 
only three final-state particles were detected.  In Ref.~\cite{bradford1} it 
has been shown that the polarization of the daughter $\Lambda$ from the 
$\Sigma^0$ decay can be measured without boosting the detected proton to the 
reference frame of the $\Lambda$.  The value of the effective weak decay 
asymmetry parameter was determined to be $\alpha_\Sigma = -0.164$, or in terms 
of the $\Lambda$ weak decay constant, $\alpha_\Sigma = -0.256 \alpha_\Lambda$, 
thus reduced from the value $-0.333 \alpha_\Lambda$. This value is independent
of $\Sigma^0$ kinematics.

As the electron beam is not 100\% polarized, the helicity term $h$ in the 
hyperon polarization must be replaced by the average longitudinal electron-beam 
polarization $P_b$ as
\begin{equation}
\label{pol1}
P_Y = P_Y^0 + P_b P_Y'.
\end{equation}

\noindent
Combining the expressions from Eqs.(\ref{ang_dist}) and (\ref{pol1}), the decay 
proton angular distribution for the two different beam helicity states can
be written
\begin{equation}
\label{ang_dist1}
\frac{dN^\pm}{d \cos \theta_p^{RF}} = N^\pm [1 + \alpha (P_Y^0 \pm P_b P_Y') 
\cos \theta_p^{RF}],
\end{equation}

\noindent
where $\alpha_\Lambda=0.642$ for the $\Lambda$ analysis and 
$\alpha_\Sigma=-0.164$ for the $\Sigma^0$ analysis.  

\subsection{Asymmetry approach}

The transferred hyperon polarization was extracted using the 
acceptance-corrected yield asymmetry for the two different electron beam 
helicity states of the form
\begin{equation}
\label{asm}
A = \frac{N^+ - N^-}{N^+ + N^-}.
\end{equation} 

\noindent
This asymmetry is formed from the $\cos \theta_p^{RF}$ yields for the three 
spin-quantization axes of the decaying hyperon (in either of the coordinate 
systems defined in Fig.~\ref{coor4}).  The terms $N^+$ and $N^-$ represent 
the acceptance-corrected decay proton yields in a given kinematic bin.  The
helicity-gated yields are given by
\begin{equation}
\label{n_pm}
N^{\pm}(\cos \theta_p^{RF}) = {\cal E} \left( \frac{d \sigma}{d \Omega} 
\right)^\pm = \sigma_0 {\cal E} [ 1 \pm P_b A_{LT'}
+ \alpha (P_Y^0 \pm P_b P_Y') \cos \theta_p^{RF}],
\end{equation}

\noindent
where ${\cal E}$ represents the CLAS detection efficiency, which is
assumed to be a helicity-independent function, and includes
the CLAS acceptance function and the beam-target luminosity factors.

As discussed in Section~\ref{sec:formalism}, this analysis was performed
by integrating the decay proton yields over all $\Phi$ angles to maximize 
the statistical precision of the measurement.  In this case the measured 
yield asymmetry becomes
\begin{equation}
A = \frac{\int_0^{2\pi} \sigma_0 {\cal E} [ 1 + P_b A_{LT'} 
    + \alpha(P_Y^0 + P_b P_Y') \cos \theta_p^{RF}] d\Phi
    - \int_0^{2\pi} \sigma_0 {\cal E} [ 1 - P_b A_{LT'} 
    + \alpha(P_Y^0 - P_b P_Y') \cos \theta_p^{RF}] d\Phi}
    {\int_0^{2\pi} \sigma_0 {\cal E} [ 1 + P_b A_{LT'} 
    + \alpha(P_Y^0 + P_b P_Y') \cos \theta_p^{RF}] d\Phi
    + \int_0^{2\pi} \sigma_0 {\cal E} [ 1 - P_b A_{LT'}
    + \alpha(P_Y^0 - P_b P_Y') \cos \theta_p^{RF}] d\Phi}.
\end{equation}

\noindent
After some simplification, this expression can be written as
\begin{eqnarray}
\label{asm2}
A = \frac{\int_0^{2\pi} \sigma_0 [ P_b A_{LT'} + \alpha P_b P_Y' 
\cos \theta_p^{RF}] d\Phi}{\int_0^{2\pi} \sigma_0 [ 1 + \alpha P_Y^0 
\cos \theta_p^{RF}] d\Phi}.
\end{eqnarray}

\noindent
As the $A_{LT'}$ term is proportional to $\sin \Phi/\sigma_0$, it integrates 
to zero for all choices of spin-quantization axes. Similarly, the term 
containing $P_Y^0$ integrates to zero along the $(x,x')$ and $(z,z')$ axes 
using the definitions in Section~\ref{sec:formalism}.

Considering our coordinate system choices (see Fig.~\ref{coor4}), it turns 
out that the $\Phi$-integrated asymmetries can only be non-zero along the 
$(x,x')$ and $(z,z')$ axes.  This can be seen as 
$\int_0^{2\pi} \sigma_0 P_Y' d\Phi = 0$ along both the $(y,y')$ axes, given 
the polarization definitions in Section~\ref{sec:formalism}.  In other 
words, when performing the $\Phi$ integration, the polarization components 
${\cal P}_{(z,z')}^0$, ${\cal P}_{(x,x')}^0$, and ${\cal P}_{(y,y')}'$ are 
all constrained to be zero.  Thus the only possible non-zero asymmetries for 
our coordinate systems will be $A_{(z,z')}$ and $A_{(x,x')}$.  Along these 
special axes, the $\Phi$-integrated asymmetries can be written as
\begin{equation}
\label{asm3}
A = \alpha P_b \left[ \frac{\int_0^{2\pi} \sigma_0
P_Y' d\Phi}{\int_0^{2\pi} \sigma_0 d\Phi} \right] \cos \theta_p^{RF} .
\end{equation}

\noindent
The quantity in brackets is equivalent to the ${\cal P}$ polarization terms 
in Table~\ref{big_def} and represents the $\Phi$-integrated hyperon 
transferred polarization, where a separate asymmetry is computed for each 
spin-quantization axis.  We therefore extract the non-zero transferred 
hyperon polarizations with respect to the different quantization axes from 
the asymmetries by fitting
\begin{equation}
\label{fit_eq}
A_\Lambda = \alpha_\Lambda P_b {\cal P}_\Lambda' \cos \theta_p^{RF} 
{~~~{\rm or}~~~} A_\Sigma = \alpha_\Sigma P_b {\cal P}_\Sigma' 
\cos \theta_p^{RF}.
\end{equation}

In forming the asymmetry of Eq.(\ref{asm}), the decay proton helicity-gated
yields are sorted for each kinematic bin of interest. We used an
event-by-event weighting factor to correct the yields for the detector
acceptance. The asymmetry method used in this analysis is relatively 
insensitive to the detailed form of the CLAS acceptance function.  This 
discussion is contained in Section~\ref{add-cuts}.

\subsection{Hyperon polarization and statistical uncertainty}
\label{sec-pol-extract}

For the general case where a given hyperon sample $Y$ is contaminated by
particle misidentification events and events from the tail of the hyperon 
$Y'$, the helicity asymmetry can be written in terms of its individual 
contributions as
\begin{eqnarray}
A_{meas} &=&\frac{(N_Y^+ + N_{Y'}^+ + N_{bck}^+) -
                 (N_Y^- + N_{Y'}^- + N_{bck}^-)}
                 {N_Y + N_{Y'} + N_{bck}}\nonumber \\
         &=& \frac{\frac{N_Y^+-N_Y^-}{N_Y} +
             \frac{N_{Y'}^+-N_{Y'}^-}{N_{Y'}} \cdot 
             \frac{N_{Y'}}{N_Y} +
	     \frac{N_{bck}^+-N_{bck}^-}{N_{bck}} \cdot \frac{N_{bck}}{N_Y}}
	     {1+\frac{N_{Y'}}{N_Y}+\frac{N_{bck}}{N_Y}},
\label{tot_asm}
\end{eqnarray}

\noindent
where $N_Y$, $N_{Y'}$, and $N_{bck}$ refer to the number of counts from the
hyperon of interest, the tail of the other hyperon, and from the background
(mostly pions misidentified as kaons), respectively, within the $Y$
identification cuts (see Section~\ref{hyp_yield}).  If we define 
$F_{Y'}=N_{Y'}/N_Y$ and $F_{bck}=N_{bck}/N_Y$, we can write
\begin{equation}
\label{fiteq}
A_{meas}=\frac{A_Y + A_{Y'} F_{Y'} + A_{bck} F_{bck}}{1 + F_{Y'} + F_{bck}}.
\end{equation}

It was observed in this analysis that the pion background asymmetry under
both hyperon peaks is consistent with zero, thus the term associated with 
$A_{bck}$ in Eq.(\ref{fiteq}) is set to zero (see Section~\ref{bck_pol}).  
The ``pure'' asymmetries, $A_\Lambda$ and $A_\Sigma$, are given by 
Eq.(\ref{fit_eq}), thus the measured asymmetry can be written in terms of 
the transferred polarizations, as well as the measured polarization within 
the hyperon mass cuts ${\cal P}'_{meas}$ as
\begin{equation}
\label{eq-Ameas}
A_{meas}=\frac{\alpha_Y P_b {\cal P}'_Y \cos\theta_p^{RF}
+ \alpha_{Y'} P_b {\cal P}'_{Y'} \cos\theta_p^{RF} F_{Y'}}
{1+F_{Y'}+F_{bck}} = \alpha_\Lambda P_b {\cal P}'_{meas} \cos\theta_p^{RF},
\end{equation}

\noindent
where $\alpha_\Lambda P_b {\cal P}'_{meas}$ is the slope extracted from the 
fit of the $\cos\theta_p^{RF}$ distribution with respect to a given 
spin-quantization axis. 

We found that the contamination of $\Sigma^0$ hyperons within the $\Lambda$
identification cuts is consistent with zero (see Section~\ref{hyp_yield}),
so for $Y=\Lambda$ and $Y'=\Sigma^0$, Eq.(\ref{eq-Ameas}) can be rearranged
to get
\begin{equation}
\label{cor_lam_pol}
{\cal P}'_\Lambda = {\cal P}'_{meas} (1 + F_{bck}).
\end{equation}

For $Y=\Sigma^0$ and $Y'=\Lambda$, Eq.(\ref{eq-Ameas}) can be rearranged to
get
\begin{equation}
\label{cor_sig_pol}
{\cal P}'_\Sigma=\frac{\alpha_\Lambda}{\alpha_\Sigma}
\left[{\cal P}'_{meas}(1+F_\Lambda+F_{bck}) - {\cal P}'_\Lambda F_\Lambda 
\right].
\end{equation}

Performing standard error propagation, the statistical uncertainty for
${\cal P}'_\Lambda$ is
\begin{equation}
\label{cor_lam_err}
\delta {\cal P}'_\Lambda = \left[ (1 + F_{bck}) (\delta {\cal P}'_{meas})^2 
+ ({\cal P}'_{meas})^2 \delta F_{bck}^2 \right]^{1/2},
\end{equation}

\noindent
where the individual uncertainties are given by
\begin{equation}
\delta F_{bck} = F_{bck} \sqrt{\frac{1}{N_{bck}}+\frac{1}{N_\Lambda}}~~~~ 
{\rm and} ~~~~ \delta {\cal P}'_{meas} = 
\frac{\delta{\rm (slope)}}{\alpha_\Lambda P_b}.
\end{equation}

\noindent
Here $\delta{\rm(slope)}$ is the uncertainty in the slope from the fit of
the $\cos\theta_p^{RF}$ distribution.  We find that the dominant
contribution to the $\Lambda$ polarization uncertainty is due to the
uncertainty in ${\cal P}'_{meas}$.

Similarly, the statistical uncertainty for ${\cal P}'_\Sigma$ is
\begin{equation}
\label{sig_err}
\delta {\cal P}'_\Sigma = \frac{\alpha_\Lambda}{\alpha_\Sigma}
\left[(1+F_\Lambda+F_{bck})^2(\delta {\cal P}'_{meas})^2
  + ({\cal P}'_{meas} - {\cal P}'_\Lambda)^2\delta F_\Lambda^2
  + ({\cal P}'_{meas}\delta F_{bck})^2 
  + (F_\Lambda \delta {\cal P}'_\Lambda)^2\right]^{1/2},
\end{equation}

\noindent
where the individual uncertainties are given by
\begin{eqnarray}
\delta F_\Lambda&=&F_\Lambda \sqrt{\frac{1}{N_\Lambda}+\frac{1}{N_\Sigma}},\\
\delta F_{bck}&=&F_{bck} \sqrt{\frac{1}{N_{bck}}+\frac{1}{N_\Sigma}},\ \  
{\rm and} \\
\delta {\cal P}'_{meas}&=& \frac{\delta{\rm (slope)}}{\alpha_\Lambda P_b}.
\end{eqnarray}

\noindent
The dominant contribution to the 
$\Sigma^0$ statistical uncertainty arises due to the uncertainty in 
${\cal P}'_{meas}$.  All other terms are at least a factor of 4 smaller in 
size.

\subsection{Depolarization factor}
\label{depol_fac}

Our formalism defines the polarization transfer as the ratio of the hyperon 
polarization to that of the electron beam. However, the electron interacts 
with the hadronic system through the exchange of a virtual photon.  Thus the 
true ``beam'' polarization is given by the product $P_b D(y)$, where $D(y)$ 
accounts for the polarization loss from the incident beam electron to the 
virtual photon.  There are a number of ways to express the factor $D(y)$.  
One form is given by~\cite{hv99} 
\begin{equation}
D(y) = \frac{y (2 -y)}{2(1-y)(1 + R_\sigma) + y^2},
\end{equation}

\noindent
where $y = E_{\gamma^*}/E_b$ is the relative energy transfer to the target
proton and $R_\sigma = \sigma_L/\sigma_T$.

With this accounting, the hyperon polarization can be rewritten from
Eq.(\ref{pol1}) as $P_Y = P_Y^0 + P_b D(y) P_Y'$.  This would lead to
slightly modified forms of the asymmetries in Eq.(\ref{fit_eq}) with the
hyperon polarizations scaled by a factor of $1/D(y)$.  This re-expression 
of the hyperon transferred polarization allows for a more direct comparison 
for experiments performed at different beam energies.  Perhaps, more 
importantly, it allows for a more direct comparison of electroproduction and 
photoproduction data sets where the ``trivial'' depolarization factor is 
accounted for in the electroproduction data.

Having made this distinction in the possible convention choice for the
hyperon transferred polarization, we have decided not to account for it
in this work, following instead the procedures in Ref.~\cite{carman03}.  The
main reason for this choice is to avoid introducing a model-dependent
uncertainty into our quoted polarizations.  Our studies have shown that
with different hadrodynamic models, the variation in $D(y)$ due to
variations in $R_\sigma$ can be up to 20\%.  

%%%%%%%%%%%%%%%%%%%%%%%%%%%%%%%%%%%%%%%%%%%%%%%%%%%%%%%%%%%%%%%%%%%%%%%%%
\begin{figure}[htbp]
\vspace{5.8cm}
\includegraphics{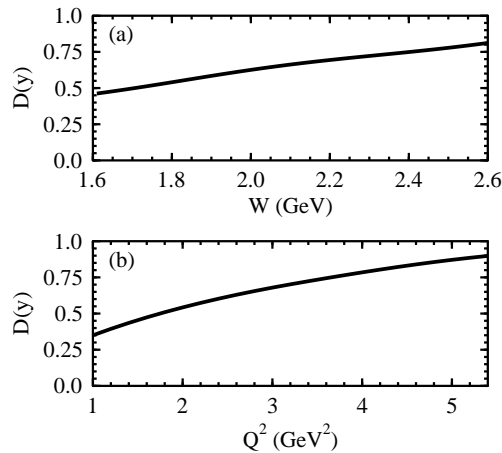}
\caption{Calculations of the depolarization factor $D(y)$ at 5.754~GeV 
using the MB model~\cite{mart_code} in representative data sorts (a)
summing over all $Q^2$ and $\cos \theta_K^{c.m.}$ and (b) summing over
all $W$ and $\cos \theta_K^{c.m.}$.}
\label{depol}
\end{figure}
%%%%%%%%%%%%%%%%%%%%%%%%%%%%%%%%%%%%%%%%%%%%%%%%%%%%%%%%%%%%%%%%%%%%%%%%%

Shown in Fig.~\ref{depol} are predictions of the depolarization factor
for several of our kinematic bins using the MB model~\cite{mart_code}.
It is seen that in our kinematics $\langle D(y) \rangle \sim 0.6$.  The
polarization transfer from the virtual photon to the hyperon is therefore 
67\% larger on average than that for the beam electron to the hyperon.
When considering the polarization data in Section~\ref{sec:results}, one
must take into account the depolarization factor when comparing to other
data and to theory. The theory calculations shown in Section~\ref{sec:results} 
match the data, namely they show the product of the depolarization factor and 
the polarization.

%
%%%%%%%%%%%%%%%%%%%%%%%%%%%%%%%%%%%%%%%%%%%%%%%%%%%%%%%%%%%%%%%%%%%%%%%%%%%
%

\section{DATA ANALYSIS}
\label{sec:analysis}

\subsection{The CLAS detector}

\begin{figure}[htbp]
\vspace{6.4cm}
\includegraphics{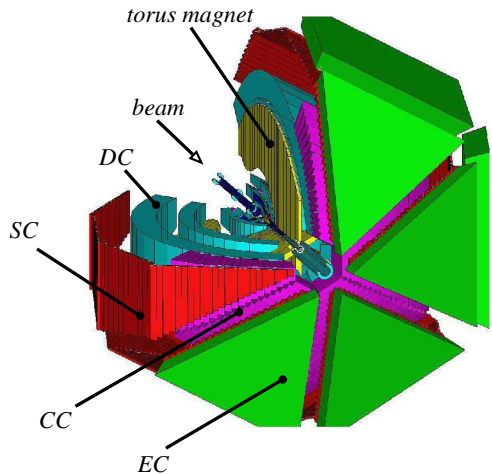}
\caption{(Color online) Three dimensional view of the CLAS detector with 
the different subsystems labeled.  A single sector of the detector has been 
cut away to enable a view of the inner subsystems.  The diameter of the CLAS 
detector is $\sim$5~m and it is $\sim$8~m long.}
\label{clas_full}
\end{figure}

All of the data shown in this analysis were collected using the CLAS
spectrometer located in Hall~B at JLab~\cite{mecking}.  The main magnetic 
field of CLAS is provided by six superconducting coils, which produce an 
approximately toroidal field in the azimuthal direction around the beam axis.  
The gaps between the cryostats are instrumented with six identical detector 
packages, as shown in Fig.~\ref{clas_full}. Each sector consists of three 
sets of drift chamber (DC) packages~\cite{dcnim} to determine the trajectories 
of the charged particles, Cherenkov counters (CC)~\cite{ccnim} for electron 
identification, scintillator counters (SC)~\cite{scnim} for charged particle 
identification, and electromagnetic calorimeters (EC)~\cite{ecnim} for electron 
identification and detection of neutral particles.  A 5-cm long liquid-hydrogen 
target was located in the center of the detector on the electron beam axis.

%%%%%%%%%%%%%%%%%%%%%%%%%%%%%%%%%%%%%%%%%%%%%%%%%%%%%%%%%%%%%%%%%%%%%%%%%
%%%%%%%%%%%%%%%%%%%%%%%%%%%%%%%%%%%%%%%%%%%%%%%%%%%%%%%%%%%%%%%%%%%%%%%%%

To reduce the electromagnetic background resulting from M{\o}ller scattering 
off atomic electrons in the target and the target cell, a small 
normal-conducting toroidal magnet (called the mini-torus) was placed 
symmetrically about the target inside of the first DC package. This magnetic 
field sweeps M{\o}ller electrons out of the detector volume.  A totally 
absorbing Faraday cup, located at the end of 
the beam line, was used to determine the integrated beam charge passing 
through the target.  The efficiency of detection and reconstruction for 
stable charged particles in the fiducial regions of CLAS is greater than 
95\%.  The solid angle coverage of CLAS is approximately 3$\pi$~sr.  The 
polar angle coverage for electrons ranges from 8$^{\circ}$ to 45$^{\circ}$, 
while for hadrons it is from 8$^{\circ}$ to 142$^{\circ}$, with an 
angular resolution of $\delta \theta, \delta \phi \sim 1$ of better than 
2~mr.  The CLAS detector was designed to track particles having momenta 
greater than roughly 200~MeV with a resolution $\delta p/p$ in the range of 
0.5 to 1\%.

The large acceptance of CLAS enabled us to detect the final-state electron
and kaon, as well as the proton from the decay of the $\Lambda$ hyperon.
Hyperon identification with CLAS relies on missing-mass reconstructions of
the reaction $e + p \to e' + K^+ + X$.  In this section, details are
provided on our procedures for particle identification, the cuts used to
isolate the $K^+\Lambda$ and $K^+\Sigma^0$ final states, the hyperon
spectrum fitting procedures, and other cuts and corrections.  

\subsection{Data set information}

The data were taken with typical electron beam currents of 5~nA at a 
luminosity of $1 \times 10^{34}$~cm$^{-2}$s$^{-1}$.  The CLAS event 
readout was triggered by a coincidence between a Cherenkov counter 
and a calorimeter detector in a single sector, generating an event rate of 
$\sim$2~kHz.  The main CLAS torus had its polarity set such that negatively 
charged particles were bent toward the electron beam line.  The electron 
beam was longitudinally polarized, with the polarization determined by a 
coincidence M{\o}ller polarimeter.  Beam polarization measurements were taken 
at regular intervals throughout the running periods and measured a stable 
electron beam polarization of 70\%.

The data in this paper were collected as part of the CLAS running periods
{\tt e1c} in early 1999 and {\tt e1-6} in late 2001/early 2002.  The {\tt
e1c} running period included data with beam energies of 2.567~GeV
(previously published in Ref.~\cite{carman03}) and 4.261~GeV acquired at
several different field settings of the main CLAS torus.  4.261 GeV
represents the luminosity-averaged beam energy for data taken with 
electron beam energies of 4.056, 4.247, and 4.462~GeV. 
Combining the data sets is justified given
the relatively small spread in the virtual photon polarization parameter
$\epsilon$ among the different energies.  The {\tt e1-6} running
period was taken with an electron beam energy of 5.754~GeV. Information 
regarding the different run periods, including the total number of triggers, 
the approximate $W$ and $Q^2$ ranges of the data, the number of hyperons in 
the different analyses detected through the $e'K^+p$ final state, and the
average beam polarization, is contained in Table~\ref{overview}.

%%%%%%%%%%%%%%%%%%%%%%%%%%%%%%%%%%%%%%%%%%%%%%%%%%%%%%%%%%%%%%%%%%%%%%%%%%%%
\begin{table}[htbp]
\begin{center}
\begin{tabular} {c|c|c|c|c|c|c} \hline
%\rule{0mm}{4mm}
$E_b$ & Triggers & $W$ & $Q^2$ & $N_\Lambda$ & $N_{\Sigma^0}$ 
& $\langle P_b \rangle$ \\ \hline
2.567 GeV &  910 M & 1.6 -- 2.15 GeV & 0.3 -- 1.5 GeV$^2$ & 42000 &  8000 & 
67\% \\
4.261 GeV & 1599 M & 1.6 -- 2.6 GeV  & 0.7 -- 3.5 GeV$^2$ & 34000 &  6500 & 
67\% \\
5.754 GeV & 5083 M & 1.6 -- 2.6 GeV  & 1.3 -- 5.4 GeV$^2$ & 82000 & 16000 & 
72\% \\
\hline
\end{tabular}
\end{center}
\caption{Information regarding the different CLAS electroproduction data sets 
associated with this work, including the number of raw triggers, the $W$ and 
$Q^2$ extents of each data set, the number of hyperons (detected via the 
$e'K^+ p$ final state), and the average longitudinal polarization of the 
electron beam.  Note that the 4.261~GeV data set sums together data acquired 
at beam energies of 4.056, 4.247, and 4.462~GeV.}
\label{overview}
\end{table}
%%%%%%%%%%%%%%%%%%%%%%%%%%%%%%%%%%%%%%%%%%%%%%%%%%%%%%%%%%%%%%%%%%%%%%%%%%%%

\subsection{Particle identification}

The first level of event reconstruction required the identification of a
viable electron candidate.  This was done by requiring that a negatively 
charged particle -- identified by its track curvature in the magnetic field 
of the spectrometer -- be matched in time and space with hits in the SC, CC, 
and EC counters.  A particle-tracking vertex cut was employed to ensure 
that the particle originated from the liquid-hydrogen target. In order to 
remove negatively charged pions from the electron candidate sample, a cut was 
placed on the ratio of the measured energy deposited in the fiducial region of 
the EC (accounting for the sampling fraction of the calorimeter) to the 
momentum of the particle.  A further reduction in pion contamination was 
achieved by placing a minimum-ionizing cut on the energy measured in the EC.

The first-level requirements for charged hadrons are that they have a track 
in the drift chamber and a matched, in-time hit in the SC in that same 
sector.  For the final-state $K^+$ and $p$ in this analysis, we require that 
the curvature for the $K^+$ and $p$ tracks be consistent with a positively 
charged particle.  We also require that the $K^+$ track originate from the 
target using a vertex cut.

\begin{figure}[htbp]
\vspace{4.8cm}
\includegraphics{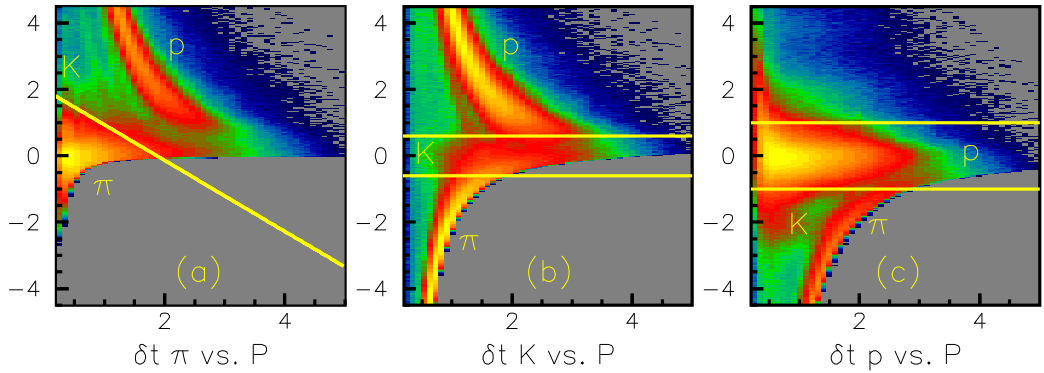}
\caption{Time difference $\delta t$ (ns) between the SC system and the flight 
time calculated for specific hadron-mass hypotheses plotted against the
hadron momentum $P$ (GeV) for (a) $\pi^+$, (b) $K^+$, and (c) $p$ at 
5.754~GeV.  Each plot shows the timing cuts imposed upon the spectra for 
the given mass hypothesis.}
\label{timing}
\end{figure}

The algorithm used for hadron identification was slightly different between
the {\tt e1c} and {\tt e1-6} data sets.  For the {\tt e1c} data set, the
final-state particles were identified with momentum-dependent cuts on the
momentum vs.~mass distribution (to account for the worsening resolution of
CLAS with increasing momentum).  For the {\tt e1-6} analysis, hadron
identification was performed using a timing cut.  The timing quantity of
interest ($\delta t = t_1 -t_2$) was the difference in the time between the
measured flight time for a particle from the event vertex to the SC system
($t_1$) and that expected for a given hadron type ($t_2$). The quantity
$t_2$ was computed for all positively charged particles assuming the mass
of the pion, kaon, and proton.  A small $\delta t$ indicates that the
correct mass hypothesis has been made. The timing cuts are defined such
that only one mass hypothesis can be satisfied for a given hadron.
Fig.~\ref{timing} shows the $\delta t$ plots used to identify final-state
$\pi^+$, $K^+$, and $p$ candidates.

%%%%%%%%%%%%%%%%%%%%%%%%%%%%%%%%%%%%%%%%%%%%%%%%%%%%%%%%%%%%%%%%%%%%%%%%%
%%%%%%%%%%%%%%%%%%%%%%%%%%%%%%%%%%%%%%%%%%%%%%%%%%%%%%%%%%%%%%%%%%%%%%%%%

Prior to imposition of final particle identification cuts and in order to
ensure an optimal resolution for the hyperon missing mass spectrum, the
reconstructed momenta for the electrons and charged hadrons in the final
state were corrected for small imperfections in the torus magnetic field map
and the drift chamber alignment by using reactions with over-determined
kinematics. The size of the momentum corrections $(\delta p /p)$ for each 
of the final-state particles is on the order of 1\%.

\subsection{Acceptance corrections}
\label{add-cuts}

In order to correct the yields for the detector acceptance, it is necessary 
to employ cuts that define the regions of CLAS where the detection efficiency 
is reasonably large and uniform.  These fiducial cuts for both electrons and 
positive hadrons depend on momentum, angle, and torus field setting.  For the 
electron, the CLAS acceptance is determined mostly by the limits of the 
azimuthal angle $\phi_e$ acceptance in each sector.  The $\phi_e$ limits are 
determined by a marked drop in the collection efficiency of the CC at the 
edges of the detector.  Additional fiducial cuts for all charged particles are
designed to exclude regions of non-uniform acceptance from attenuation
due to interactions with the mini-torus coils, the torus cryostat, or
from the edges of the drift chamber acceptance.

The acceptance correction was based on an analytic calculation that determined 
the geometrical acceptance factor on an event-by-event basis given the $\phi$ 
acceptance of each final-state particle within the defined geometrical fiducial 
region.  This factor accounted for losses due to kaon decays in-flight, bad 
scintillator paddles in the SC system, and the $\Lambda \to p \pi^-$ branching 
ratio.  Typical acceptances for the $KY$ reactions requiring detection of the 
$e'K^+p$ final state are at the level of 5\% to 20\%.  A detailed comparison 
between the nominal geometric acceptance correction and a full GEANT acceptance 
function was performed for the {\tt e1c} analysis (see Ref.~\cite{mozer}).  
Both methods were shown to have very similar functional forms.  However, the 
beauty of the asymmetry approach employed for this analysis is that the results 
are relatively insensitive to the acceptance correction. Thus a much simpler 
analytic form was chosen over a full Monte Carlo approach.    

\subsection{Hyperon yield extraction}
\label{hyp_yield}

\begin{figure}[htbp]
\vspace{5.2cm}
\includegraphics{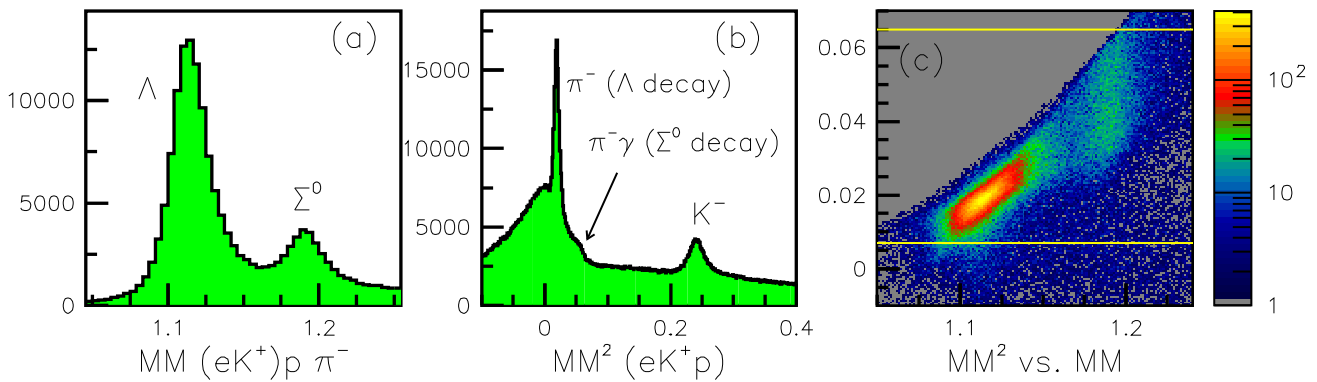} 
\caption{(Color online) $e'K^+p$ final state data from runs at 5.754~GeV. 
(a) Missing mass for $p(e,e'K^+)X$. (b) Missing mass squared for 
$p(e,e'K^+p)X$. (c) $MM^2(e'K^+p)$ vs. $MM(e'K^+)$ showing the cuts used to 
select the $K^+\Lambda$ and $K^+\Sigma^0$ event samples. (Units in GeV.)}
\label{hypfs}
\end{figure}

The reactions of interest are identified from missing-mass ($MM$) 
reconstructions of the $e'K^+$ final state.  Shown in Fig.~\ref{hypfs}(a) is 
the $MM(e'K^+)$ distribution for the $e'K^+p$ final state at 5.754~GeV.  This 
spectrum shows substantial, well-separated peaks for the ground state $\Lambda$ 
and $\Sigma^0$ hyperons. The width of the $\Lambda$ peak in this spectrum, 
summed over all $Q^2$ and $W$, is about 11~MeV.  Fig.~\ref{hypfs}(b) shows the 
$MM^2(e'K^+p)$ (missing mass squared) distribution.  Here the final-state 
proton can come from the decay of the $\Lambda$(1115) (missing $\pi^-$), the 
$\Sigma^0$(1192) (missing $\pi^- \gamma$), or the $\Lambda$(1520) (missing 
$K^-$). Fig.~\ref{hypfs}(a) requires a cut on the $MM^2(e'K^+p)$ spectrum in 
the range from 0.007 to 0.065~GeV$^2$, as shown in the correlation plot of 
Fig.~\ref{hypfs}(c), to reduce the contributions of particle misidentification 
background.  The final $K^+\Lambda$ and $K^+\Sigma^0$ yields are then extracted 
through the fitting procedures described below.

%%%%%%%%%%%%%%%%%%%% Figure : Hyperon Missing Mass %%%%%%%%%%%%%%%%%%%
%%%%%%%%%%%%%%%%%%%%%%%%%%%%%%%%%%%%%%%%%%%%%%%%%%%%%%%%%%%%%%%%%%%%%%%

The three components to the hyperon missing mass spectrum are the $K^+\Lambda$ 
final-state events, the $K^+\Sigma^0$ final-state events, and the 
particle-misidentification background (dominated by pions misidentified as 
kaons).  As discussed in Section~\ref{sec:polext}, these individual 
contributions must be determined to extract the $\Lambda$ and $\Sigma^0$ 
polarizations, and they have been measured through fits to the hyperon mass 
distributions. In this procedure, the $\Lambda$ and $\Sigma^0$ peaks were fit 
using templates derived from a phase-space GEANT Monte Carlo simulation.  The 
templates were generated with radiative effects turned on, which is necessary 
to account for the $\Lambda$ radiative tail beneath the $\Sigma^0$ peak.  The 
background contributions in each bin were studied employing two different 
procedures.  In the first, a background spectrum was derived from Monte Carlo 
using a phase space generator for multi-pion final states.  The final-state 
$\pi^+$ were then assigned the $K^+$ mass.  The resultant $MM(e'\pi^+)$
spectra were then sorted into the different analysis bins in $Q^2$, $W$, and 
$\cos \theta_K^{c.m.}$.  The second approach employed a third-order 
polynomial to fit the backgrounds in the spectra.  These two models for the 
background gave consistent answers, however the polynomial model was employed 
for the final fits as the Monte Carlo background distributions were 
statistically limited.

\begin{figure}[btp]   
\vspace{6.0cm}
\includegraphics{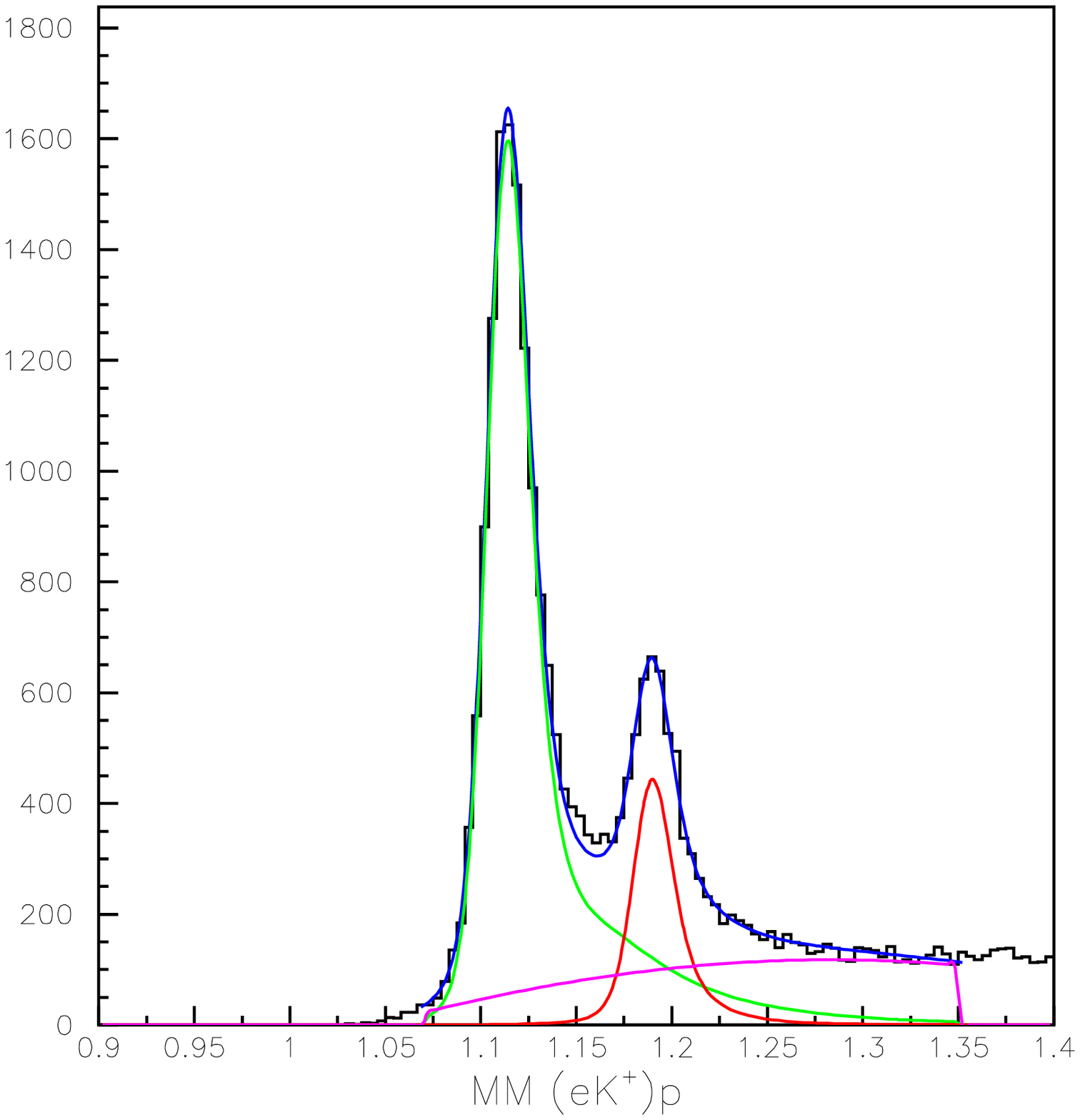}
\includegraphics{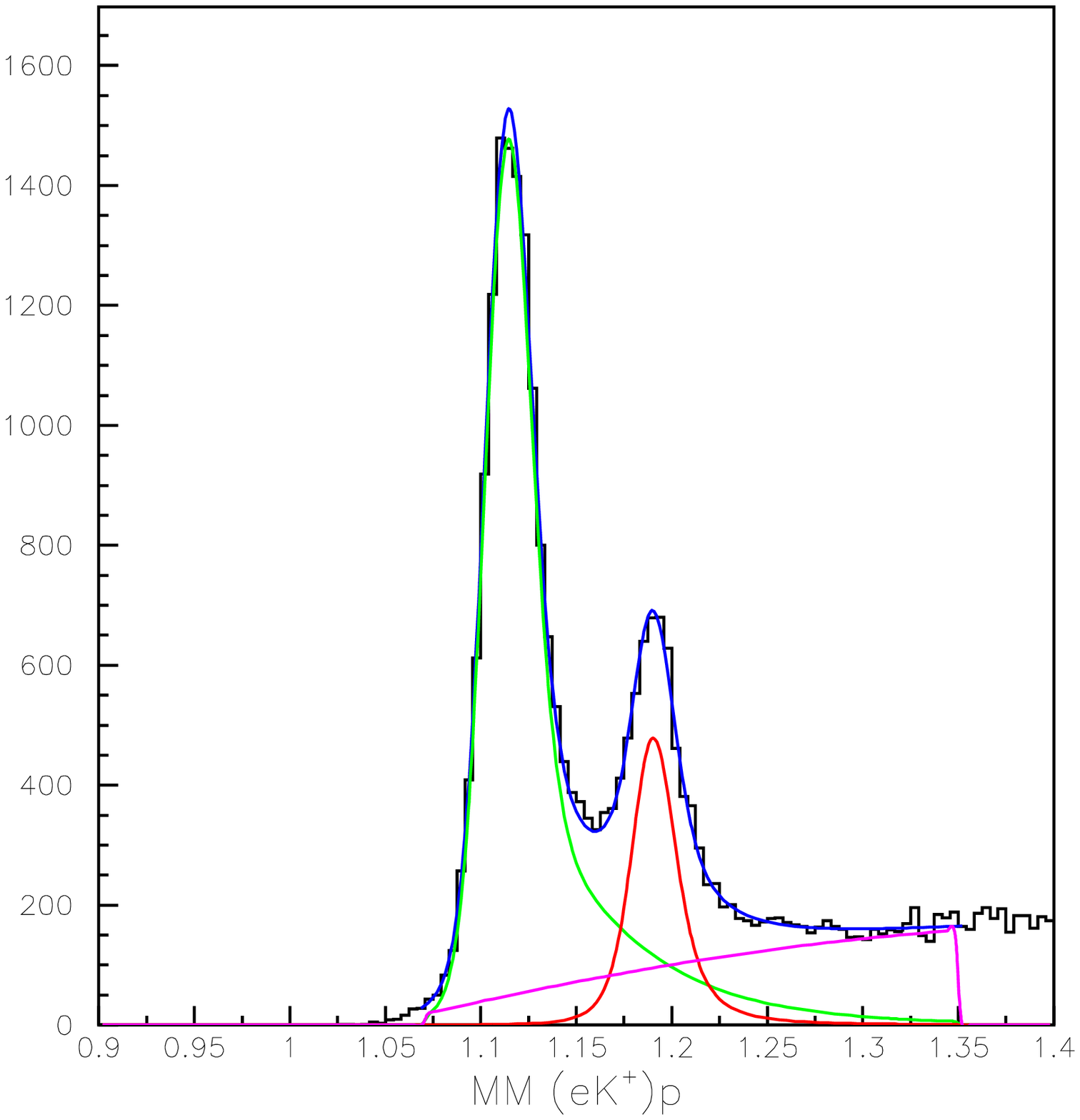}
\mbox{
\begin{picture}(-10,-10)(350,0)
\put(435,150){(a)}
\put(640,150){(b)}
\put(505,150){$Q^2$=2.32~GeV$^2$}
\put(505,138){$\chi^2/\nu$=79.59/63}
\put(707,150){$Q^2$=2.80~GeV$^2$}
\put(707,138){$\chi^2/\nu$=63.20/63}
\end{picture}}
\caption{(Color online) Sample hyperon spectrum fit results using hyperon 
templates derived from Monte Carlo ($\Lambda$:green curve, $\Sigma^0$:red
curve) and a third-order polynomial for the background (magenta curve).  The 
smoothed templates have been allowed to shift along the mass axis and 
convoluted with a Gaussian to give the best $\chi^2$ per degree of freedom
($\chi^2_\nu$) values for the fits.  These distributions at 5.754~GeV
are summed over all $W$ and $\cos \theta_K^{c.m.}$ for central $Q^2$
values as indicated. The blue curve shows the full fit result.}
\label{fit_examples}
\end{figure}

The form of the hyperon spectrum fit in each analysis bin was given by
\begin{equation}
MM = A \cdot \Lambda_{template} + B \cdot \Sigma_{template} + P(3)_{bck},
\end{equation}

\noindent
where $\Lambda_{template}$ and $\Sigma_{template}$ are the simulated
hyperon distributions with weighting factors $A$ and $B$, respectively,
and $P(3)_{bck}$ is a third-order polynomial describing the background. In 
performing these fits, the $\Lambda$ and $\Sigma^0$ Monte Carlo templates 
were allowed to shift up to $\pm$10~MeV to match the data. In addition, the 
hyperon templates were individually convoluted with a Gaussian with a width 
chosen to minimize the $\chi^2$ of the fits in each bin.  This was necessary 
as the resolution of the Monte Carlo was not a perfect match to the real data.  
Finally, the Monte Carlo templates were smoothed using a spline fit to remove 
the effects of statistical fluctuations in the simulation samples.  Fits for 
two representative bins are shown in Fig.~\ref{fit_examples}. 

%%%%%%%%%%%%%%%%%%%%%%%%%%%%%%%%%%%%%%%%%%%%%%%%%%%%%%%%%%%%%%%%%%%%%%%%%%%%%
%%%%%%%%%%%%%%%%%%%%%%%%%%%%%%%%%%%%%%%%%%%%%%%%%%%%%%%%%%%%%%%%%%%%%%%%%%%%%

\begin{figure}[tbp]   
\vspace{6.2cm}
\includegraphics{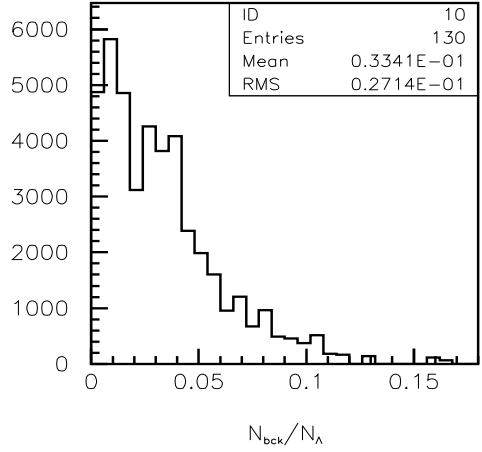}
\caption{Ratio of the number of background counts to $\Lambda$ counts in 
the $\Lambda$ analysis mass window weighted by the uncertainty in the ratio
for each of the kinematic bins in this analysis at 5.754~GeV.}
\label{lam_bck}
\end{figure}

\begin{figure}[btp]   
\vspace{5.2cm}
\includegraphics{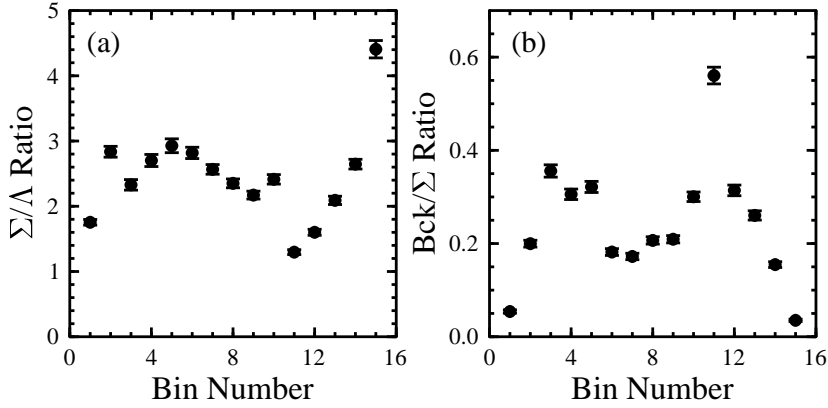}
\caption{Ratios of the number of counts in the $\Sigma^0$ analysis mass 
window for the data at 5.754~GeV.  (a) $\Sigma^0/\Lambda$ ratio. 
(b) Background/$\Sigma^0$ ratio.}
\label{ratios}
\end{figure}

Fig.~\ref{lam_bck} shows the results of the yield fits in terms of 
$F_{bck}=N_{bck}/N_\Lambda$ for events in our $\Lambda$ mass window 
($MM(e'K^+)$ from 1.080 to 1.160~GeV) for each of our data sorts.  The plot 
shows the distribution of the background ratio for each bin weighted by the 
uncertainty in the ratio.  As the particle misidentification background
was found to be relatively independent of kinematics, a single value of 
$F_{bck}$=3.3\% has been employed for all 
analysis bins based on the weighted mean of Fig.~\ref{lam_bck}. Additionally,
from the spectrum fits we have extracted the ratios 
$F_{Y'}=N_{\Sigma^0}/N_\Lambda$ and $F_{bck}=N_{bck}/N_{\Sigma^0}$ in our 
$\Sigma^0$ mass window ($MM(e'K^+)$ from 1.175 to 1.213~GeV) (see 
Fig.~\ref{ratios}).  These ratios are relatively independent of the kinematics.
However, due to the sensitivity of ${\cal P}'_\Sigma$ to the number of 
$\Lambda$ events in the $\Sigma^0$ mass window, we have employed the measured 
$\Sigma^0$, $\Lambda$, and background yields in the $\Sigma^0$ analysis
for each kinematic bin.

%%%%%%%%%%%%%%%%%%%%%%%%%%%%%%%%%%%%%%%%%%%%%%%%%%%%%%%%%%%%%%%%%%%%%%%%%%%%%
%%%%%%%%%%%%%%%%%%%%%%%%%%%%%%%%%%%%%%%%%%%%%%%%%%%%%%%%%%%%%%%%%%%%%%%%%%%%%

%%%%%%%%%%%%%%%%%%%%%%%%%%%%%%%%%%%%%%%%%%%%%%%%%%%%%%%%%%%%%%%%%%%%%%%%%%%%
%%%%%%%%%%%%%%%%%%%%%%%%%%%%%%%%%%%%%%%%%%%%%%%%%%%%%%%%%%%%%%%%%%%%%%%%%%%%

\subsection{Background polarization corrections}
\label{bck_pol}

\begin{figure}[tbhp]   
\vspace{7.5cm}
\includegraphics{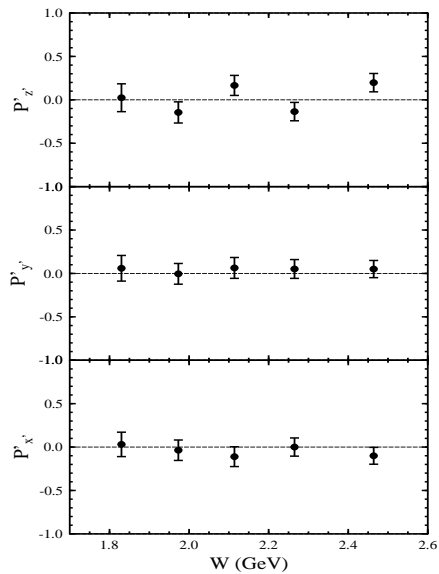}
\caption{Polarization ${\cal P}'$ vs. $W$ for the pion misidentification 
background in the $(x',y',z')$ coordinate system for the data summed over 
all $Q^2$ and $d\Omega_K^{c.m.}$.}
\label{pol_bck}
\end{figure}

Once the number of $\Lambda$, $\Sigma^0$, and background events are 
determined in each of the respective hyperon mass windows, the measured 
polarization must be corrected as discussed in Sec.~\ref{sec-pol-extract}.  
The background polarization was measured by sorting data from the 
$ep \to e'\pi^+pX$ reaction and assigning the $K^+$ mass to the $\pi^+$ 
events.  The analysis procedure then followed all of the same steps and 
procedures as for the hyperon polarization analysis.  The measured background 
polarization for one typical data sort is shown in Fig.~\ref{pol_bck} as a 
function of $W$ (summed over all other kinematic variables).  The results are 
consistent with ${\cal P}'_{bck}$=0 for all axes in all of the sorts 
investigated.  

Within the tight $\Lambda$ cuts, there was no measurable level of 
$\Sigma^0$ contamination.  However, within the $\Sigma^0$ mass window 
(1.175 to 1.213~GeV) there is significant contamination from both the 
$\Lambda$ radiative tail and pion background (see Table~\ref{statistics}).  
To correct the $\Sigma^0$ polarization for the $\Lambda$ tail, 
${\cal P}'_\Lambda$ was determined following our nominal prescription for 
determining the transferred $\Lambda$ polarization, where the $\Lambda$ 
data were binned in the same bins as the $\Sigma^0$ data.  This value of 
${\cal P}'_\Lambda$ was then used in Eq.(\ref{cor_sig_pol}) to calculate 
the corrected $\Sigma^0$ polarization.

%%%%%%%%%%%%%%%%%%%%%%%%%%%%%%%%%%%%%%%%%%%%%%%%%%%%%%%%%%%%%%%%%%%%%%%%%%%%
\begin{table}[tbhp]
\begin{center}
\begin{tabular} {c|c|c|c} \hline
%\rule{0mm}{4mm}
$W$ Bin (GeV) & $N_\Sigma$ & $N_\Lambda$ & $N_\pi$ \\ \hline
1.825 & 3067$\pm$79 & 2746$\pm$20 & ~92$\pm$27 \\
1.975 & 3872$\pm$85 & 2544$\pm$22 & 214$\pm$28 \\
2.125 & 2425$\pm$72 & 2074$\pm$20 & 329$\pm$24 \\
2.275 & 3160$\pm$80 & 2050$\pm$20 & 501$\pm$26 \\
2.470 & 3413$\pm$80 & 1758$\pm$19 & 408$\pm$19 \\ \hline
\end{tabular}
\end{center}
\caption{Results from the 5.754~GeV yield fits for $N_\Sigma$, $N_\Lambda$, 
and $N_{bck}$ with statistical uncertainties in a $\pm 2\sigma$ $\Sigma^0$ 
mass window for a sort dividing the $\Sigma^0$ analysis into 5 bins in $W$.}
\label{statistics}
\end{table}
%%%%%%%%%%%%%%%%%%%%%%%%%%%%%%%%%%%%%%%%%%%%%%%%%%%%%%%%%%%%%%%%%%%%%%%%%%%%

%%%%%%%%%%%%%%%%%%%%%%%%%%%%%%%%%%%%%%%%%%%%%%%%%%%%%%%%%%%%%%%%%%%%%%%%%%%%%
%%%%%%%%%%%%%%%%%%%%%%%%%%%%%%%%%%%%%%%%%%%%%%%%%%%%%%%%%%%%%%%%%%%%%%%%%%%%%

\subsection{Radiative corrections}

No radiative corrections have been applied to the data in this analysis.
These have purposefully been avoided by employing relatively tight cuts on
the reconstructed hyperon spectrum for the $K^+\Lambda$ and $K^+\Sigma^0$
events and by accounting for the $K^+\Lambda$ radiative tail events within 
the $K^+\Sigma^0$ event sample. This is expected to be a reasonable approach 
as the radiative effects are independent of the beam helicity and thus should 
effectively cancel out of the asymmetry calculation.  With our relatively tight 
hyperon mass cuts, the maximum radiated photon energy is only about 50~MeV, 
which has a negligible impact on our computed $\cos \theta_p^{RF}$ values with 
respect to each quantization axis.

\subsection{Bin averaging corrections}

The bin sizes for $W$, $Q^2$, and $\cos \theta_K^{c.m.}$ in this analysis
were chosen to roughly equalize the statistical precision of each
polarization data point.  To account for the finite bin sizes and the
variation of the cross section over the bins, we quote our polarization
results at the bin means.  The bin mean was determined by measuring the mean 
of the acceptance-corrected yield distribution over the kinematic bins of 
interest.  As might be expected, the largest differences between the bin mean 
and the bin center occur where the bins are larger. The kinematic
bin means for each data sort are given in Section~\ref{sec:results}.

%
%%%%%%%%%%%%%%%%%%%%%%%%%%%%%%%%%%%%%%%%%%%%%%%%%%%%%%%%%%%%%%%%%%%%%%%%%%%
%

\section{SYSTEMATIC UNCERTAINTY ANALYSIS}
\label{systematics}

In this section we examine the sources of systematic uncertainty that affect 
the extracted polarization observables for the 5.754~GeV data set.  The 
assigned systematics for the 4.261~GeV $K^+\Lambda$ data set are described in 
Ref.~\cite{cnote} and are given by $\delta {\cal P}'_{sys} < 0.084$. The 
contributions to the total systematic uncertainty belong to one of four 
general categories: Polarization extraction, beam-related factors, acceptance 
function, and background contributions.   As the statistics for the $K^+\Lambda$ 
final state dominate those for the $K^+\Sigma^0$ final state, the determination 
of the assigned systematics for both final states is based on analysis of the 
$K^+\Lambda$ data.  The exception to this is the assignment of a separate 
systematic for the background contributions.  The final systematic uncertainty 
compilation for the 5.754~GeV measurements is given in Table~\ref{systab}.  

The procedure used to assign a systematic uncertainty to each source
within a given category is to compare the measured polarization ${\cal P}'$ 
for all kinematic bins with the nominal 
analysis cuts or procedures ($nom$) to that with modified cuts or procedures 
($mod$).  The spread in the difference of the polarization over all data 
points, $\Delta {\cal P}' = {\cal P}'_{nom} - {\cal P}'_{mod}$, is used as a 
measure of the systematic uncertainty for a given source.  The estimated 
common uncertainty for a given source is the weighted root-mean-square 
(r.m.s.) of $\Delta {\cal P}'$ for all points given by
\begin{equation}
\delta {\cal P}'_{sys} = \sqrt{ \frac{\sum_{i=1}^{N} 
(\Delta {\cal P}'_i)^2/(\delta {\cal P}'_i)^2}
{\sum_{i=1}^{N} 1/(\delta {\cal P}'_i)^2}},
\end{equation}

\noindent 
where the sums are over all $N$ data points and $\delta {\cal P}'_i$ is
the statistical uncertainty of the $i^{th}$ data point.  In the studies
done for this analysis, a common systematic uncertainty is applied for
all data points as the kinematic dependence of the $\Delta {\cal P}'$ 
distributions was found to be minimal. In each of the systematic
uncertainty studies performed for this analysis, the widths of the
$\Delta {\cal P}'$ distributions were much larger than the measured
centroids, which are all consistent with zero. Thus the assignments are 
believed to be rather conservative.

\subsection{Polarization extraction}
\label{pol_sys}

The polarization has been determined by using two different analysis 
approaches.  The nominal technique is the asymmetry approach described in 
Section~\ref{sec:polext}.  An alternative approach is to extract the 
polarization from the ratio of the acceptance-corrected, helicity-gated 
yields via
\begin{equation}
R = \frac{\sigma^+}{\sigma^-} = \frac{1 + \alpha_\Lambda P_b  
{\cal P}_\Lambda' \cos \theta_p^{RF}}{1 - \alpha_\Lambda P_b  
{\cal P}_\Lambda' \cos \theta_p^{RF}}.
\end{equation}

\noindent
The difference between these two techniques resulted in an estimated systematic
uncertainty of $\delta {\cal P}'_{sys} = 0.008$.

A systematic uncertainty arises from the somewhat arbitrary choice made for 
the $\cos \theta_p^{RF}$ bin size.  Nominally the data were sorted into six 
bins in the rest frame proton angle.  A comparison of the nominal polarization 
results with the extraction from a sort with four and eight bins in this 
variable resulted in a weighted r.m.s. of $\delta {\cal P}'_{sys} = 0.018$.  
The difference in the polarization results is effectively due to the fitting 
algorithm employed in which the centroids of the $\cos \theta_p^{RF}$ bins are 
assigned to the center of the bin.  When the number of bins is reduced, the fit 
results are more sensitive to the bin content.

The final systematic uncertainty contribution in this category arises due to
the uncertainty in the weak decay asymmetry parameter $\alpha_\Lambda$.  This
uncertainty gives rise to a scale-type uncertainty on the extracted
polarization (the same for both $\Lambda$ and $\Sigma^0$ hyperons) given
by $\delta {\cal P}'_{sys} = \vert {\cal P}'_Y \vert 
\delta \alpha_\Lambda/\alpha_\Lambda = 0.02 \vert {\cal P}'_Y \vert$.

\subsection{Beam-related factors}
\label{beam_sys}

There are two possible contributions to the systematic uncertainty related 
to the beam.  The first factor is associated with the beam polarization 
measurement from the M{\o}ller polarimeter system.  This arises from the 
uncertainty in the M{\o}ller target foil polarization, the statistical 
uncertainty in the measurements, as well as a contribution from variations 
of the polarization measurements over time. These contributions have been
estimated for CLAS polarization measurements to be 4\%. The associated 
uncertainty in the hyperon polarization is 
$\delta {\cal P}'_{sys} = \vert {\cal P}'_Y \vert 
\delta P_b/P_b = 0.04 \vert {\cal P}'_Y \vert$.

The second beam-related contribution is the beam charge asymmetry that results 
from a difference in the electron beam intensity for the two beam helicity 
states.  From studies of the 5.754~GeV data set, the beam charge asymmetry 
was below the $10^{-3}$ level and no detectable difference between the 
helicity-gated live times was found, thus no systematic contribution was
assigned.

\subsection{Acceptance function}
\label{acc_func}

There are several factors that go into the systematic uncertainty
associated with the form of our acceptance correction and with the choices 
made to implement this correction, which include the specific form of the 
fiducial cuts used to define the azimuthal extent of the acceptance as a 
function of polar angle and the minimum acceptance cutoff.  
In order to assign a systematic uncertainty associated with the acceptance 
correction, we have compared the extracted polarizations with and without 
the geometric acceptance corrections.  The r.m.s. width of the difference
distribution was assigned as the systematic uncertainty for the acceptance 
correction. This value, $\delta {\cal P}'_{sys} = 0.033$, is believed 
to be a very conservative estimate.

To study the effects of the fiducial cuts employed to define the azimuthal 
acceptance for electrons and hadrons, two different sets of fiducial cuts 
were defined in the analysis.  A loose cut (the nominal cut) was designed to 
define the azimuthal acceptance edge of CLAS as a function of momentum, and 
a second cut was designed to be several degrees tighter than the nominal cut.  
Comparisons of the extracted polarizations between these two cut definitions
gave an r.m.s. width of $\delta {\cal P}'_{sys} = 0.020$, which represents
the assigned systematic uncertainty.

The minimum acceptance cutoff translates into a maximum acceptance
weight.  The minimum acceptance cutoff was nominally set at 10\% (a
somewhat arbitrary choice) for the $e'K^+p$ final state.  For our
study, we varied the acceptance cutoff by $\pm$20\% relative to the
nominal cutoff value.  The assigned systematic uncertainty, given by
the r.m.s. width of the polarization difference distribution, is
$\delta {\cal P}'_{sys} = 0.025$.

\subsection{Background contributions}
\label{bck_sys}

\subsubsection{$e'K^+\Lambda$ final state}

Our analysis of the backgrounds found no measurable level of $\Sigma^0$ 
contamination within our final $K^+ \Lambda$ event sample.  However, there 
is a few percent contamination of $\pi^+$ misidentification events that 
remain beneath the $\Lambda$ peak that serve to dilute the measured 
$\Lambda$ polarization. To estimate the systematic uncertainty associated 
with our subtraction technique, we have compared our nominal polarization 
results to results obtained assuming no pion background. Clearly, this would 
result in an overestimate of the systematic uncertainty so we have used 
one-half of the difference, or $\delta {\cal P}'_{sys}$ = 0.009. While an 
arbitrary choice, this represents a conservative estimate and is small compared 
to other sources of systematic uncertainties.

\subsubsection{$e'K^+\Sigma^0$ final state}

The uncertainties in the backgrounds from $K^+\Lambda$ and pion
misidentification have a much bigger impact on the extracted $\Sigma^0$
polarization compared to the $\Lambda$ analysis.  Therefore, it is important
to study these effects separately for this final state.  Our approach to 
assign a systematic uncertainty due to the fit uncertainties of the 
contributing backgrounds beneath the $\Sigma^0$ is to allow the extracted 
yields to vary by $\pm$10\% from the fit value and to study the effect on the 
extracted $\Sigma^0$ polarization. Variations of the background levels of 
$\pm$10\% amounted to variations on the fitted background yields of 
$N_\Lambda \pm 2 \sigma_{N_\Lambda}$ and $N_{bck} \pm 2 \sigma_{N_{bck}}$.
Our studies indicated that the maximum change in the measured $\Sigma^0$ 
polarization was $\pm$0.10, which we have assigned as the associated 
systematic uncertainty $\delta {\cal P}'_{sys}$.

\subsection{Final systematic uncertainty accounting}

Our final systematic uncertainty accounting for the 5.754~GeV $\Lambda$
and $\Sigma^0$ ${\cal P}'$ data is included in Table~\ref{systab} listing 
all of the sources discussed above. The final value for the total systematic 
uncertainty results from adding all the individual contributions in 
quadrature. (Additions in quadrature in Table~\ref{systab} are represented 
by the notation $\oplus$).

%%%%%%%%%%%%%%%%%%%%%%%%%%%%%%%%%%%%%%%%%%%%%%%%%%%%%%%%%%%%%%%%%%%%%%%%%
\begin{table}[htbp]
\begin{center}
\begin{tabular} {c|c|c} \hline
Category                 & Contribution          & Systematic Uncertainty \\ 
\hline
Polarization Extraction  & Functional Form       & 0.008 \\
                         & Bin Size              & 0.018  \\
                         & Asymmetry Parameter   & 0.02 ${\cal P}'_Y$ \\ \hline
Beam-Related Factors     & Beam Polarization     & 0.04 ${\cal P}'_Y$ \\ \hline
Acceptance Function      & Fiducial Cut Form     & 0.020 \\
                         & Acceptance Correction & 0.033 \\
                         & Acceptance Cutoff    & 0.025 \\ \hline
Background Contributions & Pion and $\Lambda$    & 0.009 ($\Lambda$), 0.100 
($\Sigma^0$) \\ 
                         & contamination & \\ \hline
\multicolumn{2} {c|} {\bf $\langle$ Total Systematic Uncertainty $\rangle$} 
& 0.051 ($\Lambda$), 0.112 ($\Sigma^0$) \\ 
\multicolumn{2} {c|} {~} & $\oplus$ $0.045 {\cal P}'_Y$ \\ \hline
\end{tabular}
\caption{Summary table of the systematic uncertainty assignments 
$\delta {\cal P}'_{sys}$ for the measured $\Lambda$ and $\Sigma^0$ 
polarizations at 5.754~GeV.  Note that the contributions from the 
uncertainties in the beam polarization and weak decay asymmetry parameter 
(given by $0.045 {\cal P}'_Y$) are added in quadrature (represented by the 
$\oplus$ notation) to the other sources.}
\label{systab}
\end{center}
\end{table}
%%%%%%%%%%%%%%%%%%%%%%%%%%%%%%%%%%%%%%%%%%%%%%%%%%%%%%%%%%%%%%%%%%%%%%%%%

\begin{figure}[htbp]
\vspace{5.5cm}
\includegraphics{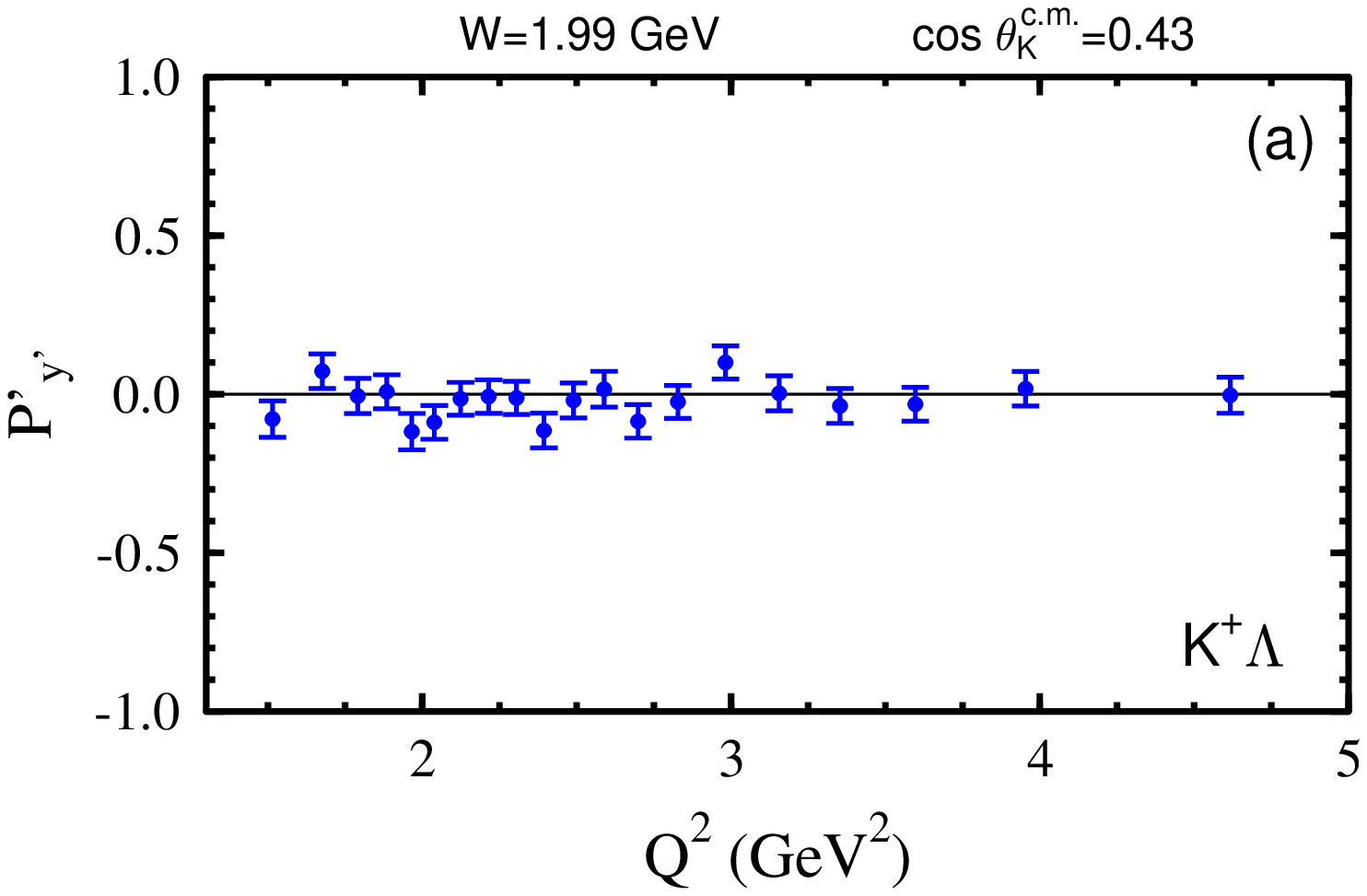}
\includegraphics{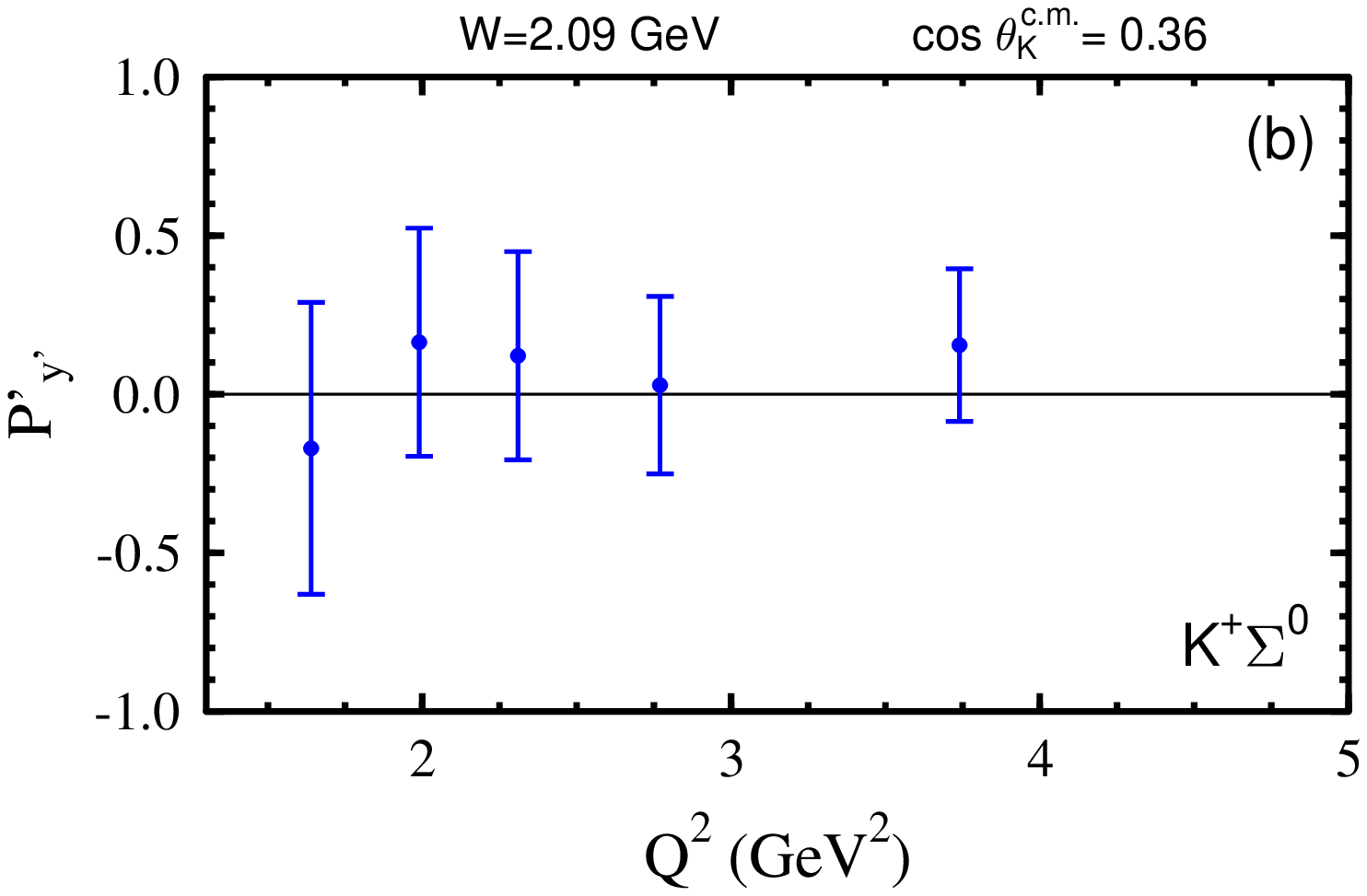}
\caption{(Color online) Distributions of the transferred $\Lambda$ (a)
and $\Sigma^0$ (b) polarization components ${\cal P}_{y'}'$ vs. $Q^2$ for 
two of our analysis bins.  The error bars shown represent the statistical
uncertainties only.}
\label{norm}
\end{figure}

One way to verify the veracity of the final systematic uncertainty assignment 
is to look at the deviations of the normal components of the extracted 
$\Lambda$ and $\Sigma^0$ polarizations (i.e. along the $y'$ and $y$ axes).  
Averaged over all analysis bins, the weighted mean of the ${\cal P}_{y'}'$ 
and ${\cal P}_y$ components for the $\Lambda$ is 0.067 and for the $\Sigma^0$ 
is 0.134.  Both of these values are consistent with our total systematic 
uncertainty assignments in Table~\ref{systab}.  The extracted normal components 
for one of our data sorts for the $\Lambda$ and $\Sigma^0$ hyperons are shown 
in Fig.~\ref{norm}.

%%%%%%%%%%%%%%%%%%%%%%%%%%%%%%%%%%%%%%%%%%%%%%%%%%%%%%%%%%%%%%%%%%%%%%%%%
%%%%%%%%%%%%%%%%%%%%%%%%%%%%%%%%%%%%%%%%%%%%%%%%%%%%%%%%%%%%%%%%%%%%%%%%%

%
%%%%%%%%%%%%%%%%%%%%%%%%%%%%%%%%%%%%%%%%%%%%%%%%%%%%%%%%%%%%%%%%%%%%%%%%%%%
%

\section{RESULTS AND DISCUSSION}
\label{sec:results}

\subsection{$\Lambda$ polarization transfer}
\label{lambda}

\begin{figure}[htbp]
\vspace{9.5cm}
\includegraphics{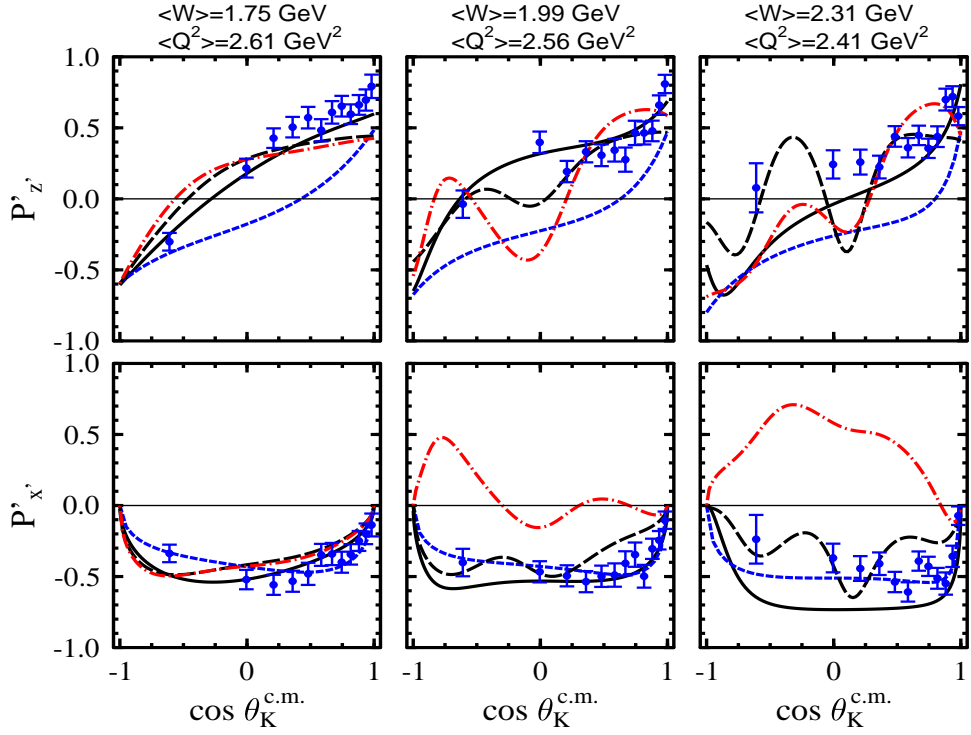}
\caption{(Color online) Transferred $\Lambda$ polarization components 
${\cal P}'$ with respect to the $(x',z')$ axes vs.~$\cos \theta_K^{c.m.}$ 
for three bin-averaged $W$/$Q^2$ values as indicated for a beam energy of 
5.754~GeV.  The curves are calculations from the MB isobar model
\cite{mart_code} (solid -- black), the GLV Regge model~\cite{guidal_code} 
(short dash -- blue), and the RPR model~\cite{ghent_code} variant including a
$P_{11}(1900)$ state (dot-dash -- red) and a $D_{13}(1900)$ state (long dash 
-- black).}
\label{dpol1}
\end{figure}

\begin{figure}[htbp]
\vspace{9.5cm}
\includegraphics{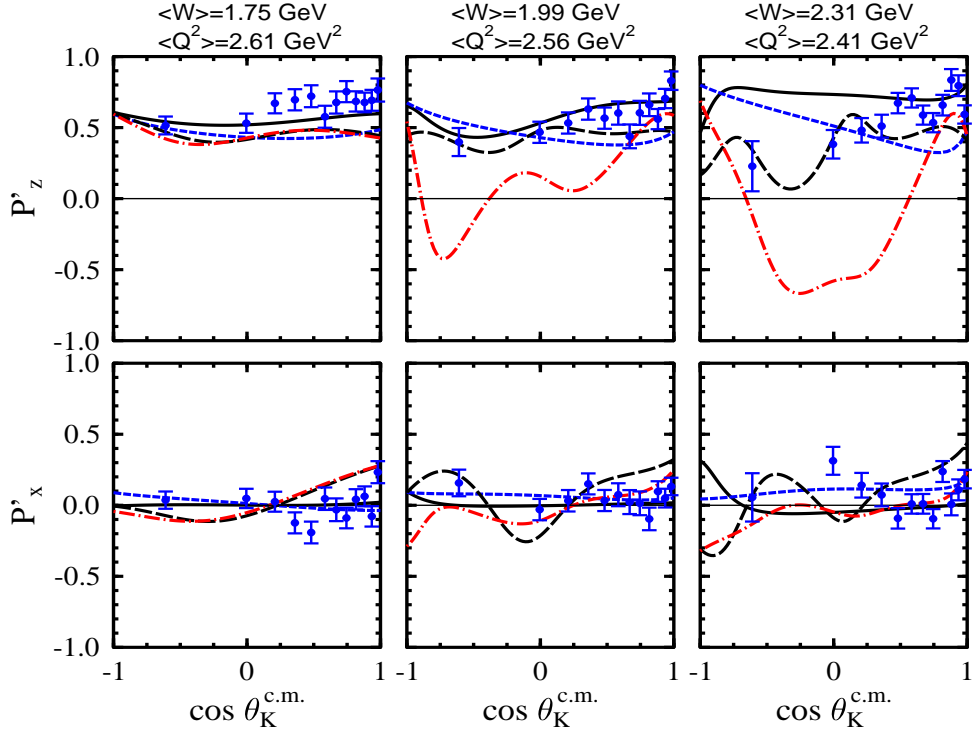}
\caption{(Color online) Transferred $\Lambda$ polarization components 
${\cal P}'$ with respect to the $(x,z)$ axes vs.~$\cos \theta_K^{c.m.}$ for 
three bin-averaged $W$/$Q^2$ values as indicated for a beam energy of 
5.754~GeV.  The model calculations are as indicated in Fig.~\ref{dpol1}.}
\label{dpol4}
\end{figure}

\begin{figure}[htbp]
\vspace{13.0cm}
\includegraphics{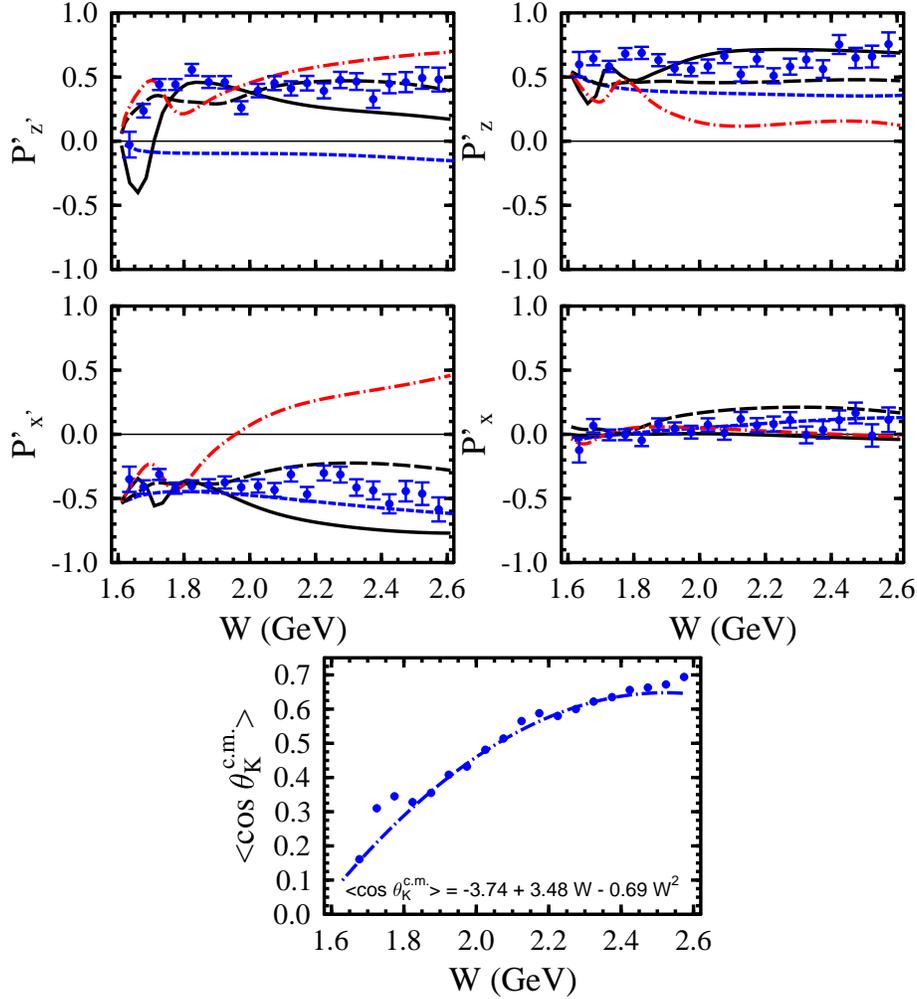}
\caption{(Color online) Transferred $\Lambda$ polarization components 
${\cal P}'$ with respect to the $(x',z')$ axes (upper left panels) and
$(x,z)$ axes (upper right panels) vs.~$W$ (GeV) for 
$\langle Q^2\rangle$=2.54~GeV$^2$ for a beam energy of 5.754~GeV. The 
bin-averaged $\cos \theta_K^{c.m.}$ values for each $W$ point are shown in 
the lower plot. A fit to $\langle \cos \theta_K^{c.m.} \rangle$ is provided 
by a second-order polynomial in $W$ as indicated on the plot.  The model 
calculations are as indicated in Fig.~\ref{dpol1}.}
\label{dpol2}
\end{figure}

\begin{figure}[htbp]
\vspace{8.5cm}
\includegraphics{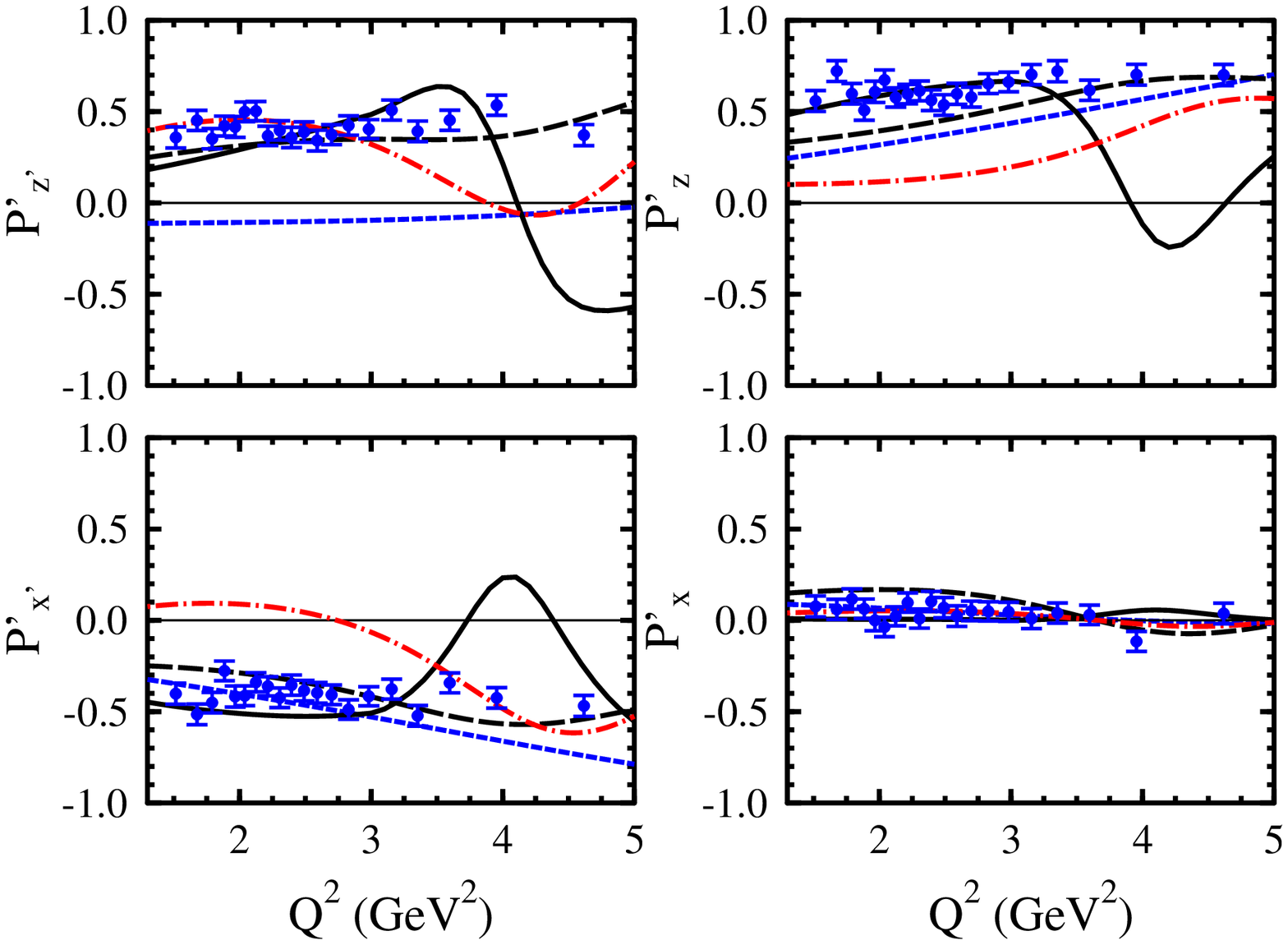}
\caption{(Color online) Transferred $\Lambda$ polarization components 
${\cal P}'$ with respect to the $(x',z')$ axes (left panels) and $(x,z)$ axes 
(right panels) vs. $Q^2$ (GeV$^2$) for $\langle W\rangle$=1.99~GeV and 
$\langle \cos \theta_K^{c.m.}\rangle$=0.43 for a beam energy of 5.754~GeV. The
model calculations are as indicated in Fig.~\ref{dpol1}.}
\label{dpol3}
\end{figure}

Our results for the transferred $\Lambda$ polarization acquired at a
beam energy of 5.754~GeV are shown in Figs.~\ref{dpol1} through \ref{dpol3} 
compared to several model calculations.  The error bars in these figures 
include statistical but not systematic uncertainties, which we estimate to 
be $0.051 \oplus 0.045 {\cal P}'_\Lambda$ on the polarization.  The full 
data set is contained in the CLAS database~\cite{database}.

Figs.~\ref{dpol1} and \ref{dpol4} show the dependence of ${\cal P}'_{x',z'}$ 
and ${\cal P}'_{x,z}$ with respect to $\cos \theta_K^{c.m.}$ for the three 
bin-averaged $W$ and $Q^2$ values indicated in the figure. Fig.~\ref{dpol1} 
shows that the value of ${\cal P}'_{z'}$ decreases smoothly with increasing 
scattering angle, whereas ${\cal P}'_{x'}$ decreases with increasing angle 
until $\cos \theta_K^{c.m.} \approx 0.8$, at which point it levels off to a 
value of about $-0.5$ over the range of $\cos\theta_K^{c.m.}$ covered by the
experiment.  The fact that the ${\cal P}'_{x'}$ data approach zero at
$\cos\theta_K^{c.m.} = 1$ is simply a result of angular-momentum conservation, 
which also requires ${\cal P}'_{x'} = 0$ at $\cos \theta_K^{c.m.} = -1$.  The 
dependence of the polarization along the $(x,z)$ axes in Fig.~\ref{dpol4} is 
qualitatively different.  The polarization along ${\cal P}'_x$ is roughly 
zero everywhere, whereas ${\cal P}'_z$ is relatively constant (at least over
the angle range $\cos \theta_K^{c.m.} > 0$ where the statistics are reasonable)
with an average value of $\sim$0.6. This may be hinting at a simple reaction 
mechanism (see Section~\ref{quark_models}).

%%%%%%%%%%%%%%%%%%%%%%%%%%%%%%%%%%%%%%%%%%%%%%%%%%%%%%%%%%%%%%%%%%%%%%%%%
%%%%%%%%%%%%%%%%%%%%%%%%%%%%%%%%%%%%%%%%%%%%%%%%%%%%%%%%%%%%%%%%%%%%%%%%%

%%%%%%%%%%%%%%%%%%%%%%%%%%%%%%%%%%%%%%%%%%%%%%%%%%%%%%%%%%%%%%%%%%%%%%%%%
%%%%%%%%%%%%%%%%%%%%%%%%%%%%%%%%%%%%%%%%%%%%%%%%%%%%%%%%%%%%%%%%%%%%%%%%%

%%%%%%%%%%%%%%%%%%%%%%%%%%%%%%%%%%%%%%%%%%%%%%%%%%%%%%%%%%%%%%%%%%%%%%%%%
%%%%%%%%%%%%%%%%%%%%%%%%%%%%%%%%%%%%%%%%%%%%%%%%%%%%%%%%%%%%%%%%%%%%%%%%%

%%%%%%%%%%%%%%%%%%%%%%%%%%%%%%%%%%%%%%%%%%%%%%%%%%%%%%%%%%%%%%%%%%%%%%%%%
%%%%%%%%%%%%%%%%%%%%%%%%%%%%%%%%%%%%%%%%%%%%%%%%%%%%%%%%%%%%%%%%%%%%%%%%%

The polarization with respect to $W$ is shown in Fig.~\ref{dpol2} for a beam
energy of 5.754~GeV. Note that there is a strong dependence of the bin-averaged 
$\cos \theta_K^{c.m.}$ value with respect to $W$ in these kinematics.  The 
central $\cos \theta_K^{c.m.}$ values extracted from the analysis are 
reasonably represented by a fit to a second-order polynomial in $W$, with the 
fit shown in Fig.~\ref{dpol2}. In the plot of ${\cal P}'_{z'}$ 
(Fig.~\ref{dpol2}) we see that the polarization rises steadily from zero near 
threshold, followed by a dip at around $1.9-2.0$~GeV, and then remains constant 
at about 0.5 over the rest of the range. The ${\cal P}'_{x'}$ data are 
relatively constant at about $-0.5$ over most of the $W$ range.  We also note 
that the magnitude of ${\cal P}'_{z'}$ is nearly equal to ${\cal P}'_{x'}$, 
although with opposite sign, indicating equal strength in the $R_{TT'}^{z'0}$ 
and $R_{TT'}^{x'0}$ responses.  With respect to the $(x,z)$ axes (see 
Fig.~\ref{dpol2}), the polarization is roughly 0.6 and relatively constant 
along $z$ and is consistent with zero along $x$.  This latter point indicates 
either a perfect cancellation of the $R_{LT'}$ response functions (see 
Table~\ref{big_def}) or that they are each nearly zero.

The polarization as a function of $Q^2$ is shown in Fig.~\ref{dpol3}.  The 
data are rather featureless and indicate almost no $Q^2$ dependence.  Both 
${\cal P}'_{z'}$ and ${\cal P}'_z$ are roughly 0.5, while ${\cal P}'_{x'} 
\approx -{\cal P}'_{z'}$, and ${\cal P}'_x$ is consistent with zero.

The ${\cal P}'_\Lambda$ data in Figs.~\ref{dpol1} through \ref{dpol3} 
are compared against the theoretical models introduced in 
Section~\ref{sec:theory}.  The MB hadrodynamic model~\cite{mart_code} is 
indicated by the solid--black lines, the GLV Regge model~\cite{guidal_code} 
is indicated by the short dash--blue lines, and the RPR model
\cite{ghent_code} is indicated by the dot-dash--red lines ($P_{11}$ model 
variant) and by the long-dash--black lines ($D_{13}$ model variant).  
The calculations qualitatively match the sign and trends of the data, but 
detailed comparisons indicate that these new polarization data can be used 
to further tune the models (e.g. the resonance parameters in the MB model 
and the RPR model) or indicate shortcomings in the dynamical description of 
the data (i.e. the pure $t$-channel description of the GLV Regge model).

Detailed comparisons of the individual models to these data are also
useful to indicate specific shortcomings of the models.  For example,
comparisons of the MB model to the data show problems with the
parameters for the resonances included below 2.0~GeV as indicated by
both ${\cal P}'_{z'}$ and ${\cal P}'_{x'}$.  The models also mostly
fail to reproduce the data as a function of $Q^2$ (Fig.~\ref{dpol3}), 
which could indicate problems with the modeling of the non-resonant 
strength with increasing $Q^2$ or the description of the $Q^2$ evolution 
of the hadronic form factors.

Comparisons of the GLV Regge model to the data indicate that a purely
$t$-channel description of the $K^+\Lambda$ reaction is not adequate to
reproduce the polarization results, even for this channel suspected to be
predominantly governed by $K$ and $K^*$ exchange~\cite{5st}.  The $s$-channel
resonance contributions still have important consequences for the
interference observables.  Whereas the GLV model produces a very smooth
behavior for ${\cal P}'$ vs. $W$, $\cos \theta_K^{c.m.}$, and $Q^2$, the
model typically underpredicts the strength and does not account for the
detailed trends in the data.  In some cases (e.g. Figs.~\ref{dpol1},
\ref{dpol2}, and \ref{dpol3}), the GLV model has the wrong sign compared to
the data or has the wrong slope.

For the RPR calculation, two model variants are compared with the data.  
One employs a $P_{11}(1900)$ state.  As was also seen in comparison 
with the CLAS $\sigma_{LT'}$ data~\cite{sltp}, the model variant with the 
$P_{11}$ (dot-dash -- red) is strongly ruled out by the polarization data.  
The second model variant (long dash -- black) employs the $D_{13}(1900)$ 
state proposed by Mart and Bennhold (see Section~\ref{sec:theory}).  This 
model provides a reasonable description of the polarization data over the 
full kinematic phase space.  The only issue with this model, which cannot be 
fully clarified by these data, is the strong interference effects seen in the 
calculations at higher $W$ (see Figs.~\ref{dpol1} and \ref{dpol4}).

\boldmath
\subsection{$\Sigma^0$ polarization transfer}
\label{sigma}
\unboldmath

\begin{figure}[htbp]
\vspace{8.5cm}
\includegraphics{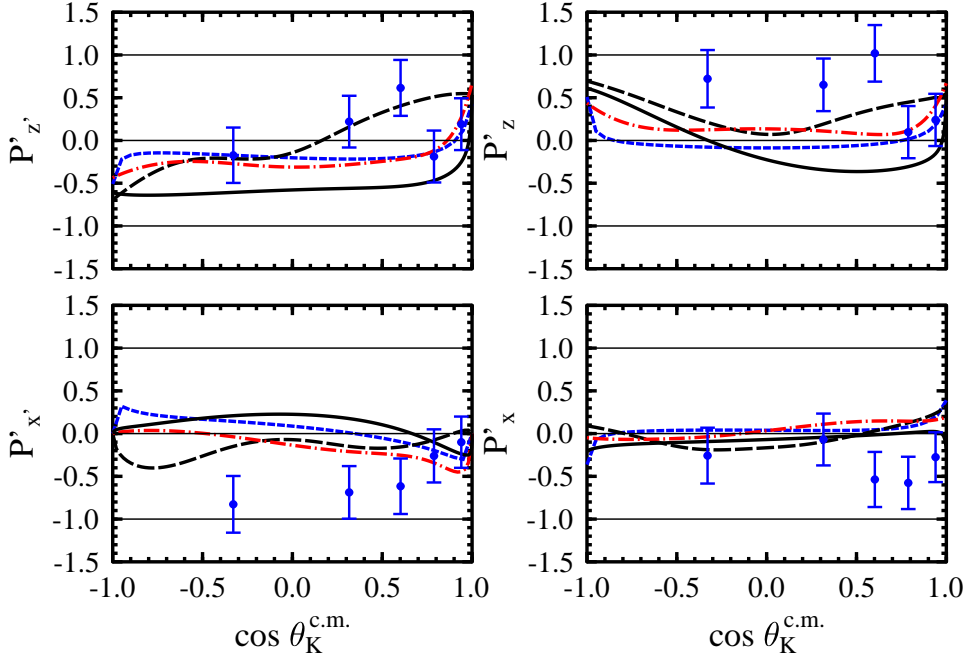}
\caption{(Color online) Transferred $\Sigma^0$ polarization components 
${\cal P}'$ with respect to the $(x',z')$ (left) and $(x,z)$ (right) axes
vs.~$\cos \theta_K^{c.m.}$ at $\langle Q^2 \rangle$=2.5~GeV$^2$ and 
$\langle W\rangle$=2.1~GeV for a beam energy of 5.754~GeV. The curves are 
calculations from the MB isobar model~\cite{mart_code} (solid -- black), the 
GLV Regge model~\cite{guidal_code} (short dash -- blue), and the RPR
model~\cite{ghent_code} with a missing $P_{11}(1900)$ state (dot-dash
-- red) and a missing $D_{13}(1900)$ state (long dash -- black).}
\label{spol14}
\end{figure}

\begin{figure}[htbp]
\vspace{13.0cm}
\includegraphics{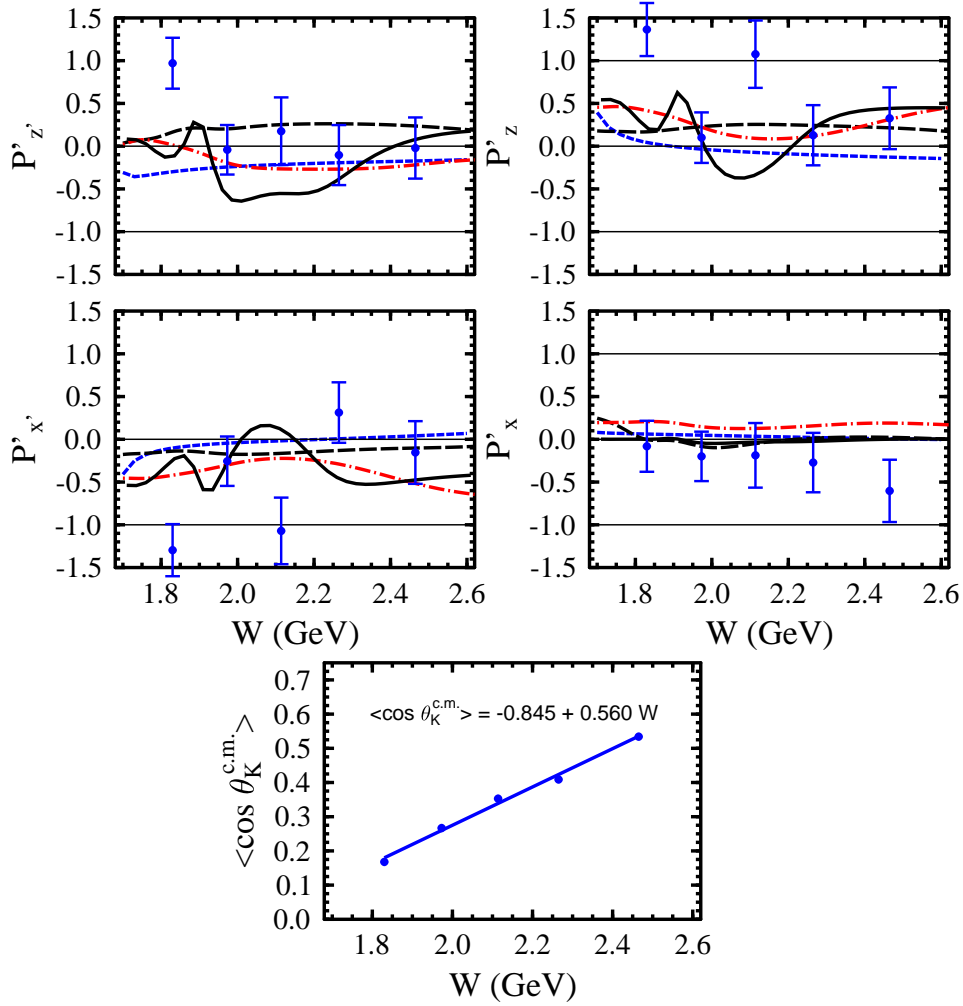}
\caption{(Color online) Transferred $\Sigma^0$ polarization components 
${\cal P}'$ with respect to the $(x',z')$ (left) and $(x,z)$ (right) axes
vs.~$W$ (GeV) at $\langle Q^2 \rangle$=2.5~GeV$^2$ for a beam energy of 
5.754~GeV.  The bin-averaged $\cos \theta_K^{c.m.}$ value for each $W$ point 
is represented by a second-order polynomial in $W$.  The model calculations 
are as indicated in Fig.~\ref{spol14}.}
\label{spol25}
\end{figure}

\begin{figure}[htbp]
\vspace{8.5cm}
\includegraphics{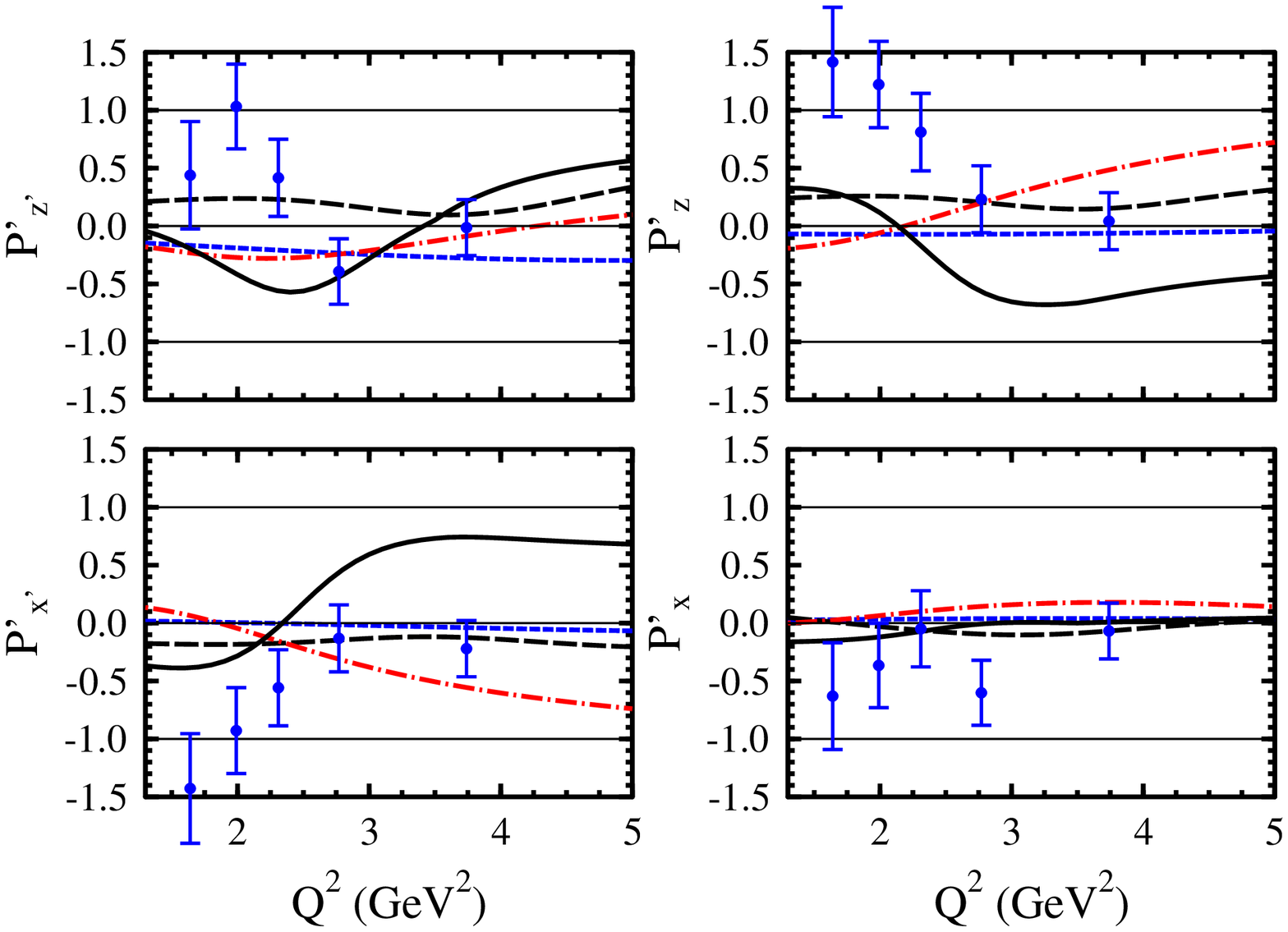}
\caption{(Color online) Transferred $\Sigma^0$ polarization components 
${\cal P}'$ with respect to the $(x',z')$ (left) and $(x,z)$ (right) axes 
vs.~$Q^2$ (GeV$^2$) at $\langle W \rangle$=2.09~GeV and 
$\langle \cos \theta_K^{c.m.} \rangle$ = 0.36 at a beam energy at 5.754~GeV.  
The model calculations are as indicated in Fig.~\ref{spol14}.}
\label{spol36}
\end{figure}

Our results for the transferred $\Sigma^0$ polarization acquired at a
beam energy of 5.754~GeV are shown in Figs.~\ref{spol14} through \ref{spol36} 
compared to several model calculations.  The error bars in these figures 
include statistical but not systematic uncertainties, which we estimate to 
be $0.112 \oplus 0.045 {\cal P}'_\Sigma$ on the polarization (see 
Section~\ref{systematics}).  The full data set is contained in the CLAS 
database~\cite{database}.

Fig.~\ref{spol14} shows the dependence of ${\cal P}'_{x',z'}$ and 
${\cal P}'_{x,z}$ with respect to $\cos \theta_K^{c.m.}$ and Fig.~\ref{spol25} 
shows the polarization with respect to $W$. As with the $\Lambda$ results, 
$\cos \theta_K^{c.m.}$ values at each $W$ point are well represented by a 
low-order polynomial in $W$ (given in Fig.~\ref{spol25}).  Fig.~\ref{spol36} 
shows the polarization with respect to $Q^2$.

%%%%%%%%%%%%%%%%%%%%%%%%%%%%%%%%%%%%%%%%%%%%%%%%%%%%%%%%%%%%%%%%%%%%%%%%%
%%%%%%%%%%%%%%%%%%%%%%%%%%%%%%%%%%%%%%%%%%%%%%%%%%%%%%%%%%%%%%%%%%%%%%%%%

%%%%%%%%%%%%%%%%%%%%%%%%%%%%%%%%%%%%%%%%%%%%%%%%%%%%%%%%%%%%%%%%%%%%%%%%%
%%%%%%%%%%%%%%%%%%%%%%%%%%%%%%%%%%%%%%%%%%%%%%%%%%%%%%%%%%%%%%%%%%%%%%%%%

%%%%%%%%%%%%%%%%%%%%%%%%%%%%%%%%%%%%%%%%%%%%%%%%%%%%%%%%%%%%%%%%%%%%%%%%%
%%%%%%%%%%%%%%%%%%%%%%%%%%%%%%%%%%%%%%%%%%%%%%%%%%%%%%%%%%%%%%%%%%%%%%%%%

An important point to note is the rather sizable statistical uncertainties
on the $\Sigma^0$ data.  This arises not due to the limitations of the
$\Sigma^0$ data sample (which has roughly 3000 counts in each analysis
bin), but rather due to the scaling by $\alpha_\Lambda/\alpha_\Sigma$ in
Eq.(\ref{sig_err}).  It should also be made clear that the {\tt e1-6}
running period at CLAS represented a very lengthy run period (extending
over 4 months) that recorded nearly 5~billion triggers.  Given the effect
of the $\Lambda$ background beneath the $\Sigma^0$ peak, CLAS will likely
not be able to provide more precise electroproduction data for this
observable.  

Figs.~\ref{spol14} through \ref{spol36} indicate that the measured $\Sigma^0$
transferred polarization tends to have the same sign as the $\Lambda$
polarization for the different axes.  More quantitative statements cannot
be made given the statistical quality of the data.  In comparing the
data to the models, with the possible exception of the MB model (solid
black), one sees that they qualitatively match the sign and trends of the
data, as was the case for the $\Lambda$ data.  However, as the 
${\cal P}'_\Sigma$ observable has not been measured before, the results can 
still serve to provide at least loose constraints on the theoretical models.
Certainly the ${\cal P}'_\Lambda$ data can be used to improve the knowledge 
of the contributing $N^*$ states by allowing for improved descriptions of the 
associated form factors and the $g_{KN\Lambda}$ coupling strengths.  These 
improvements can then be used in the $\Sigma^0$ modeling, which typically 
employ the same set of $N^*$ states for both the $K^+\Lambda$ and 
$K^+\Sigma^0$ final states.

\boldmath
\subsection{Extraction of the ratio of $\sigma_L$/$\sigma_T$}
\label{rat_ext}
\unboldmath

In order to extract $R_\sigma = \sigma_L/\sigma_T$ at $\cos \theta_K^{c.m.}=1$, 
we must first extrapolate ${\cal P}'_{z'}$ (or ${\cal P}'_z$) to 
$\cos \theta_K^{c.m.}=1$.  However, because of statistical fluctuations in
the data and finite angle resolution effects, extrapolations for 
${\cal P}'_{z'}$ give slightly different results than extrapolations for  
${\cal P}'_z$.  Following the procedure defined in 
Ref.~\cite{raue05}, the extrapolation is actually performed by summing the 
${\cal P}'_{z'}$ and ${\cal P}'_z$ components into a new quantity $R_{sum}$ 
(see Table~\ref{big_def} for component definitions) given by 
\begin{equation}
\label{eqn-Psum1}
R_{sum} \equiv \frac{{({\cal P}'_{z'}+{\cal P}'_z) \sigma_U}}{c_0} =
{\cal K}[(1+\cos\theta_K^{c.m.})R_{TT'}^{z'0}-R_{TT'}^{x'0} 
\sin\theta_K^{c.m.}].
\end{equation}
From the extrapolated value of $R_{sum}$ at $\cos \theta_K^{c.m.}=1$,
Eq.(\ref{eqn-Psum1}) can then be inverted to determine ${\cal P}'_{z'}$ (or
equivalently ${\cal P}'_z$), which in turn is used to extract $R_\sigma$
using Eq.(\ref{eqn-ratio2}). 

An additional benefit of using this form is that Eqs.(\ref{eqn_Pz1}) and
(\ref{eqn-ratio2}) provide important and useful constraints on $R_{sum}$ at
$x\equiv\cos\theta_K^{c.m.}=\pm1$.  At $x=-1$, the sum of the polarizations
must be zero, according to Eq.(\ref{eqn_Pz1}), leading to
$R_{sum}(x=-1)=0$.  Since both $\sigma_L$ and $\sigma_T$ must be positive
definite, then $R_\sigma$ (see Eq.(\ref{eqn-ratio2})) must also be positive
definite, which leads to $R_{sum}\leq 2\sigma_U$ for $x=1$. Note also that
as both $\sigma_L$ and $\sigma_T$ must be positive definite,
Eq.(\ref{eqn-ratio2}) constrains both the $\Lambda$ and $\Sigma^0$ hyperon
polarization components ${\cal P}'_z$ to be between 0 and $c_0$. This
argument confirms the sign of the hyperon polarization is correct in our
analysis.

Besides the explicit $\theta_K^{c.m.}$ dependence shown in Eq.(\ref{eqn-Psum1}) 
and in the response functions, the CGLN amplitudes contain additional 
$\theta_K^{c.m.}$ dependence (as well as $Q^2$ and $W$ dependence)
\cite{knochlein}.  This suggests that Eq.(\ref{eqn-Psum1}) can then be fit
with polynomials in $x=\cos\theta_K^{c.m.}$, provided we have prior knowledge 
of the $\sigma_U$ term.  In the case of the 4.261~GeV data, we can use the 
previously published CLAS results~\cite{5st}, while for the 5.754~GeV data, we 
have to use models to provide $\sigma_U$ at our kinematic points.

The number of terms to include in a polynomial fit to Eq.(\ref{eqn-Psum1}) 
is ultimately governed by the reaction dynamics.  The explicit 
$\theta^{c.m.}_K$ dependence alone suggests at least a third-order polynomial.  
However, given the limited number of polarization data points, the number of 
terms in any fit leading to a meaningful extrapolation to 
$\cos\theta_K^{c.m.}=1$ must also be limited.  We begin by considering 
third-order fits of the form 
\begin{equation}
\label{eqn-Psumfit}
 R_{sum} = a_0 + a_1 x + a_2 x^2 + a_3 x^3,
\end{equation}

\noindent
where $a_{i=0 \to 3}$ represent the fit coefficients. However, applying 
the constraint $R_{sum}=0$ at $x=-1$ implies $a_1= a_0 + a_2 - a_3$.

We have done a series of fits to the data points representing $R_{sum}$ in 
which we varied the number of terms in the fits, while imposing a penalty 
on the $\chi^2$ if a fit returned an unphysical value at $x=1$.  The penalty 
was chosen to be large enough to force non-negative values of $R_\sigma$.  
It should be noted that only one of the fits (5.754~GeV data at $W$=1.75~GeV 
and $Q^2$=2.61~GeV$^2$) required the imposition of a penalty.  In determining 
the optimal number of parameters in the fit for each $W$, we simply used the
number of parameters that produced the smallest minimized $\chi^2_\nu$
($\chi^2$ per degree of freedom).  All three of the 4.261~GeV fits
favored a second-order fit ($a_3=0$), while all three of the 5.754~GeV
fits favored a third-order fit.

\begin{figure}[htbp]
\vspace{8.7cm}
\includegraphics{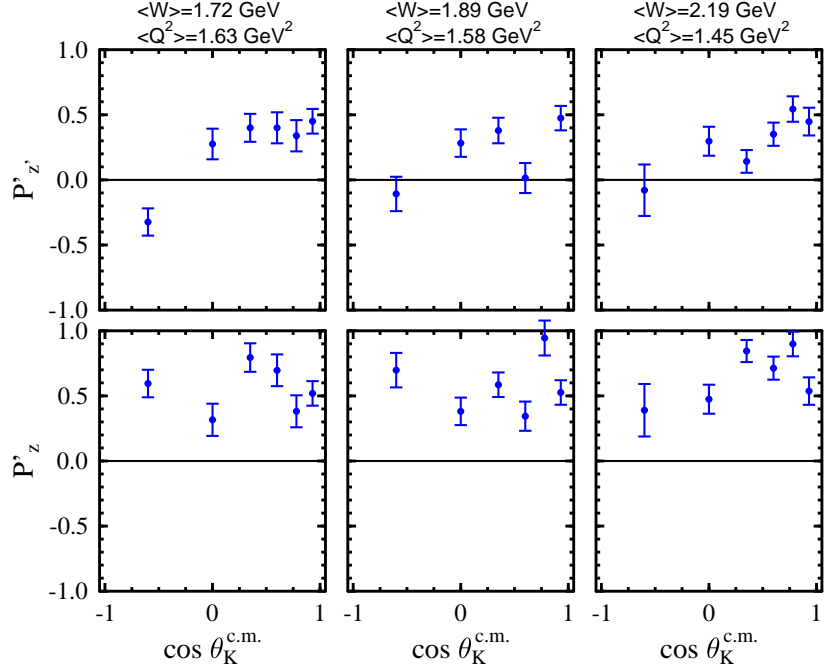}
\caption{(Color online) Transferred $\Lambda$ polarization components 
${\cal P}'$ with respect to the $(z',z)$ axes vs.~$\cos \theta_K^{c.m.}$ 
for three bin-averaged $W$/$Q^2$ values as indicated for a beam energy of 
4.261~GeV.}
\label{dpol_4gev}
\end{figure}

The ${\cal P}'_{z'}$ and ${\cal P}'_z$ data from the 4.261~GeV data set are
shown in Fig.~\ref{dpol_4gev} with respect to $\cos \theta_K^{c.m.}$. The
full set of 4.261~GeV polarization transfer data for the $(x',z')$ and
$(x,z)$ axes with respect to $W$ and $\cos \theta_K^{c.m.}$ is provided in
Ref.~\cite{database}. The ${\cal P}'_{z'}$ and ${\cal P}'_z$ data from the 
5.754~GeV data set are shown in Figs.~\ref{dpol1} and \ref{dpol4}.  We should 
point out that the ${\cal P}'_{z'}$ and ${\cal P}'_z$ results come from the 
same data.  Therefore, these observables are not independent.  They do, 
however, measure different quantities (as seen in Table~\ref{big_def}) since 
they are projections onto different axes. In adding these together to form 
$R_{sum}$, the uncertainties from ${\cal P}'_{z'}$ and ${\cal P}'_z$ were 
added together.

%%%%%%%%%%%%%%%%%%%%%%%%%%%%%%%%%%%%%%%%%%%%%%%%%%%%%%%%%%%%%%%%%%%%%%%%%
%%%%%%%%%%%%%%%%%%%%%%%%%%%%%%%%%%%%%%%%%%%%%%%%%%%%%%%%%%%%%%%%%%%%%%%%%

\begin{figure}[tbp]
\vspace{9.2cm}
\includegraphics{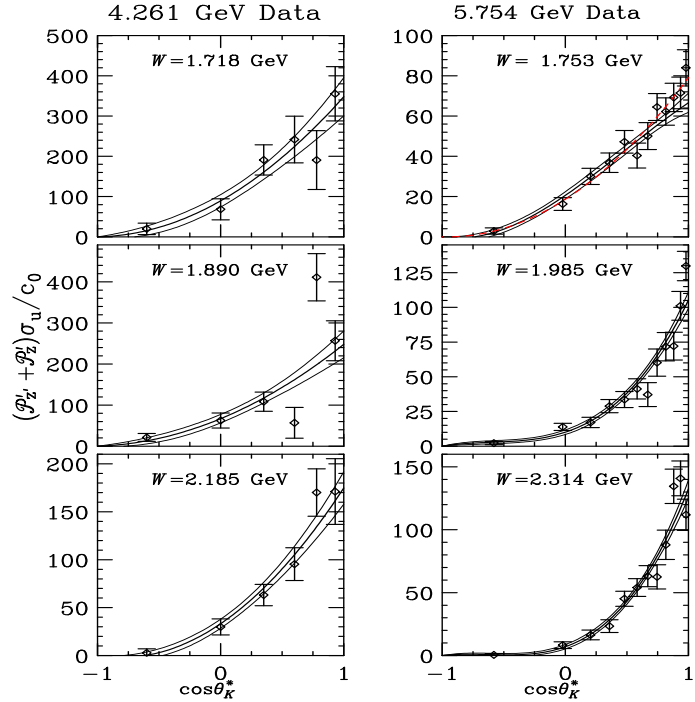}
\caption{(Color online) $R_{sum}$ (defined in Eq.(\ref{eqn-Psum1}))
vs.~$\cos\theta^{c.m.}_K$ for the 4.261~GeV and 5.754~GeV data along with
our fits (heavy solid lines) and the error band resulting from the fit
uncertainties (light solid lines).  The dashed red line in the upper right
panel indicates the result of removing the $x=1$ constraint in the fit.}
\label{fig-Rsum}
\end{figure}

The results of our fits to the 4.261~GeV and 5.754~GeV data are shown in 
Fig.~\ref{fig-Rsum} (heavy solid lines) along with an error band (light 
solid lines). The error bands include uncertainties both from the fitting of 
Eq.(\ref{eqn-Psumfit}), and, for the 4.261~GeV data, contributions from 
uncertainties in the fits of the cross section data.  The latter contribution 
to the uncertainties is about half that of the former.  The error band 
indicates that the extrapolation to $x=1$ is well constrained. For the 
5.754~GeV data we display the fit using the calculated cross section for one 
particular choice of the MB model~\cite{mart}, which allows for different 
choices of form factors and couplings.

%%%%%%%%%%%%%%%%%%%%%%%%%%%%%%%%%%%%%%%%%%%%%%%%%%%%%%%%%%%%%%%%%%%%%%%%%%%
%%%%%%%%%%%%%%%%%%%%%%%%%%%%%%%%%%%%%%%%%%%%%%%%%%%%%%%%%%%%%%%%%%%%%%%%%%%
                                                                
Table~\ref{tab-results} shows the resulting $\chi^2_\nu$, the polarization
extrapolated to $x=1$, and $R_\sigma$.  Since the 5.754~GeV data required a
model for $\sigma_U$, we repeated the fit for five different parameter set
choices within the framework of the MB model~\cite{mart}.  Thus, the
5.754~GeV results in the table reflect the average values of $\chi^2_\nu$,
the polarization extrapolated to $x=1$, and $R_\sigma$ for different
models.  We estimated the model uncertainty by using the standard deviation
of $R_\sigma$ from using the five different models.

%%%%%%%%%%%%%%%%%%%%%%%%%%%%%%%%%%%%%%%%%%%%%%%%%%%%%%%%%%%%%%%%%%%%%%%%%%%
\begin{table}[btp]
\begin{center}
\begin{tabular}{c|c|c|c|c|c} \hline
$E_b$ (GeV) &  $\langle W\rangle$ GeV & $\langle Q^2\rangle $ GeV$^2$ & 
$\chi^2_\nu$ & ${\cal P}'_{z',z}(x=1)$  & $R_\sigma$ \\ \hline
      & 1.72 & 1.63 & 0.77 &  0.451$\pm$0.066 & 0.533$\pm$0.270$\pm$0.326 \\ 
\cline{2-6}
4.261 & 1.89 & 1.58 & 5.69 &  0.440$\pm$0.063 & 0.870$\pm$0.329$\pm$0.401 \\ 
\cline{2-6}
      & 2.18 & 1.45 & 0.87 &  0.486$\pm$0.062 & 1.348$\pm$0.404$\pm$0.515 \\ 
\hline \hline
      & 1.75 & 2.61 & 1.11 &  0.607$\pm$0.070 & 0.000$\pm$0.092$\pm$0.156 \\ 
\cline{2-6}
5.754 & 1.98 & 2.56 & 2.76 &  0.610$\pm$0.065 & 0.176$\pm$0.088$\pm$0.209 \\ 
\cline{2-6}
      & 2.31 & 2.31 & 2.79 &  0.470$\pm$0.053 & 0.637$\pm$0.120$\pm$0.445 \\ 
\hline
\end{tabular}
\end{center}
\caption{Transferred polarization at $x=\cos\theta_K^{c.m.}\!=\! 1.0$
extrapolated from the fits described in the text along with the resulting
value of the ratio of longitudinal to transverse structure functions.  
Uncertainties on ${\cal P}'_{z',z}$ are the combined uncertainties arising 
from the fit to the polarization data and the uncertainties in the cross 
section data.  The first uncertainty on $R_\sigma$ is the statistical 
uncertainty (from the fit), while the second represents an estimated 
systematic uncertainty.}
\label{tab-results}
\end{table}
%%%%%%%%%%%%%%%%%%%%%%%%%%%%%%%%%%%%%%%%%%%%%%%%%%%%%%%%%%%%%%%%%%%%%%%%%%%

Inserting the extrapolated polarizations into Eq.(\ref{eqn-ratio2}), we can
determine the ratio $R_\sigma$.  These values are shown in the last column
of Table~\ref{tab-results}, along with the combined uncertainties of the
polarization and cross section fits, and an estimated systematic
uncertainty.   The systematic uncertainty includes a contribution
assuming a 10\% relative systematic uncertainty in the polarization data. 

\begin{figure}[htbp]
\vspace{5.3cm}
\includegraphics{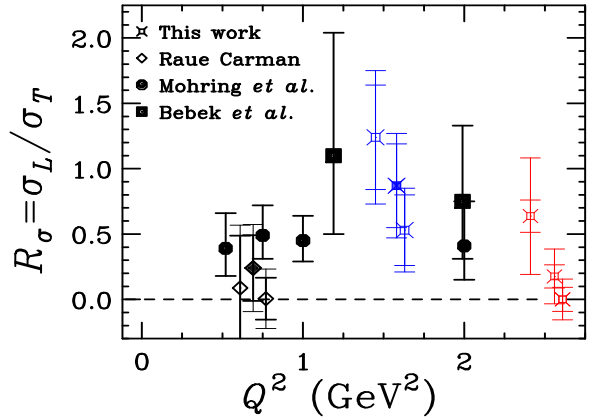}
\caption{(Color online) Ratio of longitudinal to transverse structure 
functions at $\theta_K^{c.m.}=0^\circ$ vs.~$Q^2$. The blue cross data points 
(about $Q^2$=1.5~GeV$^2$) are for 4.261~GeV and the red cross points (about 
$Q^2$=2.5~GeV$^2$) are for 5.754~GeV.  The inner error bars on our points 
represent the statistical uncertainties arising from the fit and the outer 
error bars represent the combination of statistical and estimated systematic 
uncertainties. The triangles are data from CLAS~\cite{5st}, the solid circles 
are data from Mohring {\it et al.}~\cite{mohring}, the solid squares are 
data from Bebek {\it et al.} \cite{bebek}, and the diamonds are data from Raue 
and Carman~\cite{raue05}. All of the filled points are near a common value of 
$W\approx$1.84~GeV.} 
\label{fig-RLT_ave}
\end{figure}

The resulting values for $R_\sigma$ are plotted in Fig.~\ref{fig-RLT_ave}.
For comparison, we have also included the previously published data
\cite{5st,bebek,mohring}. However, only the filled points are at or 
near the common value of $W\approx$1.84~GeV.  Other than these filled points,
one should not take any trends in the data too seriously since the data
from this analysis cover a large range in $W$ (1.72 to 2.31~GeV).

%%%%%%%%%%%%%%%%%%%%%%%%%%%%%%%%%%%%%%%%%%%%%%%%%%%%%%%%%%%%%%%%%%%%%%%%%%%
%%%%%%%%%%%%%%%%%%%%%%%%%%%%%%%%%%%%%%%%%%%%%%%%%%%%%%%%%%%%%%%%%%%%%%%%%%%

Our new results from the data sets at 4.261 and 5.754~GeV are in reasonable
accord with the existing measurements of $\sigma_L/\sigma_T$ from Bebek
{\it et al.}~\cite{bebek} and Mohring {\it et al.}~\cite{mohring}. Looking
at the results for $W\approx$1.84~GeV, the ratio rises with $Q^2$ up to
$Q^2\sim 1.5$~GeV$^2$ and thereafter seems to fall off, suggesting an
interesting and non-trivial dependence on $Q^2$, but the measurement
accuracy is not adequate to quantify this observation.  Alternatively, we
point out that our data also seem to suggest a rapid rise of
$\sigma_L/\sigma_T$ with $W$ as was suggested in our previous
publication~\cite{5st}, however again, we lack the statistical and
systematic precison to make a more definitive conclusion.  Note that the
$\sigma_L/\sigma_T$ data from Ref.~\cite{5st} at $Q^2$=1.0~GeV$^2$ cannot
be directly compared to these data as the most forward angle point in that
work is $\cos \theta_K^{c.m.}$=0.90.

The data of Fig.~\ref{fig-RLT_ave} imply that for at least a limited $Q^2$
interval, the longitudinal structure function becomes sizable. This
structure function is expected to be very sensitive to the kaon form
factor~\cite{saghai98}. A recently conducted experiment in Hall~A at
Jefferson Laboratory~\cite{markowitz} has as one of its main goals a
Rosenbluth separation at several values of momentum transfer $t$ leading to
a Chew-Low extrapolation~\cite{frazer} of the kaon form factor. However,
this method relies on having small {\it relative} uncertainties for
$\sigma_L$, which will not be the case when $\sigma_L$ is itself small.
These new results indicate that the successful extraction of the form
factor may only be possible in a limited kinematic range.

\subsection{Partonic models of the process}
\label{quark_models}

All of the models introduced thus far in this work have been used to
indicate the strong sensitivity of these polarization data to the underlying
$s$-channel resonant terms that contribute in the intermediate state of
the $\gamma^* p \to K^+Y$ process. The precision and broad kinematic coverage
of the data from Ref.~\cite{5st} have indicated that the $K^+\Lambda$
final state is dominated by $t$-channel kaon exchange. However, there are
important contributions from $s$-channel processes that must be taken into
account to describe both the cross section and polarization data in detail.

In contradistinction to the hadronic models, and as noted earlier and 
introduced in Ref.~\cite{carman03}, our data indicate the $\Lambda$ polarization 
is maximal along the virtual photon direction (see results for ${\cal P}'_z$ 
in Figs.~\ref{dpol4}, \ref{dpol2}, and \ref{dpol3}), suggesting a simple 
phenomenology.  In fact, the $\Lambda$ polarization is essentially unity if 
the virtual photon depolarization factor is taken into account (see 
Section~\ref{depol_fac}).  The lack of a strong $W$ and $Q^2$ dependence is an 
indication that the data might be more economically described in a flux-tube 
strong-decay framework.  There is growing evidence that the relevant degrees of 
freedom to describe the phenomenology of hadronic decays are constituent quarks 
bound by a gluonic flux-tube~\cite{isgur1}.  Properties of the flux-tube can 
be determined by studying $q\bar{q}$ pair production, since this is widely 
believed to produce the color field neutralization that breaks the flux-tube.  
Since the 1970's, it has been argued that a quark pair with vacuum quantum 
numbers is responsible for breaking the color flux-tube (the $^3\!P_0$ model
\cite{leyaouanc}).

%%%%%%%%%%%%%%%%%%%%%%%%%%%%%%%%%%%%%%%%%%%%%%%%%%%%%%%%%%%%%%%%%%%%%%%%%
\begin{figure}[htbp]
\vspace{7.5cm} 
\includegraphics{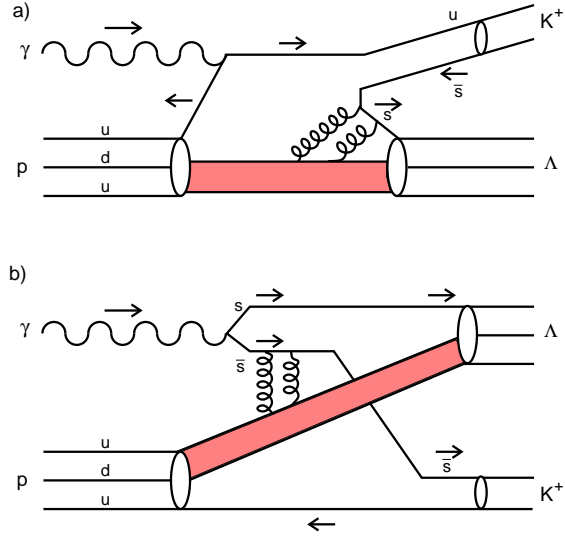}
\caption{(a) A model of the reaction where a circularly polarized virtual 
photon strikes an oppositely polarized $u$ quark inside the proton.  The 
spin of this quark flips and the quark recoils from its neighbors.  An
$s\bar{s}$ quark pair is created from a $J^\pi=0^-$ two-gluon exchange (in 
lowest order) to produce the final-state $K^+$ and $\Lambda$ hyperon.  (b) A 
model of the reaction where an $s\bar{s}$ quark pair is produced from a 
circularly polarized real photon that hadronizes such that the $s$ quark in 
the $\Lambda$ retains its full polarization after being ``precessed'' by a 
spin-orbit interaction, while the $\bar{s}$ quark ends up in the spinless 
kaon.  In both pictures the shaded band represents a spinless $u-d$ diquark 
system.}
\label{models}
\end{figure}
%%%%%%%%%%%%%%%%%%%%%%%%%%%%%%%%%%%%%%%%%%%%%%%%%%%%%%%%%%%%%%%%%%%%%%%%%

This simple phenomenology of the $\Lambda$ polarization data has led 
two groups within the CLAS Collaboration to develop semi-classical models 
based on partonic degrees of freedom to describe the associated reaction 
mechanism.  In the model of Carman {\it et al.}~\cite{carman03,courier}
(shown in Fig.~\ref{models}(a)), it is assumed that the cross section is 
dominated by photoabsorption by a $u$ quark.  Due to the helicity-conserving 
vector interaction, the $u$ quark becomes polarized along the photon direction
($+z$).  Hadronization into the $K^+\Lambda$ final state proceeds with the 
production of an $s\bar{s}$ pair that breaks the color flux-tube.  Because the 
$u$ quark hadronizes as a pseudoscalar $K^+$, the $\bar{s}$ quark spin is 
required to be opposite to that of the $u$ quark, i.e. in the $-z$ direction.  
In the non-relativistic quark model, the entire spin of the $\Lambda$ is 
carried by the $s$ quark. Since the $\Lambda$ polarization is in the $+z$ 
direction, seen by the fact that ${\cal P}'_z > 0$, it was concluded that the 
$s$ and $\bar{s}$ spins were anti-aligned when they were created, if the 
hadronization process did not flip or rotate their spins.  Note that Liang and 
Boros also posit a two-step process for the production of transversely 
polarized $\Lambda$ hyperons in the exclusive $pp \to pK^+ \Lambda$ reaction
\cite{boros}, and come to a similar conclusion that the $s\bar{s}$ quark pair 
must have been produced with spins anti-aligned.  More recently, $\Lambda$ 
polarization has been interpreted within an instanton interaction model
\cite{kochelev}, which also is assumed to occur with the production of an 
anti-aligned $s\bar{s}$ quark pair.  A dominance of spin anti-alignment for 
the $s$ and $\bar{s}$ quarks would not be consistent with the $S=1$ $^3\!P_0$ 
operator~\cite{barnes}, which predicts a 2:1 mixture of $s\bar{s}$ quarks 
produced with spins aligned vs. anti-aligned if the orbital substates are 
equally populated.  Along with other observations of failure of the $^3\!P_0$ 
model (e.g. explaining $\pi_2 \to \rho \omega$ decay~\cite{barnes}), the 
applicability of the $^3\!P_0$ model in describing all hadronic decays is 
brought into doubt if this model is appropriate.

Extensive photoproduction data for the transferred polarization for the 
$K^+\Lambda$ final state has also been published from CLAS~\cite{bradford1}.  
These data also indicate that the $\Lambda$ polarization is predominantly in 
the direction of the spin of the incoming photon, independent of the 
center-of-mass energy or meson production angle. Based on these data, 
Schumacher has introduced a different model~\cite{courier,schumacher} to 
explain the $\Lambda$ polarization results.  In this model, shown in 
Fig.~\ref{models}(b), the produced $s\bar{s}$ pair is created in a $^3S_1$ 
configuration ($J$=1, $S$=1, $L$=0, i.e. $J^\pi=1^-$).  Here, following the 
principle of vector meson dominance, the real photon fluctuates into a virtual 
$\phi$ meson that carries the polarization of the incident photon.  Therefore, 
the quark spins are in the direction of the spin of the photon before the 
hadronization interaction.  The $s$ quark of the pair merges with the 
unpolarized diquark within the target proton to form the $\Lambda$ baryon, 
and the $\bar{s}$ quark merges with the remnant $u$ quark of the proton to 
form a spinless $K^+$ meson.

The two model interpretations, while able to predict the correct sign for
the $\Lambda$ polarization transfer, nevertheless describe very different
physical processes.  Both assume that the mechanism of spin transfer to the 
$\Lambda$ hyperon involves a spectator $J^\pi=0^+$ diquark system.  The main 
difference is the role of the third quark.  Neither model specifies a detailed 
dynamical mechanism.  If we take the gluonic degrees of freedom into 
consideration, the model of Carman {\it et al.}~\cite{carman03} can be realized 
in terms of a possible mechanism in which a colorless $J^\pi=0^-$ two-gluon 
subsystem is emitted from the spectator diquark system and produces the 
$s\bar{s}$ pair as illustrated in Fig.~\ref{models}(a).  To the same order of 
gluon coupling, the model of Schumacher~\cite{schumacher} is the quark-exchange 
mechanism illustrated in Fig.~\ref{models}(b), again mediated by a two-gluon 
exchange.  The amplitudes corresponding to these diagrams may both be present 
in the production, in principle, and could contribute at different levels 
depending on the reaction kinematics.

Extending these studies to the $K^{*+}\Lambda$ exclusive final state
should be revealing.  In the Carman {\it et al.} model, the spin of the
$u$ quark is unchanged when changing from a scalar $K^+$ to a vector 
$K^{*+}$.  If the $s\bar{s}$ quark pair is produced with spins anti-aligned,
then the spin direction of the $\Lambda$ should flip.  On the other hand,
in the Schumacher model, the $u$ quark in the kaon is only a spectator; 
changing its spin direction -- and thus changing the $K^+$ to a $K^{*+}$ -- 
should not change the $\Lambda$ spin direction.  Thus there are ways to 
disentangle the relative contributions and to better understand the reaction 
mechanism and dynamics underlying the associated strangeness production
reaction.  Analyses at CLAS are underway to extract the polarization 
transfer to the hyperon in the $K^{*+}\Lambda$ final state.

In developing the quark model interpretations of polarization transfer, we
also need to consider the phenomenology of the $K^+\Sigma^0$ results.  As
shown in this work, and much more clearly in the CLAS photoproduction data
\cite{bradford1}, the $K^+\Sigma^0$ polarization transfer is very similar 
in magnitude {\em and} sign to the $K^+\Lambda$ data.  We might expect that 
when the $\Lambda$ and $\Sigma^0$ polarization transfers in these reactions 
point in opposite directions and have the same magnitudes, this would then 
give more weight to the modeling of the polarizations originating from a 
quark level interaction, in particular associated with the strange quark 
spin.  However, even though this is not what is observed in the data, we 
should realize that the spin state of the $\Sigma^0$ hyperon is not 
determined by the strange quark alone, but a combination of the $s$ quark
spin and the triplet $ud$ quark spin.  Thus the models of Fig.~\ref{models} 
are not directly applicable for $K^+\Sigma^0$ production.

Understanding a process of this sort through partonic models can shed light on 
quark-gluon dynamics in a domain usually thought to be dominated by traditional 
meson and baryon degrees of freedom.  These issues are relevant to better 
understand strong interactions and hadroproduction in general due to the 
non-perturbative nature of QCD for CLAS kinematics.  We eagerly await further 
experimental studies and new theoretical efforts to understand which 
multi-gluonic degrees of freedom dominate in quark-pair creation and their 
role in strangeness production, as well as the appropriate mechanism (or 
mechanisms) for the dynamics of spin transfer in hyperon production.  

%
%%%%%%%%%%%%%%%%%%%%%%%%%%%%%%%%%%%%%%%%%%%%%%%%%%%%%%%%%%%%%%%%%%%%%%%%%%%
%

\section{SUMMARY AND CONCLUSIONS}
\label{sec:conclusions}

In this paper we have provided extensive new data at 4.261 and 5.754~GeV 
for the beam-recoil hyperon polarization transfer for the reaction 
$p(\vec{e},e'K^+)\vec{\Lambda}$ studying its dependence on the kinematic
variables $Q^2$, $W$, and $\cos \theta_K^{c.m.}$.  These data add to the 
earlier 2.567~GeV CLAS data results from Ref.~\cite{carman03}.  In addition, 
we have provided the first-ever polarization transfer data for the reaction
$p(\vec{e},e'K^+)\vec{\Sigma}^0$.  These new data sets span a range of
momentum transfer $Q^2$ from 0.7 to 5.4~GeV$^2$, invariant energy $W$
from 1.6 to 2.6~GeV, and the full $K^+$ center-of-mass angular range.

Our data have been compared to predictions from several available theoretical 
models that have varying sensitivities to the $s$-channel resonance 
contributions.  The increased statistical precision of these new data will 
enable improved fits either for effective Lagrangian models or for 
coupled-channels model fits incorporating both photo- and electroproduction 
data that will be carried out by several groups in the near future
\cite{tshlee}, including the Excited Baryon Analysis Center (EBAC)~\cite{ebac} 
at Jefferson Laboratory.  The analysis of the full set of the world's data in 
this manner is essential to map out the full spectrum of excited states of the 
nucleon to better determine the structure of the nucleon and its associated 
degrees of freedom, both of which are necessary to better understand the strong 
interaction and QCD.

The new CLAS $\Lambda$ polarization data sets at 4.261 and 5.754~GeV have also 
been used to extract the longitudinal-to-transverse structure function ratio 
at $\theta_K^{c.m.}=0^\circ$ in the $Q^2$ range from 1.5 to 2.5~GeV$^2$, 
extending the existing CLAS measurements taken at 2.567~GeV near 
$Q^2$=1.0~GeV$^2$.  These new data, given the statistical uncertainties, could 
indicate a non-trivial $Q^2$ evolution of the structure function ratio in the 
range from $Q^2$=0.7 to 2.5~GeV$^2$, that peaks near unity at $Q^2$=1.5~GeV$^2$.
These results indicate that extraction of the kaon form factor using the 
standard Chew-Low extrapolation technique can only be carried out in the 
limited kinematic range where $\sigma_L$ is sizable.

Finally, the data have been compared to two simple semi-classical partonic
models including multi-gluon exchange that were designed to account for the 
strikingly simple phenomenology seen in the kinematic dependence of the 
polarization data.  While the two models make very different assumptions
regarding the reaction mechanism leading to production of the $K^+\Lambda$
final state and different quantum numbers of the produced $s\bar{s}$ 
pair, we have provided suggestions for testing them by comparing polarization 
data for $K^+\Lambda$ to $K^{*+}\Lambda$ final states.  Disentangling the true 
reaction dynamics in a partonic model is relevant to probe the appropriate 
quark-pair creation operator that governs the transitions to the final state 
particles and to shed light on the relevance of quark-gluon dynamics in a 
domain thought to be dominated by meson/baryon degrees of freedom.

We would like to acknowledge the outstanding efforts of the staff of 
the Accelerator and the Physics Divisions at Jefferson Lab that made 
this experiment possible.  This work was supported in part by the U.S.
Department of Energy, the National Science Foundation, the Italian Istituto 
Nazionale di Fisica Nucleare, the French Centre National de la 
Recherche Scientifique, the French Commissariat \`{a} l'Energie 
Atomique, and the Korean Science and Engineering Foundation.  The 
Southeastern Universities Research Association (SURA) operated the 
Thomas Jefferson National Accelerator Facility for the United States 
Department of Energy under contract DE-AC05-84ER40150. 

%%%%%%%%%%%%%%%%%%%%%%%%%%%%%%%%%%%%%%%%%%%%%%%%%%%%%%%%%%%%%%%%%%%%%%%%%%%%


\begin{thebibliography}{99}

\bibitem{capstick}
S. Capstick and W. Roberts, Phys. Rev. D {\bf 58}, 074011 (1998).

\bibitem{isgur} 
N. Isgur, Proceedings of the NSTAR 2000 Conference, eds. V.D. Burkert, 
L. Elouadrhiri, J.J. Kelly, and R. Minehart, (World Scientific, Singapore, 
2001), p. 403.

\bibitem{kon_isg}
R. Koniuk and N. Isgur, Phys. Rev. D {\bf 21}, 1868 (1980).

\bibitem{capstick2}
S. Capstick and W. Roberts, Phys. Rev. D {\bf 49}, 4570 (1994).

\bibitem{pdg}
C. Amsler {\it et al.}, Particle Data Group, Phys. Lett. B {\bf 667}, 1 (2008).

\bibitem{5st}
P. Ambroziewicz {\it et al.} {\it (CLAS Collaboration)}, Phys. Rev. C
{\bf 75}, 045203 (2007).

\bibitem{corthals}
T. Corthals {\it et al.}, Phys. Lett. B {\bf 656}, 186 (2007).

\bibitem{mcnabb} 
J.W.C. McNabb {\it et al.} {\it (CLAS Collaboration)}, Phys. Rev. C 
{\bf 69}, 042201(R) (2004).

\bibitem{bradford1}
R.K. Bradford {\it et al.} {\it (CLAS Collaboration)}, Phys. Rev. C {\bf 75},
035205 (2007).

\bibitem{carman03} 
D.S. Carman {\it et al.} {\it (CLAS Collaboration)}, Phys. Rev. Lett. 
{\bf 90}, 131804 (2003).

\bibitem{nikonov}
V.A. Nikonov {\it et al.}, Phys. Lett. B {\bf 662}, 245 (2008).

\bibitem{ani05a}
A.V. Anisovich {\it et al.}, Eur. Phys. J. A {\bf 24}, 111 (2005).

\bibitem{ani05b}
A.V. Anisovich {\it et al.}, Eur. Phys. J. A {\bf 25}, 427 (2005).

\bibitem{penner}
G. Penner and U. Mosel, Phys. Rev. C {\bf 66}, 055212 (2002).

\bibitem{chiang}
W. Chiang {\it et al.}, Phys. Rev. C {\bf 69}, 065208 (2004).

\bibitem{shklyar}
V. Shklyar, H. Lenske, and U. Mosel, Phys. Rev. C {\bf 72}, 015210 (2005).

\bibitem{diaz}
B. Julia-Diaz {\it et al.}, Nucl. Phys. A {\bf 755}, 463 (2005);
B. Julia-Diaz {\it et al.}, Phys. Rev. C {\bf 73}, 055204 (2006).

\bibitem{sarantsev}
A.V. Sarantsev {\it et al.}, Eur. Phys. J. A {\bf 25}, 441 (2005).

\bibitem{anisovich}
A.V. Anisovich {\it et al.}, Eur. Phys. J. A {\bf 34}, 243 (2007).

\bibitem{g13}
P. Nadel-Turonski, B.L. Berman, D. Ireland, Y. Ilieva, and A. Tkabladze,
JLab experiment E06-103, ``Kaon Production on the Deuteron Using 
Polarized Photons''.

\bibitem{frost}
F.J. Klein and L. Toder, JLab experiment E02-112, ``Search for Missing 
Nucleon Resonances in Hyperon Photoproduction''.

\bibitem{hdice}
F.J. Klein, and A.M. Sandorfi, JLab experiment E06-101, ``$N^*$ Resonances 
in Pseudoscalar Meson Photoproduction from Polarized Neutrons in 
$\vec{H}\cdot\vec{D}$ and a Complete Determination of the
$\gamma n \to K^0\Lambda$ Amplitude''.

\bibitem{raue05}
B.A. Raue and D.S. Carman, Phys. Rev. C {\bf 71}, 065209 (2005).

\bibitem{mohring}
R.M. Mohring {\it et al.}, Phys. Rev. C {\bf 67}, 055205 (2003).

\bibitem{barnes}
T. Barnes, AIP Conf. Proc. {\bf 619}, 447 (2002).

\bibitem{mart} 
H. Haberzettl {\it et al.}, Phys. Rev. C {\bf 58}, R40 (1998); T. Mart and C. 
Bennhold, Phys. Rev. C {\bf 61}, 012201 (2000).  

\bibitem{bradford}
R.K. Bradford {\it et al.} {\it (CLAS Collaboration)}, Phys. Rev. C
{\bf 73}, 035202 (2006).

\bibitem{saphir1} 
M.Q. Tran {\it et al.},  Phys. Lett. B {\bf 445}, 20 (1998).

\bibitem{saghai}
B. Saghai, AIP Conference Proceedings {\bf 594}, 421 (2001).

\bibitem{shklyar2}
V. Shklyar and U. Mosel, Eur. Phys. J. A {\bf 21}, 445 (2004).

\bibitem{mart_sul}
T. Mart and A. Sulaksono, Phys. Rev. C {\bf 74}, 055203 (2006).

\bibitem{ireland}
D.G. Ireland, S. Janssen, and J. Ryckebusch, Nucl. Phys. A {\bf 740},
147 (2004).

\bibitem{glv}
M. Guidal, J.M. Laget, and M. Vanderhaegen, Nucl. Phys. A {\bf 627},
645 (1997);  M. Guidal, J.M. Laget, and M. Vanderhaeghen, Phys. Rev. C 
{\bf 61}, 025204.

\bibitem{sltp}
R. Nasseripour {\it et al.} {\it (CLAS Collaboration)}, Phys. Rev. C
{\bf 77}, 065208 (2008).

\bibitem{knochlein}
G. Kn{\"o}chlein, D. Drechsel, and L. Tiator, Z. Phys. A {\bf 352}, 327 (1995).

\bibitem{laget}
J.M. Laget, Nucl. Phys. A {\bf 579}, 333 (1994).

\bibitem{bebek}
C.J. Bebek {\it et al.}, Phys. Rev. D {\bf 15}, 3082 (1977).

\bibitem{chew}
G.F. Chew {\it et al.}, Phys. Rev. {\bf 106}, 1345 (1957).

\bibitem{bonner}
B.E. Bonner {\it et al.}, Phys. Rev. D {\bf 38}, 729 (1988).

\bibitem{gatto}
R. Gatto, Phys. Rev. {\bf 109}, 610 (1957).

\bibitem{hv99}
E.W. Hughes and R. Voss, Ann. Rev. of Nucl. and Part. Sci. {\bf 49}, 303 
(1999).

\bibitem{mart_code}
T. Mart, code from private communication.

\bibitem{mecking}
B.A. Mecking {\it et al.}, Nucl. Inst. and Meth. A {\bf 503}, 513 (2003).

\bibitem{dcnim}
M.D. Mestayer {\it et al.}, Nucl. Inst. and Meth. A {\bf 449}, 81 (2000).

\bibitem{ccnim}
G.S. Adams {\it et al.}, Nucl. Inst. and Meth. A {\bf 465}, 414 (2001).

\bibitem{scnim}
E.S. Smith {\it et al.}, Nucl. Inst. and Meth. A {\bf 432}, 265 (1999).

\bibitem{ecnim}
M. Amarian {\it et al.}, Nucl. Inst. and Meth. A {\bf 460}, 239 (2001).

\bibitem{mozer}
M.U Mozer and D.S. Carman, CLAS-Note 02-005, see http://www.jlab.org/Hall-B/notes/.

\bibitem{cnote}
D.S. Carman and B.A. Raue, CLAS-Note 02-018, see 
http://www.jlab.org/Hall-B/notes/.

\bibitem{database}
CLAS physics database, http://clasweb.jlab.org/physicsdb.

\bibitem{guidal_code}
M. Guidal, private communication.

\bibitem{ghent_code}
T. Corthals, private communication.

\bibitem{saghai98} 
B. Saghai, Nucl. Phys. A {\bf 639}, 217 (1998).

\bibitem{markowitz}
S. Frullani, F. Garibaldi, J. LeRose, P. Markowitz, and T. Saito, JLab 
experiment E94-108, ``Electroproduction of Kaons up to $Q^2=3$~GeV$^2$''. 

\bibitem{frazer} 
W.R. Frazer, Phys. Rev {\bf 115}, 1763 (1959).

\bibitem{isgur1}
N. Isgur and J. Paton, Phys. Rev. D {\bf 31}, 2910 (1985).

\bibitem{leyaouanc}
A. LeYaouanc {\it et al.}, Phys. Rev. D {\bf 8}, 2223 (1973).

\bibitem{courier}
D.S. Carman, T.S.-H. Lee, M.D. Mestayer, and R.A.Schumacher, CERN Courier 
{\bf 47} No.7, 32 (2007).

\bibitem{boros}
Z. Liang and C. Boros, Phys. Rev. D {\bf 61}, 117503 (2000).

\bibitem{kochelev}
N. Kochelev, Phys. Rev. D {\bf 75}, 077503 (2007).

\bibitem{schumacher}
R. Schumacher, Eur. Phys. J A {\bf 35}, 299 (2008).

\bibitem{tshlee}
T.-S.H. Lee, private communication.

\bibitem{ebac}
Jefferson Laboratory Excited Baryon Analysis Center (EBAC),
see http://ebac-theory.jlab.org/main.htm.

\end{thebibliography}
\end{document}